\documentclass[12pt]{article}
\usepackage{amsmath}
\usepackage{graphicx}
\usepackage{enumerate}
\usepackage{url} 

\usepackage{etoolbox}
\usepackage{amsthm,amsfonts}

\usepackage{titlesec}

\AtBeginDocument{
  \setlength{\abovedisplayskip}{12pt plus 2pt minus 1pt}\relax
  \setlength{\belowdisplayskip}{12pt plus 2pt minus 1pt}\relax
  \setlength{\abovedisplayshortskip}{12pt plus 2pt minus 1pt}\relax
  \setlength{\belowdisplayshortskip}{12pt plus 2pt minus 1pt}\relax
  \everymath{\abovedisplayskip=12pt plus 2pt minus 1pt \belowdisplayskip=12pt plus 2pt minus 1pt}
  \everydisplay{\abovedisplayskip=12pt plus 2pt minus 1pt \belowdisplayskip=12pt plus 2pt minus 1pt}
}

\titlespacing{\section}{0pt}{*0.2}{*0.2}  
\titlespacing{\subsection}{0pt}{*0.2}{*0.2}  

\makeatletter
\patchcmd{\thm@space@setup}
  {\thm@preskip = \topsep}
  {\thm@preskip = 2pt}  
  {}{}
\patchcmd{\thm@space@setup}
  {\thm@postskip = 0pt}
  {\thm@postskip = 2pt}  
  {}{}
\makeatother

\usepackage{capt-of}

\usepackage{natbib}

\setcitestyle{authoryear,open={(},close={)}} 

\usepackage{hyperref}
\hypersetup{
  pdfstartview={XYZ null null 1.50},
   colorlinks=true,       
   citecolor=blue,        
   urlcolor=blue,
   linkcolor=blue,        
}

\usepackage{bm}
\usepackage[ruled,vlined]{algorithm2e}
\SetKwInput{KwInput}{Input}                
\SetKwInput{KwOutput}{Output}              
\usepackage{subcaption}

\usepackage{booktabs,caption}
\usepackage[flushleft]{threeparttable}
\usepackage{multirow}
\usepackage{setspace}

\DeclareFontFamily{OT1}{pzc}{}
\DeclareFontShape{OT1}{pzc}{m}{it}{<-> s * [1.10] pzcmi7t}{}
\DeclareMathAlphabet{\mathpzc}{OT1}{pzc}{m}{it}

\usepackage{placeins}

\usepackage[utf8]{inputenc}
\usepackage[english]{babel}

\begin{document}

\def\spacingset#1{\renewcommand{\baselinestretch}%
{#1}\small\normalsize} \spacingset{1}

\newtheorem{lemma}{Lemma}
\newtheorem{claim}{Claim}
\newtheorem{assumption}{Assumption}
\newtheorem{prop}{Proposition}
\newtheorem{theorem}{Theorem}



{
\title{\bf Scalable Estimation of Multinomial Response Models\\ with Random Consideration Sets
\thanks{Email: \url{chib@wustl.edu} and \url{kenichi.shimizu@ualberta.ca}.}
}
\vspace{-1in} 
\author{
\begin{tabular}[t]{c@{\extracolsep{8em}}c} 
Siddhartha Chib  & Kenichi Shimizu  \\
Olin School of Business & Department of Economics \\ 
Washington University, & University of Alberta, \\
in St.\ Louis, USA &   Canada \\
\end{tabular}
}

  \maketitle
} 

\vspace{-.50in}
\begin{abstract}
A common assumption in the fitting of unordered multinomial response models for $J$ mutually exclusive  categories is that the responses arise from the same set of $J$ categories across subjects. However, when responses measure a choice made by the subject, it is more appropriate to condition the distribution of multinomial responses on a subject-specific consideration set, drawn from the power set of $\{1,2,\ldots,J\}$. This leads to a mixture of multinomial response models governed by a probability distribution over the $J^{\ast} = 2^J -1$  consideration sets. We introduce a novel method for estimating such generalized multinomial response models based on the fundamental result that any mass distribution over $J^{\ast}$ consideration sets  can be represented as a mixture of products of $J$ component-specific inclusion-exclusion probabilities. Moreover, under time-invariant consideration sets, the conditional posterior distribution of consideration sets is sparse. These features enable a scalable MCMC algorithm for sampling the posterior distribution of parameters, random effects, and consideration sets. Under regularity conditions, the posterior distributions of the marginal response probabilities and the model parameters satisfy consistency. The methodology is demonstrated in a longitudinal data set on weekly cereal purchases that cover $J = 101$ brands, a dimension substantially beyond the reach of existing methods.   
\end{abstract}


{\it Keywords: Multinomial response, Bayesian computation, Dirichlet process mixture,  Markov chain Monte Carlo, Metropolis-Hastings algorithm, Posterior consistency}

\spacingset{1.88} 

\section{Introduction}


A common assumption when fitting unordered multinomial response models, whether applied to cross-sectional or longitudinal data, is that the responses stem from the same set of \( J \) mutually exclusive  categories across all subjects. However, this assumption may be questionable, especially when modeling the choices made by human subjects. For example, in fields such as economics and marketing, it is recognized that individuals may select from only a subset of the available alternatives, termed the ``consideration set" (\citealp{Manski1977}; \citealp{HonkaHortacsuWildenbeest2019handbook}). 
Neglecting this heterogeneity in the consideration sets can result in biased parameter estimates in the model (\citealp{BronnenbergVanhonacker1996JMR}; \citealp{ChiangChibNarasimhan1998}; \citealp{Goeree2008}; \citealp{DraganskaKlapper2011}; \citealp{DeLosSantos2018IJIO}; \citealp{Morozov2021MrkSci}; \citealp{CrawfordGriffithIaria2021}). Such biases are problematic because these models are typically employed to understand the impact of covariates on outcomes and inform decision making.

In order to fix ideas, let $\mathcal{C}_i$ represent the latent consideration set for subject $i$. When $J$ alternatives are available, $\mathcal{C}_i$ is a subset of $\{1, \ldots, J\}$, and there are $J^* = 2^J - 1$ possible consideration sets. A priori, $\mathcal{C}_i$ is assumed to be drawn from a probability mass function $\Pr(\mathcal{C}_i = c)$.
When $J$ is small, the direct approach proposed by \cite{ChiangChibNarasimhan1998} is effective. In this approach, all possible consideration sets $1, 2, \ldots, J^*$ are enumerated and assigned unknown probabilities $\pi_1, \pi_2, \ldots, \pi_{J^*}$, which can be estimated using MCMC methods under a Dirichlet prior.
However, when $J$ is large, the model has traditionally been estimated under the assumption that the distribution over consideration sets is determined by $J$ independent attention probabilities. In this framework, it is assumed that each alternative appears independently in any given consideration set \citep{BenAkivaBoccara1995, Goeree2008, ManziniMariotti2014, KawaguchiUetakeWatanabe2021, AbaluckAdams2021}. Specifically, let $q_{ij}$ denote the probability that subject $i$ considers the alternative $j$ for $j = 1, \ldots, J$. The probability that $\mathcal{C}_i = c$ given $\boldsymbol{q}_i=(q_{i1},\ldots,q_{iJ})'$ is then modeled as:
\[
\Pr\left( \mathcal{C}_i = c \mid \boldsymbol{q}_i \right) = \prod_{j \in c} q_{ij} \prod_{j \notin c} \big(1 - q_{ij}\big).
\] 
Although this model is appealing for handling the large $J$ case, the distribution over consideration sets is unrealistic and leads to model misspecification \citep{CrawfordGriffithIaria2021}.

In another approach, the consideration sets are modeled as vectors of 0-1 binary variables \citep{vanNieropPaap2010}. This vector is then modeled by a multivariate probit model (\citealp{AlbertChib1993JASA}, \citealp{ChibGreenberg1998MVP}). Although this  can generate correlation of items in consideration sets, inference is challenging because the number of parameters in the correlation matrix of the multivariate probit model increases quadratically in $J$. 

Given the significant interest in incorporating consideration set heterogeneity in various fields - such as marketing \citep{vanNieropPaap2010, Ching2014simple, KawaguchiUetakeWatanabe2021, Turlo2025discrete}, economics \citep{Goeree2008, Ching2009price,  KashaevPeerEffect2019peer, AgarwalSomaini2022Restud}, transportation science \citep{SwaitBenAkiva1987TransportationB, PaletiTransportationLetters2021}, and psychology \citep{FritsPsychology2022} - there is a pressing need to develop a scalable estimation approach for estimating such generalized multinomial response models. The importance of accounting for consideration set heterogeneity becomes even more critical as $J$ increases, which is precisely the case that current methods struggle to address.
The method we propose is based on two key components. 
The first component is a representation of the probability masses $\pi_{1}, \pi_{2}, \ldots, \pi_{J^{\ast }}$ in terms of a weighted average of products of item-specific inclusion $q_j$ and exclusion $1-q_j$ probabilities, which is based on a result from \cite{DunsonXing2009}. We refer to this approach as a \textit{mixture of independent consideration models}. 
To simulate the latent consideration sets, we introduce a straightforward and intuitive Metropolis-Hastings algorithm. It is important to highlight that, in this context, the consideration sets are latent, unlike in Dunson and Xing (2009), where the categorical variables are observed. This difference necessitates additional steps in both the theoretical derivations and the computational procedure. Another crucial feature of the method is the sparsity of the posterior distribution of the consideration sets, which occurs because sets that do not include the actual choices made by a subject must have a posterior probability of zero \citep{ChiangChibNarasimhan1998}. The scalability of the proposed approach is demonstrated through an application to marketing data involving $J=101$ brands.

We establish two key theoretical results. First, under regularity conditions, as the number of subjects increases, we demonstrate that the posterior distribution of the marginal response probabilities is consistent. Second, under certain additional identification assumptions, the posterior distribution of the model parameters also achieves consistency.

In general, this paper contributes to the expanding literature on high-dimensional demand estimation in statistics and marketing: 
(\citealp{BraunMcAuliffe2010variational}; 
\citealp{ChiongShum2019MS}; 
\citealp{SmithAllenby2019JASA};
\citealp{LoaizaNibbering2022ScalableProbit_JBES};
\citealp{Jiang2024high_MS};
\citealp{WangIaria2024_JEEA};
\citealp{Ershov2024RAND};
\citealp{Amano2018large}). 
To incorporate latent consideration sets, it is necessary to generalize the standard multinomial response model by conditioning the distribution of responses on a latent subject-specific consideration set, which is drawn from the power set of \(\{1,2,\ldots,J\}\). This results in a mixture of multinomial models based on a probability distribution over consideration sets. However, the exponential size of this power set renders the estimation of this mixture of multinomial response models computationally infeasible in general.
Moreover, the proposed method can be interpreted as a generalized multinomial logit (MNL) model, with ``structural zeros'' incorporated in the first layer of its hierarchical structure. In the field of biostatistics, methodologies have been extensively explored to estimate microbial compositions that account for the sparsity due to excessive zero counts (e.g.\ \citealp{Aitchison1982}; \citealp{Martin2015}; \citealp{Liu2020empirical}; \citealp{Cao2020multisample}; \citealp{Paulson2013differential}; \citealp{Chen2016two};
\citealp{Tang2019zero}). More recently, \cite{Zeng2023zero} introduced a zero-inflated probabilistic PCA model designed for high-dimensional, sparse microbiome data sets. Although our paper focuses on a different problem, the proposed method has the potential to be applied in similar contexts,
as we discuss in the concluding section.

The remainder of the article is structured as follows.
Section \ref{sec:approach} introduces the model, while Section \ref{sec:theory} presents the theoretical results.
Section \ref{sec:inference} discusses posterior inference and computational methods.
Section \ref{sec:simulation} reports numerical simulations, and Section \ref{sec:application} applies the methodology to a marketing dataset.
Finally, the concluding section explores the broader implications of the proposed framework.


\section{The approach}\label{sec:approach}
Suppose that we have panel (longitudinal) data with $n$ a priori independent subjects that contains multinomial (polychotomous) responses from a set 
$\mathcal{J}=\{1,\ldots ,J\}$ of $J$ mutually exclusive nominal categories/items as well as some covariates. 
Let $Y_{it}\in \mathcal{J}$ be the measured response for unit $i$ at time $t$, where $i=1,\ldots,n$ and $t=1,\ldots,T_i$. 
Let $\bm w_{it}=\{ \bm w_{ijt}\}_{j \in \mathcal{J}}$, where $\bm w_{ijt}$ is the vector of covariates characterizing the category $j$ for subject $i$ at time $t$. 
Each subject $i$ is associated with a latent consideration set $\mathcal{C}_i$, which is a subset of the entire set of alternatives $\mathcal{J}$. We model the distribution of the observed outcomes using a hierarchical approach. Specifically, we first specify the marginal distribution of the consideration sets and then define the response distribution conditional on a given consideration set. In this framework, we make the following assumptions.

\textbf{Assumption} 1: Consideration sets $\mathcal{C}_i$ vary over subjects but not over time, and the distribution over consideration sets, denoted by $\pi_c = \Pr(\mathcal{C}_i = c)$ for $c \in \mathcal{C}$, the set of all possible consideration sets minus the empty set, is free of covariates.

The assumption of time invariance is relatively mild and aids in inference. It also plays a role in the identification of model parameters. Covariates can be included in the model for consideration sets, but, as noted by \cite{ChiangChibNarasimhan1998}, a covariate-dependent model is difficult to specify without increasing the risk of model mis-specification. 

\textbf{Assumption} 2: For each  $j \in \mathcal{J}$, the responses $Y_{it}$ of subject $i$ given $\mathcal{C}_i$ and random effects $\bm b_i$ are independent over time and follow the multinomial logit model.

Based on Assumptions 1 and 2, the generalized multinomial logit model of interest has the hierarchical form:
\begin{align}
\text{Stage 1:} & \quad \mathcal{C}_i \overset{iid}{\sim} \bm{\pi},  \nonumber \\
\text{Stage 2:} & \quad \bm{b}_i \overset{iid}{\sim} \mathcal{N}(\bm{0}, \bm{D}), \label{eq:hierarchical_model} \\
\text{Stage 3:} & \quad \Pr(Y_{it} = j \mid \bm{\beta}, \bm{w}_{it}, \mathcal{C}_i, \bm{b}_i) = 
\left\{ 
  \begin{array}{ c l }
    \frac{
\exp \left( \bm{x}_{ijt}'\bm{\beta} + \bm{z}_{ijt}'\bm{b}_i \right)
}{
\sum_{\ell \in \mathcal{C}_i} \exp \left( \bm{x}_{i\ell t}'\bm{\beta} + \bm{z}_{i\ell t}'\bm{b}_i \right)
} & \quad \textrm{if } j \in \mathcal{C}_i  \\
    0                 & \quad \textrm{otherwise}
  \end{array} 
\right.  \quad t = 1, \ldots, T_i, \nonumber 
\end{align}
for $i = 1, \ldots, n$, where $\bm \pi = \{\pi_c : c \in \mathcal{C}, 0 \leq \pi_c \leq 1, \sum_{c \in \mathcal{C} } \pi_c =1 \}$  denotes
the collection of probabilities associated with all possible consideration sets, 
and $\bm b_i$ are random effects normally and independently distributed across subjects with zero mean and unknown covariance matrix $\bm{D}$. The covariates are denoted by $\bm{w}_{it} = \{ \bm{x}_{ijt}, \bm{z}_{ijt} \}_{ j \in \mathcal{J}}$, where $\bm{x}_{ijt} \in \mathbb{R}^{d_{x}}$ and $\bm{z}_{ijt} \in \mathbb{R}^{d_{z}}$. Stage 1 can be interpreted as introducing another layer of random effects, where heterogeneity arises from the random consideration sets.

Letting $\Pr (\bm Y_{i} \vert \bm \theta, \bm w_{i}, \mathcal{C}_{i}=c)$ denote the distribution of outcomes $\bm Y_i = (Y_{1i},\ldots,Y_{iT_i})$ of subject $i$ marginalized over the random effects given covariates $\bm w_i=\{\bm w_{i1},\ldots,\bm w_{iT_i}\}$, the distribution of responses takes the finite mixture form:
\[
\Pr (\bm Y_{i} \vert \bm \theta, \bm w_{i})
=
\sum_{c\in \mathcal{C}}
\pi_c 
\Pr (\bm Y_{i} \vert \bm \theta, \bm w_{i}, \mathcal{C}_i=c).  
\]
This can be seen as a generalized multinomial logit response model. 

Assumptions 1 and 2 imply  time-invariant consideration sets, conditional independence of responses, and full support of the conditional response probabilities given consideration sets. These conditions, along with additional assumptions detailed below, establish the point identification of the model parameters \citep{AguiarKashaev2024identification}  in the model that excludes random effects. Furthermore, in Theorem 2 of Section \ref{sec:theory}, we show the posterior consistency of the parameters in this case. 



\subsection{The latent consideration sets}

To fix  notation, let $\mathcal{C}$ represent the collection of all possible consideration sets, which corresponds to the power set of $\mathcal{J} = \{1, \ldots, J\}$, excluding the empty set. The consideration set for subject $i$ is indicated by $\mathcal{C}_i = c$, where $c \in \mathcal{C}$. For example, when $J = 3$, $\mathcal{C} = \{\{ 1 \}, \{2 \}, \{3\}, \{1,2\}, \{1,3\},\{2,3\}, \text{and} \, \{1,2,3\} \}$, and $c$ is one of these elements.
Furthermore, by $\bm C_i=(C_{i1},\ldots,C_{iJ})'$, we mean a $J \times 1$ multivariate binary vector where $C_{ij}=1$ if category $j$ is in the consideration set, and $0$ otherwise. In the example of $J=3$,  
$\mathcal{C}_i=\{1\}$ is equivalent to $\bm C_i=(1,0,0)'$ and 
$\mathcal{C}_i=\{1,3\}$ is equivalent to $\bm C_i=(1,0,1)'$ etc.  
In the following, we use the two notations interchangeably depending on the context. 
Researchers sometimes include an outside option in the model that is always considered by each subject. We can incorporate this into our framework by adding a $(J+1)$th category and fixing $C_{iJ+1}=1$ for all $i$.
Our goal is to put a probability distribution on $\mathcal{C}$ that
is rich enough to accommodate dependencies while maintaining scalability. 

\subsection{Dimensionality reduction via tensor decomposition }
We now review the factor decomposition technique that we employ to specify the distribution
over consideration sets.
\cite{DunsonXing2009} consider modeling large contingency tables that, for example, represent DNA sequences, each of which is defined as a collection of $J$ categorical variables, each having $d_j$ possible values $j=1,\ldots,J$, where $J$ is large. A realization of the contingency table can be
expressed as a vector $(a_1,\ldots,a_J)'$, 
where $a_j \in \{1,\ldots,d_j\}$ for $j=1,\ldots,J$. 
The true distribution of the contingency tables is a probability tensor 
$\bm \pi= \{  \pi_{a_1a_2\cdots a_J}, a_j=1,\ldots,d_j, j=1,\ldots,J \}$,  where  
$0 \leq \pi_{a_1a_2\cdots a_J} \leq 1$ and 
$\sum_{a_1=1}^{d_1}\cdots \sum_{a_J=1}^{d_J}  \pi_{a_1a_2\cdots a_J} = 1$. 
Note that consideration sets can be seen as contingency tables with $d_j=2$ for all $j$. 
Generally, there are a large number of elements in the tensor $\bm \pi$, $d_1\times \cdots \times d_J$, when $J$ is large.
\cite{DunsonXing2009} show that $\bm \pi$ can be expressed as a finite mixture of rank 1 tensors. 
We describe this result for the special case that corresponds to modeling consideration sets.

\begin{lemma}[Exact matching of consideration set probabilities]\label{lem:dunson_xing}
Let $\bm{\pi}$ be the probability mass distribution over the consideration sets: it is a collection of probabilities $\{ \pi_c = \Pr\left( \mathcal{C}_i = c  \right): c \in \mathcal{C} \}$, where 
$0\leq \pi_c \leq 1$
and 
$\sum_{c\in \mathcal{C}} \pi_c =1$.
Then there are 
$K \in \mathbb{Z}^+$, 
$
\bm \omega = (\omega_1,\ldots,\omega_K) \in \Delta^{K-1}, 
$
$
\bm q_h = (q_{h1},\ldots,q_{hJ})', \quad h = 1,\ldots,K, \quad q_{h j}\in[0,1]
$
such that for each $c\in \mathcal{C}$, 
\begin{equation}
\pi_c 
=\sum_{h=1}^K \omega_h 
\left\{ 
\prod_{j \in c} q_{h j} \prod_{ j \notin c} \big(1-q_{h j} \big)
\right\}. 
\label{eq:mixture_model}
\end{equation}
\end{lemma}
This result states that a mixture of $K$ independent consideration models can model an arbitrary distribution over the $J^{\ast} = 2^J - 1$ possible consideration sets.  Within each component $h$, items are included in or excluded from a consideration set $c$ according to an independent consideration model  defined by a vector of attention probabilities $\bm q_h=(q_{h1},\ldots,q_{hJ})'$. Therefore, the number of parameters needed to model the probabilities in $\bm{\pi}$ is reduced from $J^{\ast}$ to $K \times J + (K - 1)$, which scales linearly with $J$.

\subsection{Infinite mixture of independent consideration models}

Building on this result, we model the $J$-dimensional latent vectors \( \{ \bm{C}_i \} \) as a mixture of independent probabilities. Since the number of components \( K \) in \eqref{eq:mixture_model} is unknown, we follow \cite{DunsonXing2009} and use a Dirichlet process (DP) prior \citep{Ferguson1973BayesianNonparametrics} to induce an infinite mixture model.
One key difference from \cite{DunsonXing2009} is that their categorical variables (contingency tables) are observed, while the corresponding consideration sets are latent. This difference leads to differences in the theoretical analysis (Section \ref{sec:theory}) and in the posterior simulation approach (Section \ref{sec:inference}).

In our approach, we do not estimate $K$. This is because existing methods for consistently estimating $K$, such as those proposed by \cite{KwonMbakop2021AoS}, may not be applicable when the variables modeled by the mixture are latent. Posterior consistency in our framework only requires that the prior on $K$ has positive mass for all positive integers. Posterior inferences on model parameters and their functions (e.g., predictions) automatically account for uncertainty regarding the value of $K$.

Assume that $\{\bm C_i\}$ is i.i.d. with density $f({}\cdot{} \vert G) = \int
\prod_{j=1}^J q_j^{C_{ij}} \left( 1-q_j\right)^{1-C_{ij}} dG(\bm{q})$. 
The discrete mixing distribution $G$ is modeled by a DP prior with a concentration parameter $\alpha$ and a specified base probability measure $G_0$ that depends on a hyperparameter $ \underline{\bm \phi}_q$. 
Equivalently, by using the stick breaking construction (\citealp{Sethuraman1994constructiveDP}), we have the following representation: $\bm C_i$'s are i.i.d.\ with the density for the infinite mixture of independent consideration models: 
\begin{equation}
\Pr(\bm C_i = \bm c_i)=
\sum_{h=1}^\infty  \omega_h 
\prod_{j=1}^J 
\left\{
q_{hj}^{c_{ij}} \left( 1-q_{hj}\right)^{1-c_{ij}} 
\right\},
\label{eq:model_infinitemixture}   
\end{equation}
where 
$\bm c_i=(c_{i1},\ldots,c_{iJ})'$, 
$\omega_1= V_1,  \omega_h = V_h \prod_{\ell<h} (1-V_\ell),  h=2,\ldots,\infty,$
$V_h\overset{iid}{\sim}\text{Beta}(1,\alpha)$, and 
$\bm q_h \overset{iid}{\sim} G_0(\ \cdot \ \vert \underline{\bm \phi}_q),  h=1,\ldots,\infty$,
with 
$\bm q_h=(q_{h1},\ldots,q_{hJ})'$ being the vector of attention probabilities specific to the component $h$.
A priori, the first few weights dominate and cover most of the probability mass, which are then adjusted by the data. 
Although the model \eqref{eq:model_infinitemixture} includes infinitely many components, typically only a small number of  distinct values for $\bm{q}_h$ are imputed. 

For the baseline distribution $G_{0}$, we assume that
$q_{hj} \sim G_{0j}$ independently for $j=1,\ldots,J$ and $h=1,\ldots,\infty$.
Specifically, we assume that
$q_{hj}\sim \text{Beta}(\underline{a}_{q_j},\underline{b}_{q_j})$, independently over $j=1,\ldots,J$, for $h=1,\ldots,\infty$, and we define $\underline{\bm \phi}_q=(\underline{\bm a}_q,\underline{\bm b}_q)$ with $\underline{\bm a}_q =(\underline{a}_{q_1} ,\ldots,\underline{a}_{q_J})'$ and
$\underline{\bm b}_q =(\underline{b}_{q_1},\ldots,\underline{b}_{q_J})'$.
Note that
$\underline{\bm \phi}_q=(\underline{\bm a}_q,\underline{\bm b}_q)$ are the hyperparameters chosen by the user. We discuss this in more detail in the Supplementary Material.
We complete the model specification by assuming the prior distribution for the DP concentration parameter
$
\alpha \sim \text{Gamma}(\underline{a}_\alpha, \underline{b}_\alpha),
$
where $(\underline{a}_\alpha, \underline{b}_\alpha)$ are the hyperparameters chosen by the user. 
For smaller values of $\alpha$, $\omega_h$ decreases toward zero more rapidly as $h$ increases, so that the prior favors a sparse representation with most of the weight on a few components. We allow the data to inform about $\alpha$ and, therefore, an appropriate degree of sparsity.

\section{Theoretical results}\label{sec:theory}
We establish two key results. For simplicity, let $T_i=T$, $\forall i$ and suppose that $T\geq 1$ is fixed and $n\to \infty$. 
In Theorem \ref{thm:consistency_model_parameters} we show that the posterior of the marginal response probabilities is consistent, and in Theorem
\ref{thm:consistency_model_parameters_identified}, we show that the posterior of the model parameters is consistent when $T$ is large enough and the model does not include random effects.

Let $\bm \theta=\{ \bm \beta, \bm D \}$  denote the parameters in the response model. 
Also, recall that the distribution over the consideration sets is denoted by
$\bm \pi=\{ \pi_c: c \in \mathcal{C}\}$, where $0\leq \pi_c \leq 1$ and $\sum_{c \in \mathcal{C}}\pi_c=1$. 
Define the  probability that the sequence of items $\bm y=(y_1,\ldots,y_T)'\in \mathcal{J}^T$ is chosen 
conditional on covariates $\bm w_i =\{ \bm w_{i1},\dots,\bm w_{iT} \}$ taking some specific value $\bm w =\{\bm w_1,\ldots,\bm w_T\} \in \mathbb{R}^{TJ(d_x+d_z)}$:
\[
p_{\bm{\theta}, \bm{\pi} } (\bm y\vert \bm w)  \equiv \sum_{c\in \mathcal{C}} \pi_{c}  \Pr\left( \bm Y_{i} = \bm  y \vert \bm \theta, \bm w, c \bm \right),
\]
where the response probability given a consideration set $c$ is 
\[
\Pr (\bm Y_{i}= \bm y \vert \bm \theta, \bm w, c)
=
\int 
\prod_{t=1}^T
\Pr(Y_{it} = y_t \mid \bm{\beta}, \bm{w}_{t}, \mathcal{C}_i=c, \bm{b}_i)
\phi(\bm b_i \vert \bm 0, \bm D) d \bm b_i 
, 
\]
where the integrand is defined in 
\eqref{eq:hierarchical_model}. 
The data set contains responses $\bm y_i=\{y_{it}\}$ and covariates $\bm w_i=\{\bm w_{it} \}$ and we let $\bm D^n=\{(\bm y_i,\bm w_i): i=1,\ldots,n \}$. The covariates $\bm w_i $ are i.i.d.\ and follow an unknown distribution with density $g^*$ with support $\mathcal{W}\subset \mathbb{R}^{TJ(d_x+d_z)}$. We do not model the covariate distribution. 
Conditional on covariates, responses are generated from the collection of the data-generating response probabilities $\bm p^* =\{p_{\bm{\theta}^*, \bm{\pi}^* } (\bm y \vert \bm w) \}_{\bm y \in \mathcal{J}^T,\bm w \in \mathcal{W}}$, where $\bm \theta^*$ denotes the true response model parameter
and $\bm \pi^*=\{ \pi^*_c: c \in \mathcal{C}\}$  denotes the true probability mass function over consideration sets. We emphasize that $\bm \pi^*$ does not have to be a finite mixture. 
The joint probability measure implied by $\bm p^*$ and $g^*$ is denoted by $F_0$.
For $\varepsilon>0$, define a Kullback-Leibler neighborhood of $\bm p^*$ as 
\[
        KL_\varepsilon(\bm p^*) = 
        \left\{ 
        (\bm \theta, \bm \pi): 
        \int \sum_{\bm y \in \mathcal{J}^T} \log 
        \left(
        \frac{
        p_{\bm{\theta}^*, \bm{\pi}^* } (\bm y\vert \bm w)}{
        p_{\bm{\theta}, \bm{\pi}     } (\bm y\vert \bm w)
        } 
        \right) 
        p_{\bm{\theta}^*, \bm{\pi}^* } (\bm y\vert \bm w) g^* (\bm w) d\bm w < \varepsilon
        \right\}.
\]
It is essentially a set of $(\bm \theta,\bm \pi)$ that makes $p_{\bm{\theta}, \bm{\pi}}$ close to $p_{\bm{\theta}^*, \bm{\pi}^*}$. 

Given a $K\in \mathbb{Z}^+$, define $\bm \phi_{1:K}=\{\omega_h,\bm q_h: h=1,\ldots,K\}$, the collection of all component-specific parameters, where $\bm q_h=(q_{h1},\ldots, q_{hJ})'$. 
Note that by Lemma \ref{lem:dunson_xing}, there exist $\{K, \tilde{\bm \phi}_{1:K}\}$, which may not be unique, such that  
$
\pi_c^* 
=\sum_{h=1}^{K} \tilde{\omega}_h 
\left\{ 
\prod_{j \in c} \tilde{q}_{h j} \prod_{ j \notin c} \big(1-\tilde{q}_{h j} \big)
\right\},$
$\text{ for all } c\in \mathcal{C}, 
$
and 
the KL divergence is zero at $\{\bm \theta^*, K, \tilde{\bm \phi}_{1:K} \}$. 
In the following lemma, we establish that the KL divergence can be made arbitrarily small in  sufficiently small neighborhoods of $(\bm \theta^*, \tilde{\bm \phi}_{1:K})$. 
Define the model induced probability for a consideration set $c\in \mathcal{C}$:
$
\pi(c\vert K, \bm \phi_{1:K}) = \sum_{h=1}^K \omega_h \prod_{j\in c} q_{hj} \prod_{j \notin c}(1- q_{hj}),
$
and the model induced marginal response probability as
    \[
     p(\bm y \vert \bm w; \bm \theta, K, \bm \phi_{1:K})
     =\sum_{c\in \mathcal{C}} 
     \pi(c \vert K, \bm \phi_{1:K}) 
     \Pr(\bm Y_{i}= \bm y \vert \bm \theta, \bm w, c).
    \]
\begin{lemma}\label{lem:KL_continuity}
Suppose:  
    (i)  $\bm \beta^* \in \text{interior}(\mathcal{B})$, where $\mathcal{B}$ is a compact subset of $\mathbb{R}^{d_x}$ and $\bm D^*$ is positive definite, and 
    (ii) $\mathcal{W}$ is compact.  
Then $\forall \varepsilon>0$, 
$\exists$ an open neighborhood $\mathcal{O}$ of $\bm \theta^*$, 
$K \in \mathbb{Z}^+$, and an open  neighborhood $\mathcal{P}^K$ 
such that for any $\bm \theta \in \mathcal{O}$ and $\bm \phi_{1:K} \in \mathcal{P}^K$,
\[
    \int \sum_{\bm y \in \mathcal{J}^T} \log 
    \left(
    \frac{
    p_{\bm{\theta}^*, \bm{\pi}^* } (\bm y\vert \bm w)}{
    p(\bm y\vert \bm w; \bm \theta,K,\bm \phi_{1:K})
    } 
    \right) 
    p_{\bm{\theta}^*, \bm{\pi}^* } (\bm y\vert \bm w) g^* (\bm w) d\bm w < \varepsilon.
\]
\end{lemma}
The proof can be found in the Appendix. 
\noindent Let $\Pi(\cdot)$ denote the prior  for  the response model parameter $\bm \theta$ and the distribution of 
consideration sets $\bm \pi$.  

\begin{theorem}\label{thm:consistency_model_parameters}
Suppose conditions (i) and (ii) of Lemma \ref{lem:KL_continuity}. 
Suppose  (iii) for  any open neighborhood  $\mathcal{O}$ of $\bm \theta^*$, 
and for any $K, \bm \phi_{1:K}$, and an open neighborhood $\mathcal{P}^K$ of $\bm \phi_{1:K}$, 
$\Pi(\bm \theta \in \mathcal{O}, \bm \phi_{1:K} \in \mathcal{P}^K, K)>0$. 
Then, for all weak neighborhoods $\mathcal{U}$ of $\bm p^*$,  as $n\to \infty$,
$
\Pi\left( \mathcal{U} \vert  \bm D^n \right)\to 1 \text{ a.s. } F_0^\infty.
$
\end{theorem}

\begin{proof}[Proof of Theorem \ref{thm:consistency_model_parameters}]
%
By Schwartz's theorem (\citealp{GhosalVaart2017fundamentals}, ch.6), the result follows if we show that $\Pi( KL_\varepsilon(\bm p^*))>0$. 
By Lemma \ref{lem:KL_continuity}, there exist open neighborhoods 
$\mathcal{O}$ and $\mathcal{P}^K$ on which the KL divergence can be made sufficiently small. 
The lemma combined with a prior that places positive mass on open neighborhoods (condition iii) implies that $\Pi(KL_\varepsilon(\bm p^*))>0$. 
\end{proof}

This result shows that the model-induced response probability in the limit converges to the true data-generating process. 
A similar result is proved in \cite{DunsonXing2009}, Theorem 2, but for the case in which the categorical variables are observed 
and there are no covariates. Because our setup relaxes both of those conditions, we have a more involved proof that involves  
the KL-divergence (Lemma \ref{lem:KL_continuity}).  \cite{NoretsShimizu2024} also establish a related result for semiparametric dynamic discrete choice models, but our proof strategy is different, due to the random effects, continuous covariates, and a different model. Last, the compactness assumption (ii) is common in Bayesian nonparametric estimation, 
and condition (iii) of Theorem 1 is satisfied by our DP prior for $\omega_h$'s and the Beta prior for $q_{hj}$'s, following Dunson and Xing (2009). 

We now address the possibility that multiple parameter pairs \((\bm{\theta}, \bm{\pi})\) may be consistent with the true response probabilities. This relates to the issue of \textit{partial identification} \citep{MasatliogluNakajimaOzbay2012revealed, CattaneoMaMasatliogluSuleymanov2020, BarseghyanCoughlinMolinariTeitelbaum2021, Lu2022}, where point identification holds only under specific conditions \citep{DardanoniManziniMariottiTyson2020ECMA, AbaluckAdams2021, BarseghyanMolinariThirkettle2021discrete}. 
Following \citet{AguiarKashaev2024identification}, we impose the assumption that the panel is sufficiently \textit{long} and that \textit{random effects are absent}. Under these conditions, we show that the two sources of variation in responses—differences in utility and differences in consideration sets—can be separately identified. Formally, we show in the next theorem that the posterior distribution contracts to within an arbitrarily small ball around $(\bm \beta^*, \bm \pi^*)$ under the distance function $d((\bm \beta, \bm \pi), (\bm \beta', \bm \pi'))=\max \{||\bm \pi - \bm \pi'||_1, ||\bm \beta-\bm \beta'||_2 \}$.

\begin{theorem}\label{thm:consistency_model_parameters_identified}
Suppose (i) the model does not contain random effects; (ii) the parameter $\bm{\beta}$ belongs to $\mathcal{B}$, a compact subset of $\mathbb{R}^{d_x}$, with $\bm{\beta}^* \in \text{interior}(\mathcal{B})$; (iii) $\mathcal{W}$ is compact; and (iv) for any open neighborhood $\mathcal{O}$ of $\bm{\beta}^*$, any $K$, any $\bm{\phi}_{1:K}$, and any open neighborhood $\mathcal{P}^K$ of $\bm{\phi}_{1:K}$, it holds that
$\Pi(\bm{\beta} \in \mathcal{O}, \bm{\phi}_{1:K} \in \mathcal{P}^K, K) > 0$.
Then, if the number of periods $T$ satisfies $\lfloor (T-3)/2 \rfloor \geq J$, we have that for all $\varepsilon > 0$, as $n \to \infty$,
$\Pi\left( (\bm{\beta}, \bm{\pi}): d((\bm{\beta}, \bm{\pi}), (\bm{\beta}^*, \bm{\pi}^*)) < \varepsilon \mid \bm{D}^n \right) \to 1 \quad \text{a.s. } F_0^\infty.
$
\end{theorem}
\begin{proof}[Proof of Theorem \ref{thm:consistency_model_parameters_identified}]
The proof is by Schwartz's theorem. 
The identification assumption together with Assumptions 1-2 ensures that $p_{\bm \beta,\bm \pi} \ne p_{\bm \beta',\bm \pi'}$ whenever $(\bm \beta,\bm \pi)\ne(\bm \beta',\bm \pi')$ (\citealp{AguiarKashaev2024identification}). 
Identifiability, continuity of $p_{\bm \beta,\bm \pi}$ in $(\bm \beta,\bm \pi)$ for the total variation norm (Lemma SA.3), and compactness of the parameter space ensure the existence of consistent tests (\citealp{vanderVaart2000asymptotic}, Lemma 10.6). 
The approximation result (Lemma \ref{lem:KL_continuity})   without random effects can be established as a special case, and together with the regularity conditions on the prior distribution, the KL-support condition holds. 
\end{proof}

%


We remark that in Theorem 2 we suppose a model without random effects,  though we use random effects in our modeling. The complication in having both is that latent consideration sets in our model operate similarly to random effects and introduce dependence across time. Disentangling these two sources of dependence at a theoretical level requires a stronger condition on $T$, though the precise details are not straightforward to establish. We leave this extension for future work. Nonetheless, the numerical experiments in the Supplementary Material indicate that the convergence described in the theorem holds more generally, as we observe convergence to the true values even in the presence of random effects.

\section{Inference}\label{sec:inference}

Let $\bm{Y}_i=(Y_{i1},\ldots,Y_{iT_i})'$ and $\bm{y}_i=(y_{i1},\ldots,y_{iT_i})'$ be the sequence of random responses made by unit $i$ over $T_i$ periods and its observed counterpart. 
Define 
\begin{equation}
p(\bm{Y}_i=\bm{y}_i \vert  \bm{\beta}, \bm{b}_i, \bm w_{i}, \bm{C}_i ) = \prod_{t=1}^{T_i} \Pr(Y_{it}=y_{it} \vert  \bm \beta, \bm{b}_i, \bm w_{it},\mathcal{C}_i ), \label{eq:choice_prob_T}
\end{equation}
where 
$\bm w_i = \{\bm w_{i1},\ldots, \bm w_{iT_i}\}$ 
and 
$\Pr(Y_{it}=y_{it} \vert  \bm \beta, \bm{b}_i, \bm w_{it},\mathcal{C}_i )$ 
is 
\begin{equation}
\Pr (Y_{it}=j \vert \bm \beta, \bm b_i, \bm w_{it}, \mathcal{C}_{i})
=
\frac{
\exp \left( \bm x_{ijt}'\bm \beta+\bm z_{ijt}'\bm b_i\right) 
}{
\sum_{\ell \in \mathcal{C}_i}\exp 
\left( \bm x_{i\ell t}'\bm \beta+\bm z_{i\ell t}'\bm b_i\right) 
}
\text{ \ if $j\in \mathcal{C}_i$, and $0$ otherwise}.
\label{eq:mixed_logit_given_bi}
\end{equation}
Note that $\bm C_i$ is the conditioning variable on the left side of \eqref{eq:choice_prob_T}, while $\mathcal{C}_i$ is on the right side. Although the two objects represent the same information, the $J$-dimensional vector $\bm C_i$ is easier to use when we discuss posterior sampling of individual consideration sets. Hence, we use $\bm C_i$ to define the individual's contribution to the likelihood. 
Let $\bm Y=\{\bm{Y}_1,\ldots,\bm{Y}_n\}$ and $\bm y=\{\bm{y}_1,\ldots,\bm{y}_n\}$ denote the random and observed sequences of the responses made by all units, and let
$\bm W=\{\bm w_1,\ldots,\bm w_n\}$ be the observed covariates. 
Then the likelihood  
conditional on 
the common fixed-effects $\bm \beta$, 
the random effects  $\bm b = (\bm{b}_1,\ldots,\bm{b}_n)'$, 
the covariates $\bm W$, 
and the latent consideration sets $\bm C=(\bm{C}_1,\ldots,\bm{C}_n)$ is  given by 
\begin{equation}
p(\bm Y= \bm y \vert  \bm{\beta}, \bm b, \bm W, \bm C) =
\prod_{i=1}^n p(\bm{Y}_i=\bm{y}_i \vert  \bm{\beta}, \bm{b}_i, \bm w_i, \bm C_i  ). \label{eq:choice_prob_Tn}
\end{equation}

We complete the model by specifying standard prior distributions for the parameters in the response model:
$\bm{\beta} \sim \mathcal{N}_{d_x}\left( \bm{0}, \underline{\bm{V}}_\beta \right)$ and 
$\bm{D}^{-1} \sim \text{Wishart} \left( \underline{v}, \underline{\bm{R}}\right)$, indepdently, 
a normal distribution for  $\bm{\beta}$, and
an inverse Wishart distribution for $\bm D$ with degrees-of-freedom parameter $\underline{v}$ and scale matrix $\underline{\bm{R}}$. 
The hyperparameters $( \underline{\bm{V}}_\beta, \underline{v}, \underline{\bm{R}})$ are chosen by the user. 

\subsection{Posterior distribution}
For the mixture model on the latent consideration sets $\bm C=(\bm{C}_1,\ldots,\bm{C}_n)$, let $S_i \in\{1,2,\ldots\}$ be the latent cluster assignment such that
$C_{ij} \vert S_i=h\sim \text{Bernoulli}(q_{hj})$, independently $j=1,\ldots,J$, for $i=1,\ldots,n$.
We have the latent consideration sets $\bm C$, 
the common fixed-effects $\bm \beta$, 
the random effects $\bm b$,
the corresponding covariance matrix $\bm D$,  
the DP parameters $\bm V =(V_1,V_2,\ldots)$ as well as $\bm Q=(\bm q_1,\bm q_2,\ldots)$,  
the DP cluster assignment variables $\bm S =(S_1,\ldots,S_n)$, and 
the DP concentration parameter $\alpha$. 
Let $\pi(\cdot)$ denote the prior density.  
Then, from the Bayes theorem, the posterior density of interest is 
\begin{equation}
p\big( \bm C,  \    \bm S, \bm V, \bm Q, \alpha,  \bm \beta, \bm b, \bm D \big\vert \bm y , \bm W \big)
\propto 
p\big(\bm y \big\vert   \bm \beta, \bm b, \bm W, \bm C  \big) 
\cdot 
p( \bm \beta, \bm b, \bm D)
\cdot 
p\big( \bm C,   \bm S, \bm Q,\bm V,  \alpha  \big)  ,
 \label{eq:joint_distribution}
\end{equation}
where the first term is given by  \eqref{eq:choice_prob_Tn} and 
only the last term  is associated with the DP prior. 

We sample from the posterior distribution using a tailored Markov Chain Monte Carlo (MCMC) algorithm. 
The method is designed for scalability and consists of simple and intuitive steps. 
Posterior inference is then based on the sampled values
\begin{equation}
\big\{ V_h^{(g)}, \bm q_h^{(g)}, S_i^{(g)}, \alpha^{(g)}, \bm C_i^{(g)}, \bm \beta^{(g)}, \bm b_i^{(g)}, \bm D^{(g)} \big\}, \quad g = 1,\dots,G,
\end{equation}
where $G$ is the number of MCMC draws beyond a suitable burn-in period.

\subsection{Simulation of consideration sets}\label{sec:conditional_CS}
We now focus on sampling the conditional distribution of consideration sets. 
The other steps in the MCMC simulation follow from standard calculations and
are given in the Supplementary Material.
From Equation \eqref{eq:joint_distribution}, the full conditional distribution of  $\bm C_i$  is 
\begin{equation}
\pi( \bm C_i \vert \bm \beta,\bm  b_i, \bm q_{S_i}, S_i, \bm y_i , \bm w_i) 
\propto \  
p\big(\bm Y_{i}=\bm y_{i} \big\vert \bm \beta, \bm b_i, \bm w_i, \bm C_i \big) 
\cdot
\prod_{j=1}^J q_{S_ij}^{C_{ij}} (1-q_{S_ij})^{1-C_{ij}}    , 
\label{eq:conditional_C1}    
\end{equation}
where the proportionality sign is with respect to $\bm C_i$, and the first term is defined in \eqref{eq:choice_prob_T}. 
Importantly, consideration sets that exclude any observed response made by subject $i$ receive zero posterior probability (see Table \ref{table:increasingT} for an example). 
This is because the first term on the left-hand side of \eqref{eq:conditional_C1} is zero for these consideration sets. 
This desirable feature of our approach is based on \cite{ChiangChibNarasimhan1998}. 
In contrast, in many existing methods, every consideration set receives a strictly positive probability, as pointed out by \cite{CrawfordGriffithIaria2021}. 
%
%
Now, due to the independence structure in \eqref{eq:conditional_C1} over $j=1,\ldots,J$, 
\[
\pi( C_{ij} \vert \bm  C_{i}\setminus \{ j \}, \bm  \beta,\bm  b_i,\bm  q_{S_i}, S_i,\bm  y_i, \bm w_i ) 
\propto \ 
p\big(\bm  Y_{i}=\bm y_{i} \big\vert  \bm  \beta,\bm  b_i, \bm w_i, \bm C_i \big) 
\cdot 
q_{S_ij}^{C_{ij}} (1-q_{S_ij})^{1-C_{ij}}    , 
\]
where $\bm  C_{i}\setminus \{ j \}$ denotes $\bm C_i$ without the coordinate $j$.
To sample from this distribution, we employ the Metropolis-Hastings (M-H) algorithm \citep{ChibGreenberg1995understandingMH}. 
An effective implementation of this approach is detailed in Algorithm \ref{alg:MH_for_CS_main}.
\FloatBarrier
\begin{algorithm}[ht]
\caption{M-H step for Sampling Consideration Sets}
\KwInput{The current draws at the $g$th iteration $\big\{\bm  C_i^{(g)} \big\}, \{\bm  q_h^{(g)} \}, \{ S_i^{(g)}=h \}, \bm \beta^{(g)}, \big\{\bm b_i^{(g)} \big\}$}
\KwOutput{The updated consideration sets $\big\{\bm  C_i^{(g+1)} \big\}$}
\For{$i \in \{1,\ldots,n\}$}{
\For{$j \in \{1,\ldots,J\}$}{
    1) Propose $\tilde{C}_{ij} \sim \text{Bernoulli}(q_{hj}^{(g)})$ and define  $\bm C_{i}^{(1)}=(C_{i1}^{(g+1)},\ldots,C_{ij-1}^{(g+1)},\tilde{C}_{ij},C_{ij+1}^{(g)},\ldots,C_{iJ}^{(g)})'$  

  \quad \\
    2) Accept  $\tilde{C}_{ij}$ with  probability 
\[
\min\left\{  \frac{p\big(\bm  Y_{i}=\bm  y_{i} \big\vert  \bm  \beta^{(g)},\bm  b_i^{(g)}, \bm w_i, \bm C_i^{(1)} \big) }{ p\big(\bm  Y_{i}=\bm y_{i} \big\vert \bm \beta^{(g)}, \bm b_i^{(g)},  \bm w_i,\bm C^{(0)}_i \big)  } ,1   \right\},
\]
where $\bm C_{i}^{(0)}=(C_{i1}^{(g+1)},\ldots,C_{ij-1}^{(g+1)},C_{ij}^{(g)},C_{ij+1}^{(g)},\ldots,C_{iJ}^{(g)})'$. \\
Otherwise, set $C_{ij}^{(g+1)}=C_{ij}^{(g)}$
}
}
\label{alg:MH_for_CS_main}
\end{algorithm}

In Step 1 of Algorithm \ref{alg:MH_for_CS_main}, we generate a proposal from a one-dimensional Bernoulli distribution. In Step 2, given the current state $\bm{C}_i^{(0)}$ and the proposed state $\bm{C}_i^{(1)}$, the acceptance probability is computed as the ratio of the likelihood contributions for subject $i$. This Metropolis-Hastings step is valid because the likelihood $p\big(\bm Y_i = \bm y_i \mid \bm \beta, \bm b_i, \bm w_i, \bm C_i \big)$ is uniformly bounded. See \cite{ChibGreenberg1995understandingMH} (p.\ 330, the third algorithm) for more discussion. 
In practice, we update the states in a random order within each MCMC iteration. In addition, the computational burden is minimized by parallelizing the loop on the $n$ subjects. 

Finally, the proposed Metropolis-Hastings step exhibits an important sparsity property. 
Suppose that an alternative $j$ was not chosen by the subject $i$ in any period (otherwise, it must be in the consideration set for $i$ and $C_{ij}=1$). 
Depending on the current $C_{ij}^{(g)}$, and the proposed $\tilde{C}_{ij}$, there are four possible moves in the  M-H step. 
First, if $\tilde{C}_{ij}=C_{ij}^{(g)}=1$ or $\tilde{C}_{ij}=C_{ij}^{(g)}=0$, then the proposed value is accepted with probability one. 
Second, if  $\tilde{C}_{ij}=0$ and $C_{ij}^{(g)}=1$, then the proposed value is also accepted with probability one. 
In other words, the algorithm ``prefers" a smaller consideration set. 
This sparsity-inducing property is proven below. 
Lastly, when the proposed consideration set adds an alternative $j$ that is not in the current consideration set, that is, 
$\tilde{C}_{ij}=1$ and $C_{ij}^{(g)}=0$, the acceptance probability is between 0 and 1 and is determined by the likelihood ratio. 

\begin{prop}[Sparsity-inducing property]
Consider the M-H step  described in Algorithm \ref{alg:MH_for_CS_main}. 
Let $j$ be an alternative that is not observed to be chosen by the subject $i$.
If the step proposes to exclude $j$ from the consideration set of $i$, it is accepted with probability 1.
\end{prop}
\begin{proof}
Let the consideration set for the $i$th subject at iteration $g$ be $ \mathcal{C}_i^{(g)}$. 
Suppose that a category $j \in  \mathcal{C}_i^{(g)}$ is proposed to be removed so that $\tilde{\mathcal{C}}_i = \mathcal{C}^{(g)}_i \setminus \{ j \}$. 
The acceptance probability is 
\[
\min\left\{ \frac{p\big(\bm  Y_{i}=\bm  y_{i} \big\vert  \bm  \beta^{(g)},\bm  b_i^{(g)}, \bm w_i,\tilde{\mathcal{C}}_i \big) }{ p\big(\bm  Y_{i}=\bm y_{i} \big\vert  \bm \beta^{(g)}, \bm b_i^{(g)},  \bm w_i, \mathcal{C}^{(g)}_i \big)  } ,1   \right\}
=
\min\left\{  
\frac{ 
\prod_t
\sum_{\ell \in \mathcal{C}^{(g)}_{i}}\exp \left(V_{i\ell t} \right)
 }{ 
\prod_t
\sum_{\ell \in \tilde{\mathcal{C}}_{i}}\exp \left(V_{i\ell t} \right)
} ,1   \right\}
=1, 
\]
where $V_{ij t}=\bm x_{ijt}' \bm \beta^{(g)} +\bm z_{ijt}' \bm b_i^{(g)} $, 
and the last equality is due to the fact that the ratio is larger than 1.
Hence,  $\tilde{\mathcal{C}}_i$ is accepted with probability 1.
\end{proof}

\subsection{Numerical illustration}

We  illustrate  posterior probabilities of consideration sets on synthetic panel data with $n=100$ subjects observed over $T\in \{1,2,\ldots,15 \}$
time periods. We let $J=4$ and give the $2^{J}-1=15$ consideration sets in the first column of Table \ref{table:increasingT}. 
In the table we report the posterior probabilities of each possible consideration set for a randomly chosen subject $i$ 
whose true consideration set is $\mathcal{C}_{i}^*=\{1,3,4 \}$. 
\begin{table}[ht]
\centering
\caption{Posterior probabilities of consideration sets for unit $i$}
\resizebox{\linewidth}{!}{%
\begin{threeparttable}

\begin{tabular}{ c | c c c c c c c c c c c c c c c}
& $T=1$ & $T=2$ & $T=3$ & $T=4$ & $T=5$ & $T=6$ & $T=7$ & $T=8$ & $T=9$ & $T=10$ & $T=11$ & $T=12$ & $T=13$ & $T=14$ & $T=15$\\ 
\toprule 

$\{1\}$ & 0.059 & 0.602 & 0 & 0 & 0 & 0 & 0 & 0 & 0 & 0 & 0 & 0 & 0 & 0 & 0 \tabularnewline
$\{2\}$ & 0 & 0 & 0 & 0 & 0 & 0 & 0 & 0 & 0 & 0 & 0 & 0 & 0 & 0 & 0 \tabularnewline
$\{3\}$ & 0 & 0 & 0 & 0 & 0 & 0 & 0 & 0 & 0 & 0 & 0 & 0 & 0 & 0 & 0 \tabularnewline
$\{4\}$ & 0 & 0 & 0 & 0 & 0 & 0 & 0 & 0 & 0 & 0 & 0 & 0 & 0 & 0 & 0 \tabularnewline
$\{1, 2\}$ & 0.199 & 0.15 & 0 & 0 & 0 & 0 & 0 & 0 & 0 & 0 & 0 & 0 & 0 & 0 & 0 \tabularnewline
$\{1, 3\}$ & 0.046 & 0.033 & 0 & 0 & 0 & 0 & 0 & 0 & 0 & 0 & 0 & 0 & 0 & 0 & 0 \tabularnewline
$\{1, 4\}$ & 0.065 & 0.12 & 0.691 & 0.738 & 0.746 & 0.78 & 0.834 & 0.864 & 0.884 & 0.844 & 0 & 0 & 0 & 0 & 0 \tabularnewline
$\{2, 3\}$ & 0 & 0 & 0 & 0 & 0 & 0 & 0 & 0 & 0 & 0 & 0 & 0 & 0 & 0 & 0 \tabularnewline
$\{2, 4\}$ & 0 & 0 & 0 & 0 & 0 & 0 & 0 & 0 & 0 & 0 & 0 & 0 & 0 & 0 & 0 \tabularnewline
$\{3, 4\}$ & 0 & 0 & 0 & 0 & 0 & 0 & 0 & 0 & 0 & 0 & 0 & 0 & 0 & 0 & 0 \tabularnewline
$\{1, 2, 3\}$ & 0.113 & 0.01 & 0 & 0 & 0 & 0 & 0 & 0 & 0 & 0 & 0 & 0 & 0 & 0 & 0 \tabularnewline
$\{1, 2, 4\}$ & 0.243 & 0.048 & 0.164 & 0.138 & 0.113 & 0.119 & 0.058 & 0.052 & 0.042 & 0.035 & 0 & 0 & 0 & 0 & 0 \tabularnewline
$\{\bm 1, \bm 3, \bm 4\}$ & 0.048 & 0.028 & 0.118 & 0.1 & 0.114 & 0.091 & 0.101 & 0.074 & 0.072 & 0.114 & 0.969 & 0.99 & 0.98 & 0.993 & 0.992 \tabularnewline
$\{2, 3, 4\}$ & 0 & 0 & 0 & 0 & 0 & 0 & 0 & 0 & 0 & 0 & 0 & 0 & 0 & 0 & 0 \tabularnewline
$\{1, 2, 3, 4\}$ & 0.227 & 0.009 & 0.027 & 0.024 & 0.027 & 0.01 & 0.007 & 0.01 & 0.002 & 0.007 & 0.031 & 0.01 & 0.02 & 0.007 & 0.008 \tabularnewline
$y_{iT}$ & 1 & 1 & 4 & 1 & 4 & 1 & 4 & 4 & 1 & 1 & 3 & 3 & 1 & 1 & 4 \tabularnewline
$Acc.Rate.$ & 0.876 & 0.77 & 0.647 & 0.618 & 0.641 & 0.64 & 0.598 & 0.591 & 0.605 & 0.588 & 0.662 & 0.685 & 0.693 & 0.696 & 0.709 \tabularnewline

\bottomrule 
\end{tabular}

\begin{tablenotes}
\small
\item 
The results are based on a synthetic panel data with $J=4$ and $n=100$.
The true consideration set is $\mathcal{C}_{i}^*=\{1,3,4 \}.$
The row $y_{i,T}$ shows the actual response made by subject $i$ at  time  $T$.
Acc. Rate denotes the acceptance rate of consideration sets in the M-H step. 
\end{tablenotes}

\end{threeparttable}
}

\label{table:increasingT}

\end{table}

The first column  ($T=1$) shows the results for the initial period given the observed outcome of $1$. 
Consideration sets that do not include item 1 have a posterior probability of zero. 
As $T$ increases, the posterior concentrates on the true consideration set $\{1,3,4\}$. 

\section{Monte Carlo Simulation}\label{sec:simulation}

We demonstrate the sampling performance of the proposed approach through simulation studies first with $J=4$ alternatives, where it is possible to enumerate all the support points in $\bm \pi$, and then extend the study to a high-dimensional case with $J = 100$. 
The goal is to empirically validate the findings of Theorem 2 and demonstrate that the proposed approach can effectively assess consideration dependence. 
In the Supplementary Material, we conduct additional experiments under autocorrelated covariates, random effects, and time-varying true consideration sets. In general, the experiments show posterior consistency in the estimation of $\bm\theta=(\bm \beta, \bm D)$ and $\bm \pi$, and that the restrictive approach with $K=1$ produces larger root mean squared errors and biases. 

\subsection{$J = 4$}
\begin{figure}[ht] 
\centering
  \begin{subfigure}[b]{0.25\linewidth}
    \centering
    \includegraphics[width=\linewidth]{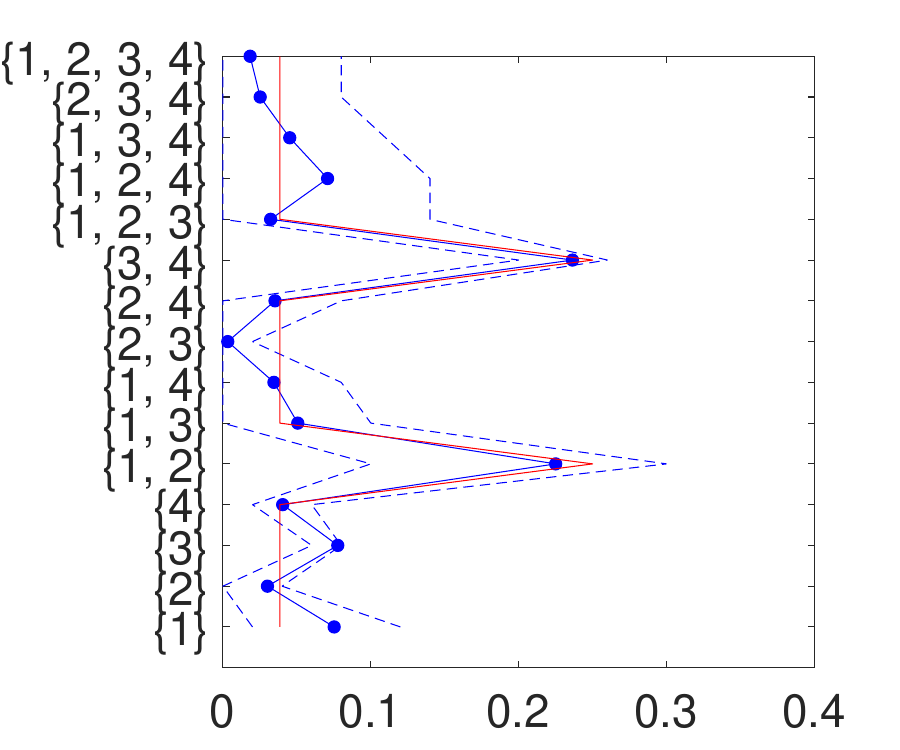} 
\caption{\small{ $K=\infty, n=50$ }}
  \end{subfigure}
  \begin{subfigure}[b]{0.25\linewidth}
    \centering
    \includegraphics[width=\linewidth]{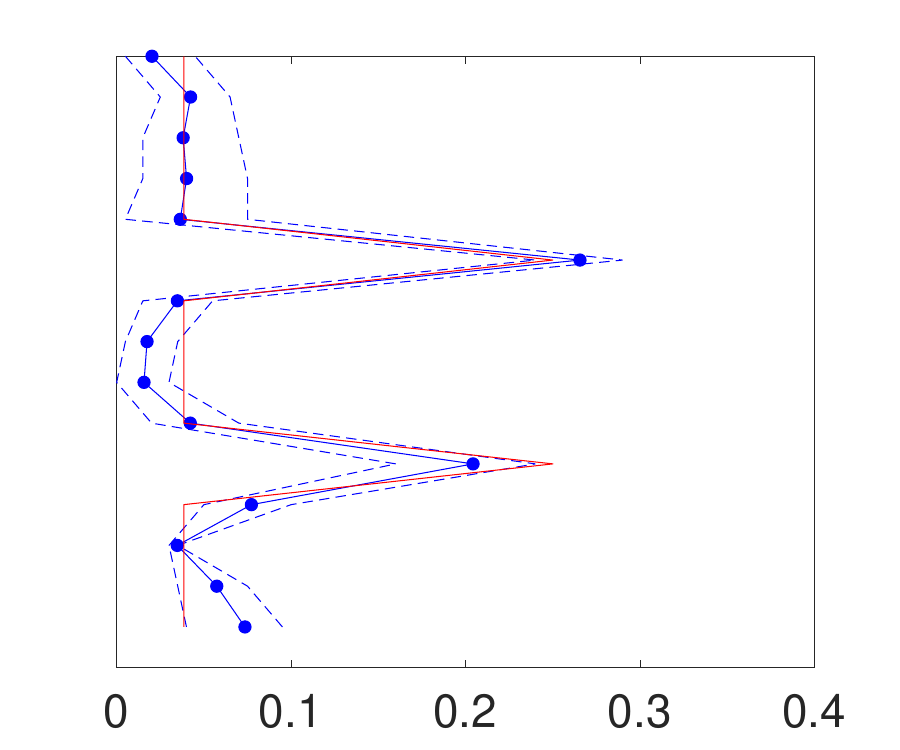} 
\caption{\small{ $K=\infty, n=200$ }}
  \end{subfigure}
  \begin{subfigure}[b]{0.25\linewidth}
    \centering
    \includegraphics[width=\linewidth]{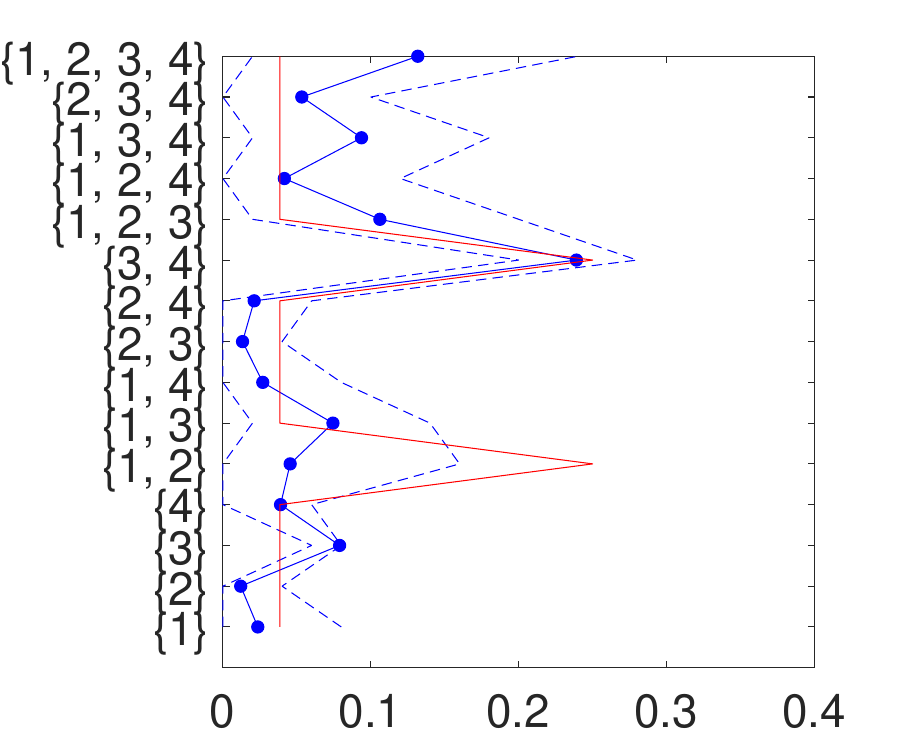} 
\caption{\small{ $K=1, n=50$ }}
  \end{subfigure}
  \begin{subfigure}[b]{0.25\linewidth}
    \centering
    \includegraphics[width=\linewidth]{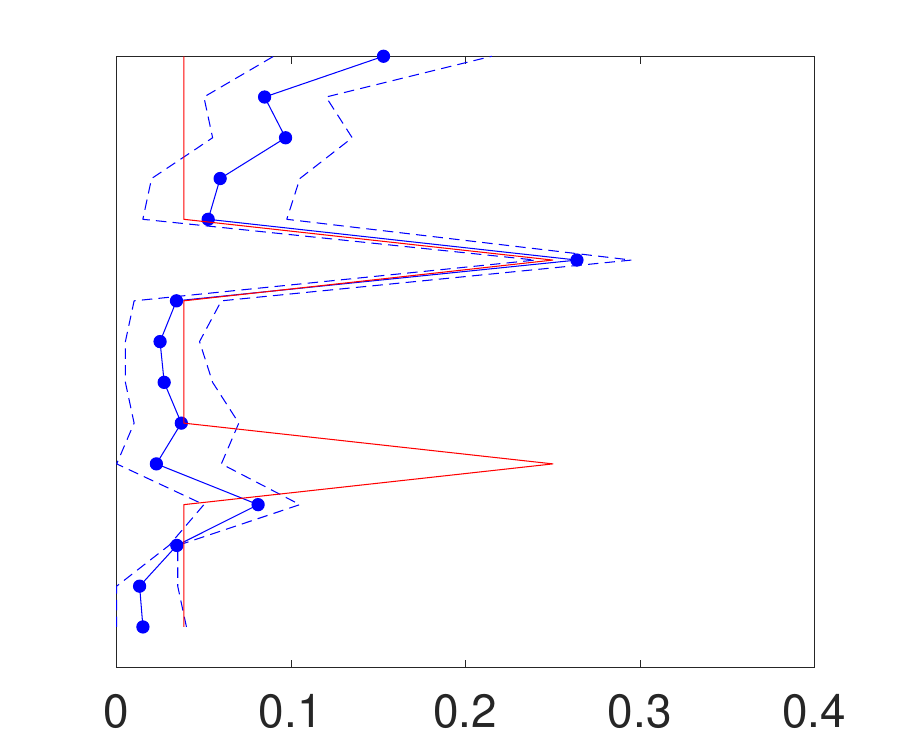} 
\caption{\small{ $K=1, n=200$ }}
  \end{subfigure}
  
\caption{\small{
The true distribution over consideration sets (solid, red), 
posterior mean (solid with dots, blue),  95\% equal-tailed credible interval (dashed, blue). 
Each plot is based on  one realization of simulated data. $J=4$, $T=5$. 
 }}
\label{fig:simulation_SmallJ_marginalCS}
\end{figure}

We let $T_i = T$ for all $i$. 
In one case we set $T=5$ and in the other $T=15$. The latter satisfies the length condition of Theorem \ref{thm:consistency_model_parameters_identified}. In simulating the data, we first specify the distribution of the consideration sets $\bm \pi^*=\{ \pi^*_c = \Pr(\mathcal{C}_i=c): c \in \mathcal{C}\}$. We induce dependence in product consideration by letting the first two and last two products have a relatively high probability of being considered together: $\pi^*_{\{1,2\}}=\pi^*_{\{3,4\}}=0.25$. As motivation, the first two products might represent non-vegetarian options, and the last two vegetarian. The other 13 consideration sets $c \in \mathcal{C}$ are given a probability of 0.0385. Figure \ref{fig:simulation_SmallJ_marginalCS} shows $\bm \pi^*$ in red. 
%
Given this $\bm \pi^*$, we generate the true consideration sets $ \mathcal{C}_i^*$, for $i=1,\ldots,n$. We then generate outcomes from the logit model with
$V_{ijt} =\delta^*_j+ \beta^*  x_{ijt}, $
letting $(\delta_1^*,\delta_2^*,\delta_3^*,\delta_4^*)'=(1.0, 0.5, -1.0, 0)'$ and $\beta^*=1$, and $x_{ijt} \overset{iid}{\sim} N(0,1)$. We let $n\in\{50,100\}$. 

%
We compare the performance between the proposed infinite mixture of independent consideration models ($K=\infty$) and the model that assumes independent consideration ($K=1$) over 200 replicated data sets. The results are given in Table \ref{tab:simulation_SmallJ_pref} where we report 
the root mean squared error (RMSE) for the response parameter $\bm \beta=(\delta_1,\delta_2,\delta_3, \beta)$ 
as well as the $L_1$ norm between the posterior mean and the truth for the distribution of the consideration sets $\bm \pi$ (L1-error), 
their Monte Carlo errors (MCE), 
the posterior standard deviation (SD), 
the empirical standard deviation (ESD),  
the empirical coverage of the equal-tailed 95\% credible intervals (Cov), 
and the computational time. 
The MCE  quantifies the precision for the performance criterion. The MCEs are negligible, allowing for valid comparisons based on the $200$ replications. 
As $n$ increases, the posterior of $\bm \beta$ and  $\bm \pi$ contracts to the true values, even when $T = 5$, indicated by the smaller  RMSEs and L1-errors as well as SDs. 
In contrast, when $K=1$, we do not observe sufficient evidence of posterior consistency. The RMSEs and L1-error are much larger in some cases than those under $K=\infty$, due to misspecification. 
When $T$ increases to $15$, a value that satisfies the identifying condition in Theorem \ref{thm:consistency_model_parameters_identified}, the RMSEs/L1-errors/SDs become smaller for both $K=\infty$ and $K=1$, but for $K=1$, they are larger, and there are distortions in the coverage. 
Finally, our approach ($K=\infty$) delivers good coverages in general. 
The SDs are similar to ESDs, indicating that the posterior standard deviations provide a good representation of  the sampling variability of the posterior means. 
\begin{table}[ht]
\caption{Simulation results  with $J=4$}
\centering
\resizebox{1.1\columnwidth}{!}{%
\begin{threeparttable}
\begin{tabular}{ l  l | l l l    l l l     l l l    l l l    l l l    l  }
$(K,T)$ & $n$ &   \multicolumn{3}{c}{$\beta$}  &  \multicolumn{3}{c}{$\delta_1$} &  \multicolumn{3}{c}{$\delta_2$} &  \multicolumn{3}{c}{$\delta_3$}  &  \multicolumn{3}{c}{$\pi$} & Time \\ 
\hline

 &  &   
 \multicolumn{3}{l}{RMSE (MCE) \  \ SD (ESD)  \ \ \ \ \ \ Cov}  &  
 \multicolumn{3}{l}{RMSE (MCE) \  \ SD (ESD)  \ \ \ \ \ \ Cov} &  
 \multicolumn{3}{l}{RMSE (MCE) \  \ SD (ESD)  \ \ \ \ \ \ Cov} &  
 \multicolumn{3}{l}{RMSE (MCE) \  \ SD (ESD)  \ \ \ \ \ \ Cov}  &  
 \multicolumn{3}{l}{L1-error (MCE)   SD (ESD)   \ \ \ \ \ \ Cov} & \\

\hline
\multirow{2}{*}{$(\infty, 5)$} 
&$50$ & 0.168 ( 0.01 ) & 0.16 ( 0.167 ) & 0.94 & 0.464 ( 0.024 ) & 0.44 ( 0.442 ) & 0.94 & 0.47 ( 0.023 ) & 0.44 ( 0.442 ) & 0.94 & 0.328 ( 0.018 ) & 0.34 ( 0.329 ) & 0.96 & 0.446 ( 0.01 ) & 0.03 ( 0.028 ) & 0.94 & 1.84 \tabularnewline
&$100$ & 0.116 ( 0.005 ) & 0.11 ( 0.115 ) & 0.98 & 0.361 ( 0.021 ) & 0.31 ( 0.32 ) & 0.91 & 0.359 ( 0.021 ) & 0.31 ( 0.324 ) & 0.92 & 0.235 ( 0.01 ) & 0.25 ( 0.227 ) & 0.96 & 0.366 ( 0.006 ) & 0.02 ( 0.022 ) & 0.96 & 3.37 \tabularnewline

\hline 
\multirow{2}{*}{$(\infty, 15)$} 
&$50$ & 0.092 ( 0.004 ) & 0.09 ( 0.093 ) & 0.93 & 0.194 ( 0.01 ) & 0.21 ( 0.189 ) & 0.96 & 0.185 ( 0.01 ) & 0.2 ( 0.179 ) & 0.95 & 0.171 ( 0.009 ) & 0.17 ( 0.172 ) & 0.96 & 0.358 ( 0.005 ) & 0.03 ( 0.023 ) & 0.96 & 2.51 \tabularnewline
&$100$ & 0.062 ( 0.003 ) & 0.06 ( 0.062 ) & 0.97 & 0.132 ( 0.007 ) & 0.14 ( 0.13 ) & 0.96 & 0.136 ( 0.007 ) & 0.14 ( 0.136 ) & 0.97 & 0.117 ( 0.006 ) & 0.12 ( 0.116 ) & 0.97 & 0.294 ( 0.003 ) & 0.02 ( 0.018 ) & 0.93 & 4.84 \tabularnewline

\hline
\multirow{2}{*}{$(1, 5)$} 
&$50$ & 0.161 ( 0.008 ) & 0.15 ( 0.158 ) & 0.94 & 0.883 ( 0.028 ) & 0.4 ( 0.442 ) & 0.49 & 0.943 ( 0.029 ) & 0.4 ( 0.473 ) & 0.47 & 0.33 ( 0.019 ) & 0.33 ( 0.331 ) & 0.96 & 0.847 ( 0.006 ) & 0.03 ( 0.024 ) & 0.5 & 1.62 \tabularnewline
&$100$ & 0.112 ( 0.005 ) & 0.11 ( 0.104 ) & 0.91 & 0.949 ( 0.024 ) & 0.28 ( 0.336 ) & 0.18 & 0.981 ( 0.027 ) & 0.28 ( 0.355 ) & 0.18 & 0.243 ( 0.012 ) & 0.24 ( 0.239 ) & 0.94 & 0.876 ( 0.007 ) & 0.02 ( 0.02 ) & 0.17 & 2.81 \tabularnewline

\hline
\multirow{2}{*}{$(1, 15)$} 
&$50$ & 0.092 ( 0.004 ) & 0.09 ( 0.092 ) & 0.93 & 0.283 ( 0.016 ) & 0.22 ( 0.23 ) & 0.92 & 0.272 ( 0.017 ) & 0.22 ( 0.218 ) & 0.92 & 0.182 ( 0.01 ) & 0.18 ( 0.178 ) & 0.94 & 0.714 ( 0.002 ) & 0.02 ( 0.018 ) & 0.89 & 2.24 \tabularnewline
&$100$ & 0.062 ( 0.003 ) & 0.06 ( 0.062 ) & 0.96 & 0.206 ( 0.01 ) & 0.15 ( 0.151 ) & 0.87 & 0.199 ( 0.012 ) & 0.15 ( 0.159 ) & 0.85 & 0.133 ( 0.007 ) & 0.12 ( 0.122 ) & 0.95 & 0.706 ( 0.001 ) & 0.01 ( 0.013 ) & 0.82 & 4.24 \tabularnewline

\end{tabular}
\begin{tablenotes}
\small
\item 
For $\beta$ and $\delta$, for each case, 
we show  the estimated root mean squared error (RMSE), using 
the posterior means as  point estimator. 
In parenthesis, 
the jackknife estimate of Monte Carlo Error (MCE) for the RMSE is presented. 
Next, the average of the posterior standard deviations (SD) is shown with  the empirical standard deviation (ESD) of the posterior mean in the parenthesis.
Third, the empirical coverage (Cov) of 95\% credible interval is given. 
\item 
Letting $\hat\theta_r$ denote the posterior mean from replication $r$ $(r=1,\ldots,R)$ and letting $\theta^\ast$ denote the true value, we summarize finite-sample accuracy and variability using: 
\begin{itemize}
    \item $RMSE=\sqrt{\frac{1}{R}\sum_{r=1}^R (\hat{\theta}_r-\theta^*)^2}$, 
    \item $MCE(\widehat{RMSE})=\sqrt{\frac{R-1}{R}\sum_{r=1}^R(RMSE_{(-r)}-\overline{RMSE_{(-r)}})^2}$,  where $RMSE_{(-r)}$ is the RMSE estimated with the $r$th replicate removed and 
    $\overline{RMSE_{(-r)}}=\frac{1}{R}\sum_{r=1}^R RMSE_{(-r)}$,  
    \item $ESD=\sqrt{\frac{1}{R-1}\sum_{r=1}^R(\hat{\theta}_r-\bar{\hat{\theta}})^2}$, where $\bar{\hat{\theta}}=\frac{1}{R}\sum_{r=1}^R\hat{\theta}_r$.
\end{itemize}
\item For $\pi$, we show the average of $L_1$ norm between the posterior mean and $\bm \pi^*$ (L1-error). 
In the parenthesis, we show its jackknife estimate of MCE. 
The SDs, ESDs, and Covs are averaged over the 15 elements in $\pi$. 

\item Time is the average seconds taken for sampling 1,000 MCMC draws in Matlab on a desktop with a 4.9GHz processor and 64GB RAM. 
The study is based on $R=200$ replications. 
2,000 MCMC draws are obtained for each replication. The average of the inefficiency factors is around 6.6 with standard deviation 1.2.
\end{tablenotes}
\end{threeparttable}
}
\label{tab:simulation_SmallJ_pref}
\end{table}

The vertical axes of Figure \ref{fig:simulation_SmallJ_marginalCS} list the 15 consideration sets, with the true distribution of the consideration sets, $\bm \pi^*$, highlighted in red. Each panel of the figure displays the posterior mean (solid with dots, blue) along with the 95\% credible intervals (dashed, blue), based on one realized data set. The first two panels illustrate that under the proposed approach ($K=\infty$), as the sample size $n$ increases, the discrepancy between the posterior mean and the true distribution diminishes. In contrast, the right two panels show that when $K=1$, even as $n$ increases, the posterior does not adequately converge to the truth. This is because the model does not account for the true consideration dependence.

\subsection{$J=100$}\label{sec:simulation_LargeJ}

We now consider a high-dimensional scenario with $J=100$ alternatives. 
One mechanism by which the dependence of consideration among categories can be induced is through multiple latent subpopulations of subjects having different probabilities of consideration. Within a subpopulation, considerations are independent across categories. However, marginalizing out the latent subpopulation indicator, one obtains dependence in those category considerations. We generate the data with two subpopulations. To generate the true consideration set of a given subject, we used a Bernoulli distribution with attention probability 0.05 for each category except for categories 10, 30, 50, 70, and 90 for the first subpopulation ($i=1,\ldots,n/2$)  where the  attention probability was set to 0.8. For the remaining subjects in the second subpopulation ($i=n/2+1,\ldots,n$), the Bernoulli probability was set at 0.05 except for categories 20, 40, 60, 80, and 100 where the probability was set to 0.8. 
Conditional on the true consideration sets, we generated the responses as in the case with $J=4$ with $\delta_j^*=0, \ j=1,\ldots,J-1$.

Because in this case there are $2^{100}-1$ support points in $\bm \pi$, it is not possible to show the entire distribution as in the case of $J=4$. Also, there are 99 $\delta_j$'s to estimate. Hence, in Table \ref{tab:simulation_LargeJ_pref}, we report the results only for the slope $\beta$ as well as $\delta_{97}, \delta_{98},$ and $\delta_{99}$. The results are based on 200 replications. 
The general observations from the small $J$ simulation still hold: as $n$ increases, the RMSEs/SDs of $\bm \beta$ become smaller even when $T=5$, supporting posterior consistency. For $T = 200$, a span that approximately satisfies the identification condition in Theorem \ref{thm:consistency_model_parameters_identified}, our approach also results in good coverages.

\FloatBarrier
\begin{table}[ht]
\caption{Simulation results  with $J=100$}
\centering
\resizebox{1.1\columnwidth}{!}{%
\begin{threeparttable}
\begin{tabular}{ l  l | l l l    l l l     l l l    l l l      l  }
$(K,T)$ & $n$ &   \multicolumn{3}{c}{$\beta$}  &  \multicolumn{3}{c}{$\delta_{97}$} &  \multicolumn{3}{c}{$\delta_{98}$} &  \multicolumn{3}{c}{$\delta_{99}$}   & Time \\ 
\hline

 &  &   
 \multicolumn{3}{l}{RMSE (MCE) \  \ SD (ESD)  \ \ \ \ \ \ Cov}  &  
 \multicolumn{3}{l}{RMSE (MCE) \  \ SD (ESD)  \ \ \ \ \ \ Cov} &  
 \multicolumn{3}{l}{RMSE (MCE) \  \ SD (ESD)  \ \ \ \ \ \ Cov} &  
 \multicolumn{3}{l}{RMSE (MCE) \  \ SD (ESD)  \ \ \ \ \ \ Cov}  &  
 \\

\hline
\multirow{2}{*}{$(\infty, 5)$} 
&$50$ & 0.101 ( 0.004 ) & 0.08 ( 0.083 ) & 0.85 & 0.82 ( 0.036 ) & 0.94 ( 0.803 ) & 0.95 & 0.811 ( 0.049 ) & 0.91 ( 0.783 ) & 0.95 & 0.85 ( 0.048 ) & 0.93 ( 0.837 ) & 0.90 & 3.32 \tabularnewline
&$100$ & 0.09 ( 0.004 ) & 0.06 ( 0.057 ) & 0.79 & 0.777 ( 0.037 ) & 0.79 ( 0.741 ) & 0.93 & 0.793 ( 0.037 ) & 0.79 ( 0.765 ) & 0.93 & 0.816 ( 0.04 ) & 0.81 ( 0.8 ) & 0.92 & 7.29 \tabularnewline


\hline
\multirow{2}{*}{$(\infty, 200)$} 
&$50$ & 0.014 ( 0.001 ) & 0.01 ( 0.014 ) & 0.92 & 0.261 ( 0.014 ) & 0.29 ( 0.239 ) & 0.84 & 0.232 ( 0.01 ) & 0.25 ( 0.211 ) & 0.86 & 0.231 ( 0.014 ) & 0.35 ( 0.218 ) & 0.90 & 216.4 \tabularnewline
&$100$ & 0.01 ( 0.001 ) & 0.01 ( 0.009 ) & 0.97 & 0.145 ( 0.01 ) & 0.14 ( 0.128 ) & 0.93 & 0.127 ( 0.006 ) & 0.13 ( 0.11 ) & 0.96 & 0.165 ( 0.011 ) & 0.14 ( 0.153 ) & 0.92 & 417.9 \tabularnewline

\end{tabular}
\begin{tablenotes}
\small
\item 
For $\beta$ and $\delta$, for each case, 
we show  the estimated root mean squared error (RMSE), using 
the posterior means as  point estimator. 
In parenthesis, 
the jackknife estimate of Monte Carlo Error (MCE) for the RMSE is presented. 
Next, the average of the posterior standard deviations (SD) is shown with  the empirical standard deviation (ESD) of the posterior mean in the parenthesis.
Third, the empirical coverage (Cov) of 95\% credible interval is given. 
Time is the average minutes taken for sampling 1,000 MCMC draws. 
The study is based on $R=200$ replications. 
3,000 MCMC draws are obtained for each replication. The average of the inefficiency factors is around 3.26 with standard deviation 0.79.
\end{tablenotes}
\end{threeparttable}
}
\label{tab:simulation_LargeJ_pref}
\end{table}
\FloatBarrier

\section{Application to Cereal Consumption in Midwest}\label{sec:application}

\begin{figure}[ht] 
\centering
    \includegraphics[width=0.8\linewidth]{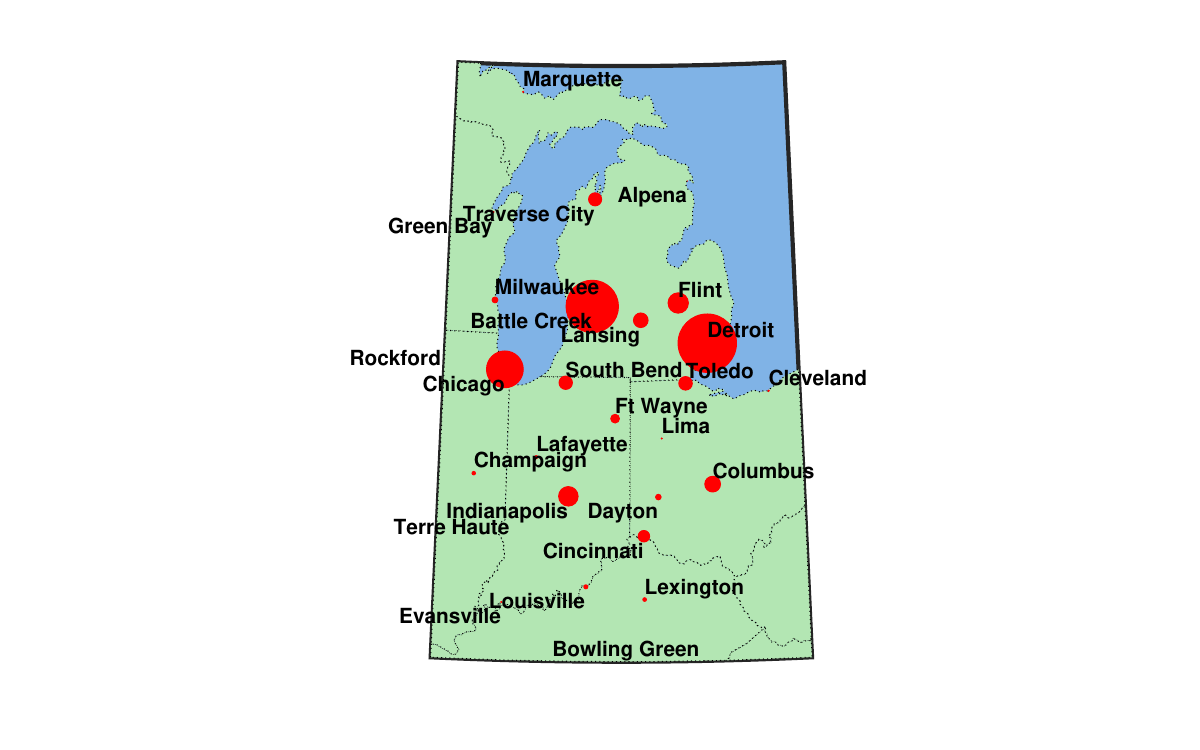} 
\caption{\small{
Locations of the 239 stores under the chosen retailer. Circle sizes correspond to  purchases (percentages).  
 }}
\label{fig:application_map}
\end{figure}

In this section, we apply our approach to a manually constructed longitudinal data set that includes $J=101$ cereal brands, a size that is significantly beyond the feasibility of existing methods. For comparison, $J$ was 4 in \cite{ChiangChibNarasimhan1998}, 10 in \cite{vanNieropPaap2010}, and 5 in \cite{AguiarKashaev2024identification}. We constructed the data set by integrating Nielsen Consumer Panel data with Retail Scanner Data, focusing on weekly shopping trips in 2019 in stores operated by a single anonymous retailer primarily based in the United States Midwest. Although data from 2020 are available, we chose to use the most recent pre-pandemic year to avoid potential biases introduced by pandemic-related shopping behavior. This particular retailer was selected because it consistently stocked more than 100 cereal brands throughout the sample period. Furthermore, we limited our analysis to a single retailer to prevent inconsistencies in brand definitions between different retailers, which would have required speculative alignment of brand names from various sources. The final data set includes $J=101$ brands and $n=1880$ households, covering 25,849 purchases in 239 stores during the 52-week period in 2019. See Figure \ref{fig:application_map}, for the locations of these stores with relative purchase volumes, and Table \ref{tab:application_ProdSpecific}, for the list of the brands. The average number of shopping trips per household ($T_i$) is 13.7, and the price $P_{ijt}$ of each brand $j \in {1,\ldots,J}$ is represented by a size-weighted price index constructed from prices at the UPC level. For the analysis, we used the first 10 months of data for the estimation and reserved the last two months for the prediction outside the sample. Further details on data preparation are provided in the Supplementary Material.

Conditional on the consideration set $\{\mathcal{C}_i\}$, in the most general version of the model, we enter the fixed effects and random effects in the MNL model $V_{ijt} =\delta_j + P_{ijt}(\beta + b_i)$,
where $i \in \{1, \ldots, 1880\}$ indexes households, and $t \in \{1, \ldots, T_i\}$ indexes purchase occasions. In this model, $\delta_j$ represents the brand-specific fixed effect for brand $j$, with the normalization $\delta_J = 0$. The parameter $\beta$ is the common fixed effect, and $b_i \sim \mathcal{N}(0, D)$ is the random effect for household $i$.
We consider four variants of the  MNL, differentiated by the inclusion of random effects and/or consideration set heterogeneity, as detailed in models (1)–(4) of Table \ref{tab:estimates}. In addition, models (5) and (6) assume an independent consideration structure (i.e.,\ $K=1$). Each of these cases is estimated using the simulation method developed in Section \ref{sec:inference}, by omitting the components not present in the full hierarchical model (MNL\_RC). 


\subsection{Empirical Results}

We obtained 20,000 MCMC draws for each of the six models in Matlab on a desktop with a 4.9GHz processor and 64GB RAM. The average of the inefficiency factors is around 7.38 with standard deviation 2.41, indicating that the MCMC output mixes well. 
Broadly speaking, the estimated parameters of the response model from the approaches (1)-(4) shown in Table \ref{tab:estimates} 
are similar to those in the literature. 
For instance, when consideration set heterogeneity is incorporated, the magnitude of the slope parameter $\beta$ on price increases and 
the number of significant brand-specific terms $\delta_j$'s decreases (See Table \ref{tab:application_ProdSpecific} for the list of estimated $\delta_j$'s under the MNL\_RC model).
These patterns are consistent with previous studies based on smaller models, including \citet{Stopher1980}, \citet{SwaitBenAkiva1986}, \citet{ChiangChibNarasimhan1998}, and \citet{vanNieropPaap2010}.
However, without the scalable fitting methodology developed in this paper, it was unclear if those patterns 
would persist in a model of the scale we have estimated.

Moreover, when we control for consideration sets, the posterior mean of $D^{1/2}$ decreases, 
which aligns with the findings in \citealp{ChiangChibNarasimhan1998}, \citealp{Morozov2021MrkSci} that random effect heterogeneity is overestimated in models that omit consideration set heterogeneity. 

\begin{table}[ht]
\caption{Estimation results}  

\centering
\resizebox{1\columnwidth}{!}{%

\begin{threeparttable}
\begin{tabular}{ccccccccccccccccc}
    \toprule
    \toprule
    &
      \multicolumn{2}{c}{(1)} &
      \multicolumn{2}{c}{(2)} &
      \multicolumn{2}{c}{(3)} &      
      \multicolumn{2}{c}{(4)} &
      \multicolumn{2}{c}{(5)} &      
      \multicolumn{2}{c}{(6)} \\

    \multirow{1}{*}{} &
      \multicolumn{2}{c}{MNL} &
      \multicolumn{2}{c}{MNL\_R} &
      \multicolumn{2}{c}{MNL\_C} &      
      \multicolumn{2}{c}{MNL\_RC} &
      \multicolumn{2}{c}{MNL\_C\_K1} &      
      \multicolumn{2}{c}{MNL\_RC\_K1} \\
      \cmidrule(lr){2-3}\cmidrule(lr){4-5}\cmidrule(lr){6-7}\cmidrule(lr){8-9}\cmidrule(lr){10-11}\cmidrule(lr){12-13}
       & {mean} & {s.d.} &  {mean} & {s.d.}  &  {mean} & {s.d.}  &  {mean} & {s.d.} &  {mean} & {s.d.} &  {mean} & {s.d.} \\

\midrule
\textit{Random effects on price}\\
$\beta$               & -0.69***    & 0.02     & -0.77***  & 0.03  & -0.73*** & 0.02 & -0.82***  & 0.04 & -0.73*** & 0.02 & -0.85***&0.04\\
$D^{1/2}$  & {------} & {------} & 0.99 & 0.02   & {------} & {------} & 0.97  & 0.03  & {------}& {------}& 1.06& 0.04\\
\midrule
\textit{Brand-specific fixed-effects}\\
\# of ``significant'' params.   & \multicolumn{2}{c}{97}  & \multicolumn{2}{c}{98} & \multicolumn{2}{c}{75} & \multicolumn{2}{c}{67}
& \multicolumn{2}{c}{73}& \multicolumn{2}{c}{69}  \\
\midrule
\textit{Computational time}\\
min. per 1,000 MCMC iters.  & \multicolumn{2}{c}{66}  & \multicolumn{2}{c}{67} & \multicolumn{2}{c}{120} & \multicolumn{2}{c}{124} 
& \multicolumn{2}{c}{118}& \multicolumn{2}{c}{122} \\
\midrule
\textit{Test for indep. consid.} & \multicolumn{2}{c}{--}  & \multicolumn{2}{c}{--} & \multicolumn{2}{c}{Reject $H_0$} & \multicolumn{2}{c}{Reject $H_0$} & \multicolumn{2}{c}{--}& \multicolumn{2}{c}{--}  \\
\midrule 
random effects & \multicolumn{2}{c}{No}  & \multicolumn{2}{c}{Yes} & \multicolumn{2}{c}{No} & \multicolumn{2}{c}{Yes} & \multicolumn{2}{c}{No}& \multicolumn{2}{c}{Yes}  \\
Consideration sets & \multicolumn{2}{c}{No}  & \multicolumn{2}{c}{No} & \multicolumn{2}{c}{$K=\infty$} & \multicolumn{2}{c}{$K=\infty$} & \multicolumn{2}{c}{$K=1$}& \multicolumn{2}{c}{$K=1$}  \\

\bottomrule
\bottomrule
\end{tabular}          
\begin{tablenotes}
\small
\item 
$K=\infty$ ($K=1$) refers to the proposed infinite mixture of independent consideration models (the model under the independent consideration).
The first panel shows posterior means of the mean  $\beta$ of the random effects on price and the standard deviations $D^{1/2}$ with their posterior standard deviations. 
Three stars indicate that the corresponding 99\% credible interval does not include 0. 
The second panel shows the number of brand-specific fixed effects whose 95\% posterior credible intervals do not include 0 (out of 100 terms). 
See Table \ref{tab:application_ProdSpecific} for the estimated $\delta_j$'s under MNL\_RC. 
The third panel shows computational time (minutes) on a desktop with a 4.9GHz processor and 64GB RAM. 
The fourth panel shows the results for the test for independent consideration with the null hypothesis $H_0:\text{independent consideration}$, which we discuss in the Supplementary Material in detail.
The results are based on 20,000 posterior draws. We discard the first 6,000 draws as a burn-in sample and use the remaining 14,000 draws for the analysis. 
The average of the inefficiency factors is around 7.38 with standard deviation 2.41. 
\end{tablenotes}
\end{threeparttable}
  }
\label{tab:estimates}
\end{table}

Under the independent consideration assumption ($K=1$), i.e.,\ (5) and (6), the estimated parameters 
are similar to the proposed flexible approach i.e.,\ (3) and (4) except that the estimated $D^{1/2}$ under (6) is slightly larger than (2), which contradicts with the previous studies. 
In general, it is possible that the obtained estimates under $K=1$ are biased, as shown in simulation studies in Section \ref{sec:simulation}. 
We conduct the test for independent consideration, which is introduced and studied in Supplementary Material.  
Under both (3) and (4), the estimated posterior probability of the alternative hypothesis (dependent consideration) is very close to one, 
and we conclude that the considerations of cereal products in this particular market are dependent. 

Table \ref{tab:estimates} also shows the computational time per 1,000 MCMC draws. 
The extra burden of estimating latent consideration sets using our proposed approach is reasonable. 
For instance, when consideration sets are estimated along with random effects, the computational time  roughly  doubles (67 mins.\ for MNL\_R and 124 mins.\ for MNL\_RC). Not surprisingly, compared to the fully flexible estimator, the estimators that assume independent consideration take  less computational time but only slightly.

\subsection{Estimated parameters in the mixture model}
We begin by reporting in Table \ref{tab:application_ProdSpecific} the posterior mean and standard deviation (s.d.) of the 100  brand fixed effect parameters. The 95\% posterior credibility intervals of most of these brand-specific intercepts exclude zero indicating that these brands are endowed with significant brand equity.  We next investigate the clustering of households according to the proposed mixture model. 
The posterior mode of the number of nonempty clusters under the full-specification (MNL\_RC) is six.
The Supplementary Material shows further estimation results on the number of clusters and the DP concentration parameter $\alpha$.

\begin{table}[h]
\caption{MNL\_RC Model on Cereal Market data: Estimates of the 100 brand fixed-effects}  
\centering
\resizebox{0.82\columnwidth}{!}{%
\begin{threeparttable}
\scalebox{1}{   
\begin{tabular}{llrlllrl}
& brand & mean & s.d & & brand & mean & s.d \\
\toprule
1 & BEAR NAKED FIT GRN & -0.02  & 0.29 & 51 & KELLOGGS FROOT LOOPS & -0.27* & 0.07 \\
2 & BEAR NAKED GRN & 0.48* & 0.19 & 52 & KELLOGGS FROOT LOOPS MARSHMALLOW & -1.38* & 0.18 \\
3 & BETTER OATS & -0.8* & 0.21 & 53 & KELLOGGS FROSTED FLAKES & 0.19* & 0.07 \\
4 & CREAM OF WHEAT & 0.24* & 0.11 & 54 & KELLOGGS FROSTED MINI\-WHEATS & 0.61* & 0.06 \\
5 & CTL BR & -0.34* & 0.06 & 55 & KELLOGGS FROSTED MINI\-WHT LTTLE BTS & -0.41* & 0.08 \\
6 & GENERAL MILLS APPLE CINNAMON CHEERIOS & -1.09* & 0.15 & 56 & KELLOGGS KRAVE & 0.64* & 0.09 \\
7 & GENERAL MILLS BLUEBERRY CHEX & -0.27  & 0.14 & 57 & KELLOGGS RAISIN BRAN & -0.02  & 0.07 \\
8 & GENERAL MILLS BREAKFAST PACK & -1.62* & 0.52 & 58 & KELLOGGS RAISIN BRAN CRUNCH & 0.08  & 0.07 \\
9 & GENERAL MILLS CHEERIOS & 0.2* & 0.06 & 59 & KELLOGGS RICE KRISPIES & -0.32* & 0.08 \\
10 & GENERAL MILLS CHEERIOS OAT CRUNCH CNMN & -0.38* & 0.11 & 60 & KELLOGGS RICE KRISPIES TREATS & 0.11  & 0.32 \\
11 & GENERAL MILLS CHOCOLATE CHEERIOS & -1.53* & 0.22 & 61 & KELLOGGS SPECIAL K & -0.49* & 0.17 \\
12 & GENERAL MILLS CHOCOLATE CHEX & -0.43* & 0.13 & 62 & KELLOGGS SPECIAL K CHOCOLATY DELGHT & 0.18  & 0.11 \\
13 & GENERAL MILLS CHOCOLATE PNUT BTR CHEERIO & -1.01* & 0.19 & 63 & KELLOGGS SPECIAL K CINNAMON PECAN & -0.57* & 0.17 \\
14 & GENERAL MILLS CINNAMON CHEX & -1.14* & 0.18 & 64 & KELLOGGS SPECIAL K FRUIT \& YOGURT & 0.25* & 0.12 \\
15 & GENERAL MILLS CINNAMON TOAST CRUNCH & 0.06  & 0.06 & 65 & KELLOGGS SPECIAL K PROTEIN & -0.13  & 0.11 \\
16 & GENERAL MILLS CINNAMON TOAST CRUNCH CHRS & -1.51* & 0.18 & 66 & KELLOGGS SPECIAL K RED BERRY & 0.24* & 0.08 \\
17 & GENERAL MILLS COCOA PUFFS & -0.55* & 0.09 & 67 & KELLOGGS SPECIAL K VANILLA ALMOND & 0.05  & 0.12 \\
18 & GENERAL MILLS COOKIE\-CRISP & -0.96* & 0.16 & 68 & KELLOGGS STBY KRISPIES US OLYMPC TM & -1.5* & 0.19 \\
19 & GENERAL MILLS CORN CHEX & -0.36* & 0.11 & 69 & M\-O\-M BERRY COLOSSAL CRN & -0.67* & 0.33 \\
20 & GENERAL MILLS FIBER ONE & 0.03  & 0.23 & 70 & M\-O\-M CINNAMON TOASTERS & -0.03  & 0.28 \\
21 & GENERAL MILLS FIBER ONE HONEY CLUSTERS & 0.42* & 0.21 & 71 & M\-O\-M COCOA DYNO\-BITES & -0.2  & 0.27 \\
22 & GENERAL MILLS FROSTED CHEERIOS & -1.67* & 0.3 & 72 & M\-O\-M FROSTED FLAKES & -0.25  & 0.31 \\
23 & GENERAL MILLS GOLDEN GRAHAMS & -0.06  & 0.08 & 73 & M\-O\-M FROSTED MINI SPOONERS & 0.72* & 0.29 \\
24 & GENERAL MILLS HONEY NUT CHEERIOS & 0.26* & 0.06 & 74 & M\-O\-M FRUITY DYNO\-BITES & 0.01  & 0.25 \\
25 & GENERAL MILLS HONEY NUT CHEX & -0.93* & 0.17 & 75 & M\-O\-M GOLDEN PUFFS & 0.37* & 0.18 \\
26 & GENERAL MILLS LUCKY CHARMS & 0.21* & 0.06 & 76 & M\-O\-M TOOTIE FRUITIES & -0.28  & 0.26 \\
27 & GENERAL MILLS MPL CHEERIOS CLC DSS FNDTN & -0.63* & 0.11 & 77 & POST COCOA PEBBLES & -0.42* & 0.15 \\
28 & GENERAL MILLS MULTIGRAIN CHEERIOS & 0.1  & 0.08 & 78 & POST FRUITY PEBBLES & -0.26* & 0.1 \\
29 & GENERAL MILLS NATURE VALLEY GRN PROTEIN & -0.35  & 0.37 & 79 & POST GOLDEN CRISP & -1.35* & 0.18 \\
30 & GENERAL MILLS RAISIN NUT BRAN & 0.26  & 0.17 & 80 & POST GRAPE\-NUTS & -0.21  & 0.21 \\
31 & GENERAL MILLS REESE'S PUFFS & 0.31* & 0.07 & 81 & POST GRAPE\-NUTS FLAKES & 0.31  & 0.29 \\
32 & GENERAL MILLS RICE CHEX & -0.33* & 0.1 & 82 & POST HONEY BUNCHES OF OATS & 0.37* & 0.07 \\
33 & GENERAL MILLS VANILLA CHEX & -0.74* & 0.16 & 83 & POST HONEY BUNCHES OF OATS GRN & -2.15* & 0.73 \\
34 & GENERAL MILLS VERY BERRY CHEERIOS & -0.86* & 0.15 & 84 & POST HONEY\-COMB & -0.97* & 0.14 \\
35 & GENERAL MILLS WHEAT CHEX & -0.56* & 0.16 & 85 & POST OREO OS & -1.63* & 0.36 \\
36 & GENERAL MILLS WHEATIES & 0.16  & 0.18 & 86 & POST RAISIN BRAN & -0.5* & 0.22 \\
37 & KASHI CINNAMON HARVEST & -0.55* & 0.29 & 87 & POST SELECTS GREAT GRAINS & -0.14  & 0.12 \\
38 & KASHI GO LEAN & -0.87* & 0.16 & 88 & POST SHRD WHT 'N BRN SP SZ & 0.46* & 0.16 \\
39 & KASHI GO LEAN CRUNCH! & -1.36* & 0.28 & 89 & POST SHREDDED WHEAT & -0.92* & 0.35 \\
40 & KASHI ORGANIC BLUEBERRY CLST & -1.4* & 0.28 & 90 & QUAKER & -0.05  & 0.06 \\
41 & KELLOGGS AL JS CN PS FRFL FTLP CKSP & -3.02* & 0.53 & 91 & QUAKER CAP'N CRN & -0.76* & 0.13 \\
42 & KELLOGGS ALL\-BRAN & -0.53  & 0.3 & 92 & QUAKER CAP'N CRN CRN BRY & -0.89* & 0.12 \\
43 & KELLOGGS ALL\-BRAN COMPLETE WHT FLK & 0.59  & 0.45 & 93 & QUAKER CINNAMON LIFE & -0.47* & 0.09 \\
44 & KELLOGGS APPLE JACKS & -0.25* & 0.09 & 94 & QUAKER GRN & -0.91  & 0.67 \\
45 & KELLOGGS CHOCOLT FRS FLKS TN TH TGR & -1.68* & 0.26 & 95 & QUAKER LIFE & -0.58* & 0.1 \\
46 & KELLOGGS COCOA KRISPIES & -0.64* & 0.12 & 96 & QUAKER OATMEAL SQUARES & -0.1  & 0.11 \\
47 & KELLOGGS CORN FLAKES & -0.37* & 0.1 & 97 & QUAKER OVERNIGHT OATS & -2.33* & 0.26 \\
48 & KELLOGGS CORN POPS & -0.5* & 0.12 & 98 & QUAKER PROTEIN & -0.49* & 0.15 \\
49 & KELLOGGS CRACKLIN' OAT BRAN & 0.49  & 0.24 & 99 & QUAKER REAL MEDLEYS & -1.06* & 0.22 \\
50 & KELLOGGS CRISPIX & -0.09  & 0.1 & 100 & QUAKER SELECT STARTS & -0.88* & 0.14 \\
\bottomrule 
\end{tabular}
}
\begin{tablenotes}
\small
\item 
The stars indicate that the corresponding 95\% credible interval does not include 0. 
The ``other'' option -specific fixed-effect is normalized to 0 for identification. 
The results are obtained under the MNL\_RC model. 
\end{tablenotes}
\end{threeparttable}
  }
\label{tab:application_ProdSpecific}
\end{table}
To understand how households are clustered, 
we computed the posterior mean of the event that 
a given pair of households $(i,k)$ are clustered together i.e.\ $\{S_i = S_{k}\}$. 
This results in a $n \times n$ similarity matrix, which can be found in the Supplementary Material.

An examination of how households are clustered reveals interesting points. 
Take household A as an example whose actual choices consist of $\{4,37,62,64,73\}$. 
Define an estimator $\hat{\mathcal{C}}_i$ of the consideration set for household $i$ as the set of brands $j$ whose posterior probability that $C_{ij}=1$ is greater than 0.2658,  the prior median of $ q_{hj}$.  
This results in the estimated set $\hat{\mathcal{C}}_{A}=\{4,26,37,59,60,61,62,64,68,73,79 ,  101\}$.
The upper panel of Table \ref{tab:application_SimilarityMatrix} lists the three households with the highest posterior similarity to subject $A$. 
There are several observations.

\begin{table}[ht]
\caption{Households clustered with $i\in\{A,B\}$.}  

\centering
\resizebox{1\columnwidth}{!}{%

\begin{threeparttable}
\begin{subtable}[ht]{0.95\textwidth}
\centering
\begin{tabular}{ c  l l l}
  $k$ &  Similarity  & Chosen brands &  $\hat{\mathcal{C}}_{k}$ \\  
 \toprule

\textit{Household $i=A$}\\ 
 $k=A$  & 1.00 
 & $\{4,37,62,64,73 \}$  
 &$\{4,26,37,59,60,61,62,64,68,73,79 ,  101  \}$ \\  
 $k=1357$  & 0.36
 & $\{8,11,73,89\}$ 
 &$\{3, 8,11,26,59,60,61,68,73,79,89,   101 \}$ \\  
 $k=226$  & 0.36
 & $\{5,25,33,39,53,63,77,79,81,89,92\}$ 
 &$\{3, 5,25,33,39,53,60,63,68,77,79,81,89,92 ,  101\}$ \\  
 $k=105$  & 0.35
 & $\{3, 5,26 ,  101\}$ 
 &$\{3, 5,25,26,59,60,61,63,66,68,73,79,   101\}$ \\  
  \midrule

\textit{Household $i=B$}\\  
 $k=B$  & 1.00 
 & $\{25,26,49,77,79,81,   101\}$  
 &$\{3,25,26,49,59,60,63,68,73,77,79,81,   101 \}$ \\  
 $k=1689$  & 0.71
 & $\{7,33,49,51,68,69,79,81,96 ,  101\}$ 
 &$\{3, 7,33,49,51,68,69,79,81,96 ,  101 \}$ \\ 
 $k=481$  & 0.65
 & $\{5, 7, 8,12,25,26,...,51,...,68,69,73,77,78,79,81,89,90,   101\}$ 
 & $\{5, 7, 8,12,25,26,...,51,...,68,69,73,77,78,79,81,89,90 ,  101\}$ \\   
 $k=1019$  & 0.62
 & $\{12,26,27,51,60,67,68,72,79,80,81,86\}$ 
 &$\{3,12,25,26,27,51,59,60,63,67,68,72,73,79,80,81,86,   101\}$ \\

\bottomrule
\end{tabular} 
\end{subtable}

\begin{tablenotes}
\small
\item 
Similarity is defined as the posterior mean  of $1\{S_{k}=S_{i}\}$, which corresponds to the ith row (equivalently ith column) of the similarity matrix presented in the Supplementary Material. The estimated consideration set $\hat{\mathcal{C}}_{i}$ is defined as the set of brands $j$ whose posterior probability that $ C_{ij}=1$ is greater than 0.2658, the prior median of  $q_{hj}$. The result is from the MNL\_{RC} model. 
\end{tablenotes}
\end{threeparttable}
  }
\label{tab:application_SimilarityMatrix}
\end{table}

First, the actual choices of the households tend to overlap within a cluster; each household purchased at least one of  brands 5, 73, or 89. 
Second, the estimated consideration sets $\hat{\mathcal{C}}_{k}$ are similar between households in a cluster. For example, household A did not choose brands 26, 79, and 101, but other households  did, and they are in $\hat{\mathcal{C}}_{A}$. 
Third, the stronger the purchase overlap, the higher the chance of being in the same cluster.  
The lower panel of Table \ref{tab:application_SimilarityMatrix} shows the results for household B. 
In this cluster, brands 79 and 81 were purchased by all the four households, brands 26, 51, 68, and 101 were each purchased by three households, and we see higher similarity scores ($\geq 0.60$).
In this way, our algorithm discovers the probabilistic grouping patterns in the choice data. 

\subsection{\textbf{Price sensitivity of demand}}

\begin{figure}[h!]
    \centering
    \includegraphics[width=0.45\linewidth]{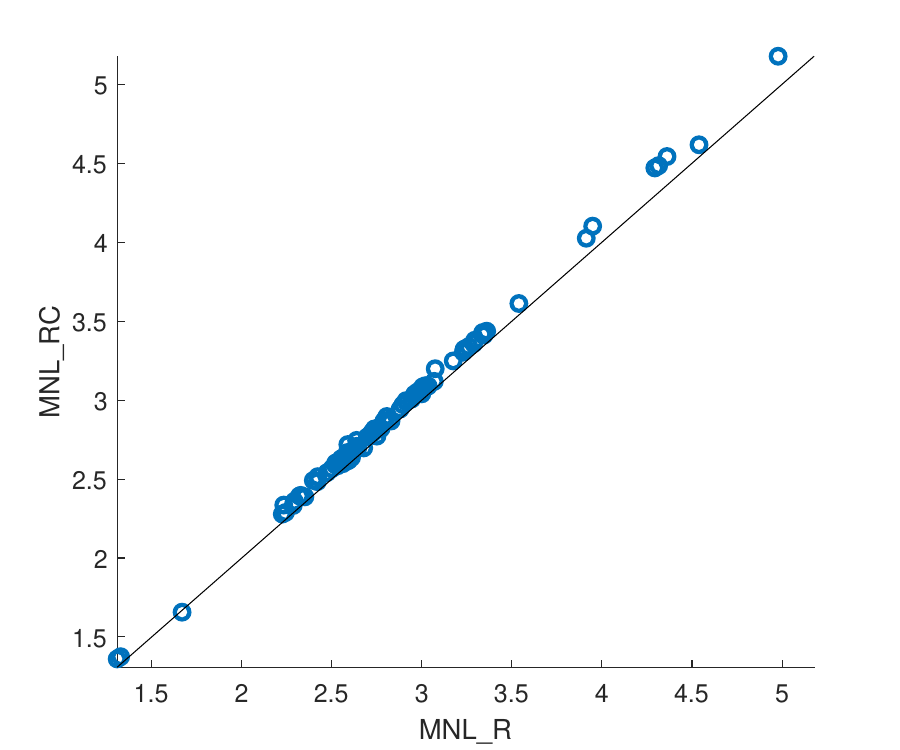}
    \caption{Price sensitivity of aggregate demand based on a random set of 100 households. Each circle represents the percentage decrease in demand of brand $j$ when its price increases by 1\%, $j=1,\ldots,J$ under the MNL\_R and MNL\_RC models. The  45-degree line is plotted as a  solid line.}
    \label{fig:application_elasticity}
\end{figure}

To analyze household shopping behavior, we randomly select 100 units and report in Figure \ref{fig:application_elasticity}
the percentage decrease in aggregate demand when the price of a brand increases by 1\% under the MNL\_R and MNL\_RC models. For all but one brand, this sensitivity is higher under consideration set heterogeneity, in conformity with previous findings that were derived in a small $J$ setting.

\subsection{Predictive performance}

\begin{figure}[ht] 
\centering
\includegraphics[width=0.4\linewidth]{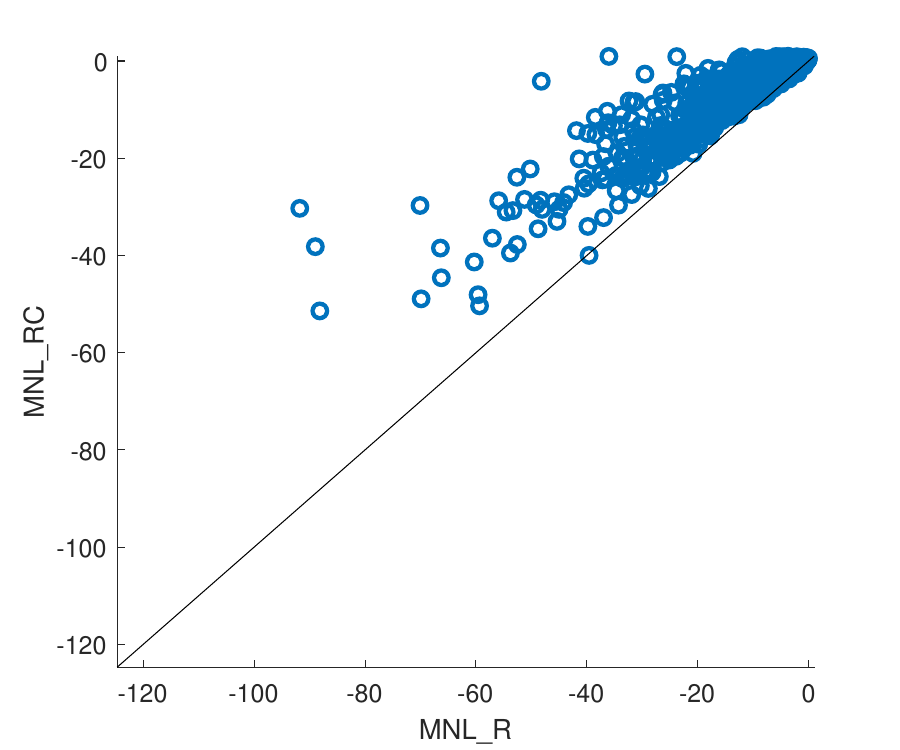} 
\caption{\small{Log-predictive likelihoods (circles) for the 1079 households that made purchases in the out-of-sample period. The x-coordinate of each circle is the log-predictive likelihood under MNL\_R, and the y-coordinate is under MNL\_RC. The  45-degree line is plotted as a  solid line.}}
\label{fig:application_LogPredLike}
\end{figure}

We next assess the predictive performance of the proposed model using the last two months of data as an out-of-sample period. 
Let $\mathcal{O}\subset \{1,\ldots, n\}$ denote the set of subjects who made purchases in the out-of-sample period. This set contains 1079 subjects.  
For each $i \in \mathcal{O}$, we predict $\bm Y^f_i = \{ Y_{iT_i+s}:  s=1,\ldots,h_i \}$, given the covariates $\bm w^f_i = \{ \bm w_{iT_i+s}:  s=1,\ldots,h_i \}$, where $h_i$ denotes the forecast horizon for the subject $i$. 
Let $\bm y^f_i=\{ y_{iT_i+s}:  s=1,\ldots,h_i \}$ be the actual set of responses for the subject $i \in \mathcal{O}$.
Then, as a measure of predictive performance, we calculate the predictive likelihoods 
\begin{align*}
p(\bm y^f_i \vert \bm y, \bm w,\bm w^f_i  )&=\int \prod_{s=1 }^{h_i} 
\Pr (Y_{iT_i+s}=y_{iT_i+s}|\bm \delta,\bm \beta, \bm b_{i},\bm w_{iT_i+s} , \mathcal{C}_i)d\pi(\bm \delta,\bm \beta, \{\bm b_{i}\}, \{\mathcal{C}_i \} \vert \bm y, \bm w) \nonumber \\
& \approx
\frac{1}{G}\sum_{g=1}^G  \prod_{s=1 }^{h_i}   \Pr (Y_{iT_i+s}=y_{iT_i+s}|\bm \delta^{(g)},\bm \beta^{(g)}, \bm b_{i}^{(g)},\bm w_{iT_i+s}, \mathcal{C}_i^{(g)}) \, , i \in \mathcal{O}\,  \label{eq:est_pred_like_household},     
\end{align*}
where the response probability conditional on a consideration set is given in \eqref{eq:mixed_logit_given_bi}.
Figure \ref{fig:application_LogPredLike} gives the log-predictive likelihood for each household under the (MNL\_R) and (MNL\_RC) models.  
The higher predictive likelihood under the latter model shows that including consideration set heterogeneity tends to improve predictive
performance. More details about this are given in the Supplementary Material.  

\section{Discussion}

In this concluding section, we discuss the broader relevance of the work, especially to the modeling of excess zeros in high-dimensional sparse microbiome data sets. In a microbiome dataset with $n$ samples and $J$ taxa, 
let $u_{ij}$ denote the measured count for taxon $j$ in sample $i$, and $T_i = \sum_{j=1}^J u_{ij}$ represent the total count over  taxa in the $i$th sample, where $i = 1, \ldots, n$ and $j = 1, \ldots, J$. Typically it is assumed that the vector $\bm{u}_i = (u_{i1}, \ldots, u_{iJ})'$ follows a multinomial distribution with index $T_i$ and a vector of probabilities $\bm{\rho}_i = (\rho_{i1}, \ldots, \rho_{iJ})'$, where $0 < \rho_{ij} < 1$ and $\sum_{j=1}^J \rho_{ij} = 1$. In our notation, $u_{ij}$ relates to $Y_{it}$ through  $u_{ij}=\sum_{t=1}^{T_i}1(Y_{it}=j)$.
To address the high dimensionality and sparsity of such datasets, \cite{Zeng2023zero} propose the following hierarchical model translated in our terminology as
\begin{align*}
    & \Pr(\bm C_i = \bm c_i )=q_{j}^{c_{ij}} \left( 1-q_{j}\right)^{1-c_{ij}}, 
    \quad f_{i1}, \ldots, f_{ik} \overset{\text{ind}}{\sim} \mathcal{N}(0, 1), \\
    & \bm{u}_i \vert \bm{\rho}_i, T_i \overset{\text{ind}}{\sim} \text{MN}(\bm{\rho}_i, T_i), \quad
    \rho_{ij} = \frac{C_{ij} \exp\left(\beta_{0j} + \bm{f}_i' \bm{\beta}_j \right)}
    {\sum_{\ell=1}^J C_{i\ell} \exp\left(\beta_{0\ell} + \bm{f}_i' \bm{\beta}_\ell \right)},
\end{align*}
where $\bm C_i = (C_{i1},\ldots,C_{iJ})'$, and $1-C_{ij}$ are latent indicators for excess zeros, and the $q_{j}$ are the corresponding probabilities. 
The $\bm{f}_i$ are latent factors, and the $\bm{\beta}_j$ denote the loadings of the associated factors. 
Note that the excess zeros across taxa are independent in this modeling. 
In other words, the model above corresponds to the independent consideration model that we review in the  introduction. 
In this context, complex dependency patterns across taxa in the excess zeros can be captured 
by our modeling. To do this, we would let $\bm C_i$ be correlated vectors and i.i.d.\ with the density for the infinite mixture of independent consideration models (3):
\[
\Pr(\bm C_i = \bm c_i )=
\sum_{h=1}^\infty  \omega_h 
\prod_{j=1}^J 
\left\{
q_{hj}^{c_{ij}} \left( 1-q_{hj}\right)^{1-c_{ij}} 
\right\}. 
\]
We can adapt the MCMC framework of this paper to estimate this model. 
Our approach for updating  the $\bm C_i$'s and the mixture parameters can be used in conjunction with existing approaches for simulating the factor-related objects.

Another key issue in practice is variable selection when many subject-level covariates are available. This challenge can be addressed using shrinkage priors. A natural extension of our framework involves modeling consideration sets that change at one or two points in time due to learning from past choices. This would require incorporating the learning process into the model and modifying the theoretical analysis accordingly. We leave this promising direction for future work. 



\appendix

\section{Proof of Lemma 2}

\begin{proof}[Proof of Lemma \ref{lem:KL_continuity}]
Recall that 
$
p_{\bm{\theta}, \bm{\pi} } (\bm y\vert \bm w)  \equiv \sum_{c\in \mathcal{C}} \pi_{c} 
\Pr\left( \bm Y_{i} = \bm y \vert \bm \theta, \bm w, c \bm \right).
$
For any $\bm y \in \mathcal{J}^T$, if  $  p_{\bm{\theta}^*, \bm{\pi}^* } (\bm y\vert \bm w) =0$, the integrand in the KL divergence is $\log(0) 0$ 
which is defined to be zero. 
Therefore, without loss of generality, suppose that for all $\bm y$, there is  $c_y\in \mathcal{C}$ that contains all the elements of $\bm y$ and have $\pi^*_{c_y}>0$.
By Lemma \ref{lem:dunson_xing}, we can find a finite mixture of independent consideration models that exactly matches the true distribution of consideration sets; i.e. $\exists (K,\tilde{\bm \phi}_{1:K})$ such that $\pi^*_c=\sum_{h=1}^K\tilde{\omega}_h \prod_{j\in c} \tilde{q}_{hj} \prod_{j \notin c} (1-\tilde{q}_{hj})$ for each $c\in \mathcal{C}$.    
Hence we have
\begin{align*}
    &\int \sum_{\bm y}
    \log 
    \left(
    \frac{
    p_{\bm{\theta}^*, \bm{\pi}^* } (\bm y\vert \bm w)}{
    p(\bm y\vert \bm w; \bm \theta,K,\bm \phi_{1:K})
    } 
    \right) 
    p_{\bm{\theta}^*, \bm{\pi}^* } (\bm y\vert \bm w) g^* (\bm w) d\bm w \\
    &=
    \int \sum_{\bm y} 
    \left\{
    \log 
    \left(
    \frac{
    p_{\bm{\theta}^*, \bm{\pi}^* } (\bm y \vert \bm w)}{
    p(\bm y \vert \bm w; \bm \theta^*,K,\tilde{\bm \phi}_{1:K})
    } 
    \right) 
    +
    \log 
    \left(
    \frac{
    p(\bm y \vert \bm w; \bm \theta^*, K,\tilde{\bm \phi}_{1:K}) }{
    p(\bm y \vert \bm w; \bm \theta,K,\bm \phi_{1:K})
    } 
    \right) 
    \right\}
    p_{\bm{\theta}^*, \bm{\pi}^* } (\bm y \vert \bm w) g^* (\bm w) d\bm w  ,  
\end{align*}
and the first term in the brackets  is zero. Hence, it suffices to show that the integral of the second term is continuous in $(\bm \theta, \bm \phi_{1:K})$ 
at $(\bm \theta^*, \tilde{\bm \phi}_{1:K})$. 
In the  Supplementary Material, we prove that the response probability is continuous in $\bm \phi_{1:K}$ (Lemma SB1) and it is continuous also in $\bm \theta$ (Lemma SB2). 
Let $(\bm \theta^m,\bm \phi_{1:K}^m)$ be a sequence of parameter values converging to $(\bm \theta^*, \tilde{\bm \phi}_{1:K})$. Then 
\[
\lim_{m\to \infty} \log\left( 
\frac{
p(\bm y \vert \bm w; \bm \theta^*, K, \tilde{\bm \phi}_{1:K} )
}{
p(\bm y \vert \bm w; \bm \theta^m, K, \bm \phi^m_{1K} )    
}
\right)=0.
\]
The result will follow from the dominated convergence theorem if there is an integrable (with respect to $p_{\bm \theta^*,\bm \pi^*}(\bm y \vert \bm w) g^*(\bm w)$) upper bound of $|\log p(\bm y \vert \bm w; \bm \theta^m, K, \bm \phi_{1:K}^m)|$. Note that 
\begin{align*}
p(\bm y \vert \bm w; \bm \theta^m, K, \bm \phi_{1:K}^m)
&=\sum_{c\in \mathcal{C}} 
\pi(c \vert K, \bm \phi^m_{1K}) 
\Pr(\bm Y_{i}=\bm y \vert \bm \theta^m, \bm w, c)\\
&\geq
\pi(c_y \vert K, \bm \phi^m_{1K}) 
\Pr(\bm Y_{i}=\bm y \vert \bm \theta^m, \bm w, c_y)      ,      
\end{align*}
where $\pi(c_y \vert K, \bm \phi^m_{1K})=\sum_{h=1}^K \omega_h^m \prod_{\ell \in c_y} q^m_{h\ell} \prod_{\ell \notin c_y} (1-q^m_{h\ell})$.
First, since $\bm \phi^m_{1K} \to \tilde{\bm \phi}_{1:K}$ and $\pi^*_{c_y}=\sum_{h=1}^K \tilde{\omega}_h \prod_{\ell \in c_y} \tilde{q}_{h\ell} \prod_{\ell \notin c_y} (1-\tilde{q}_{h\ell})>0$, the first term is bounded below by some $\ell_1(\bm y)>0$ for sufficiently large $m$. 
Second, since $\bm \theta^m\to \bm \theta^*$ and 
$\Pr(\bm Y_{i}=\bm y \vert \bm \theta^*, \bm w, c_y) >0$ (as $\bm \beta^*$ is in a compact set, $\bm D^*$ is positive definite, and $\mathcal{W}$ is compact), 
$\Pr(\bm Y_{i}=\bm y \vert \bm \theta^m, \bm w, c_y)$      
is  bounded below by some $\ell_2(\bm y,\bm w)>0$ for sufficiently large $m$.
Finally, 
$
1\geq 
p(\bm y \vert \bm w; \bm \theta^m, K, \bm \phi_{1:K}^m)
\geq
\inf_{\bm w\in \mathcal{W}} \min_{\bm y \in \mathcal{J}^T}
\ell_1(\bm y) \ell_2(\bm y,\bm w)>0,
$
for all $(\bm y,\bm w) \in \mathcal{J}^T \times \mathcal{W}. $
\end{proof}

\begin{spacing}{1.00}
\bibliography{references}
\end{spacing}


\clearpage 
\bigskip
\begin{center}
{\large\bf Acknowledgments}
\end{center}
We thank the editor and the review panel for excellent comments and suggestions that have improved the manuscript. 
We benefited from  helpful comments from 
Dimitris Korobilis,  
Andriy Norets, 
Nail Kashaev, 
Alessandro Iaria,
Ao Wang, 
Zhentong Lu, 
Toru Kitagawa,  
Yao Luo, 
Kosuke Uetake, 
Victor Aguiar, 
Laura Liu, 
Paola Manzini, 
Francesca Molinari, 
Elena Manresa,
Martin Weidner,  
Asim Ansari, 
Frank Schorfheide,  
Victor Aguirregabiria, 
Andrew Ching, 
Matthijs Wildenbeest, 
Mitsuru Igami,
and the participants at 
the economics and statistics seminars at the University of Alberta, 
the 2023 NBER-NSF Seminar in Bayesian Econometrics and Statistics (SBIES) at the Federal Reserve Bank of Philadelphia,
the 2023 INFORMS Annual Meeting  in Phoenix, 
the 2024 International Industrial Organization Conference in Boston,  
the 2025 Marketing Science Conference in Washington DC, 
and 
the 2025 Econometric Society World Congress in Seoul. 
Kenichi Shimizu gratefully acknowledges financial support from the Canadian Social Sciences and Humanities Research Council Insight Development Grant. 
Disclaimer: Researchers' own analyses calculated (or derived) based in part on data from Nielsen Consumer LLC and marketing databases provided through the NielsenIQ Datasets at the Kilts Center for Marketing Data Center at The University of Chicago Booth School of Business. The conclusions drawn from the NielsenIQ data are those of the researchers and do not reflect the views of NielsenIQ. NielsenIQ is not responsible for, had no role in, and was not involved in analyzing and preparing the results reported herein.

\clearpage
\begin{center}
{\large\bf SUPPLEMENTARY MATERIAL}
\end{center}

\appendix
\numberwithin{equation}{section}
\counterwithin{figure}{section}
\counterwithin{table}{section}
\counterwithin{lemma}{section}

Section \ref{sec:proofs} provides intermediate results used to prove Theorems 1 and 2 of the main paper, and their proofs.
Section \ref{sec:conditionals} presents the conditional posterior distributions of the parameters other than the consideration sets. 
Section \ref{sec:prior_on_cs} illustrates the impact of the prior choice for the attention probabilities on the prior on the distribution of consideration sets. 
Section \ref{sec:simulation_additional} shows additional simulation results. 
Section \ref{sec:additional_application} provides additional results from the empirical application. 

\renewcommand{\thesection}{S\Alph{section}}


\section{Intermediate theoretical results and proofs}\label{sec:proofs}

\subsection{Intermediate results in the proof of Lemma 2}

The following two lemmas are used to prove Lemma 2. 
They state that the marginal response probability is continuous with respect to the mixture parameters as well as the parameters in the response model. 
We prove the intermediate results for the case of $T=1$. The extensions to the $T>1$ case can be done similarly but at the expense of proof simplicity.

\begin{lemma}[Continuity of response probabilities wrt mixture  parameters]\label{lem:continuity_ccp_wrt_mixtures}
    Let $\bm \theta$ and $\bm w\in \mathcal{W}$. Then for each $j\in \mathcal{J}$, 
    $\forall \varepsilon>0$ and $\bm \phi_{1:K}^{(1)}$,
    $\exists \delta>0$ such that 
    for any $\bm \phi_{1:K}^{(2)}$ satisfying 
    $\sum_{j=1}^J |q_{hj}^{(1)}- q_{hj}^{(2)} |< \delta$ and 
    $|\omega_h^{(1)}-\omega_h^{(2)}|<\delta$,        
    for $h=1,\ldots, K$, we have 
    \[
    \left| p(j \vert \bm w; \bm \theta, K, \bm \phi_{1:K}^{(1)})
    -
    p(j \vert \bm w; \bm \theta, K, \bm \phi_{1:K}^{(2)})    
    \right| < \varepsilon.
    \]        
\end{lemma}

\begin{proof}[Proof of Lemma \ref{lem:continuity_ccp_wrt_mixtures}]
We have
\begin{align*}
    \left| 
    p(j \vert \bm w; \bm \theta, K, \bm \phi_{1:K}^{(1)})
    - 
    p(j \vert \bm w; \bm \theta, K, \bm \phi_{1:K}^{(2)})
    \right|
    \leq 
    \sum_{c\in \mathcal{C}} 
    \left| 
    \pi(c\vert K,\bm \phi_{1:K}^{(1)})
    -
    \pi(c\vert K,\bm \phi_{1:K}^{(2)})  
    \right|
    \Pr(Y_{it}=j \vert \bm \theta, \bm w_t, c)   
\end{align*}
where $\Pr(Y_{it}=j \vert \bm \theta, \bm w_t, c)   \leq 1$.
The term in the absolute value is 
\begin{align*}
    &\left| 
    \sum_{h=1}^K \omega_h^{(1)} \prod_{j \in c} q_{hj}^{(1)} \prod_{j \notin c} (1-q_{hj}^{(1)})
    -
    \sum_{h=1}^K \omega_h^{(2)} \prod_{j \in c} q_{hj}^{(2)} \prod_{j \notin c} (1-q_{hj}^{(2)})
    \pm 
    \sum_{h=1}^K \omega_h^{(1)} \prod_{j \in c} q_{hj}^{(2)} \prod_{j \notin c} (1-q_{hj}^{(2)})
    \right| \\
    &\leq
    \sum_{h=1}^K \omega_h^{(1)} 
    \left| 
    \underbrace{    
    \prod_{j \in c} q_{hj}^{(1)} \prod_{j \notin c} (1-q_{hj}^{(1)})
    -
    \prod_{j \in c} q_{hj}^{(2)} \prod_{j \notin c} (1-q_{hj}^{(2)})
    }_{I}
    \right|
    +
    \sum_{h=1}^K \left| \omega_h^{(1)}-\omega_h^{(2)} \right|
\end{align*}
The term $I$ equals to 
\begin{align*}
    &\prod_{j \in c} q_{hj}^{(1)} \prod_{j \notin c} (1-q_{hj}^{(1)})
    -
    \prod_{j \in c} q_{hj}^{(2)} \prod_{j \notin c} (1-q_{hj}^{(2)})
    \pm
    \prod_{j \in c} q_{hj}^{(2)} \prod_{j \notin c} (1-q_{hj}^{(1)}) \\
    &=
    \left( \prod_{j \in c} q_{hj}^{(1)}-\prod_{j \in c} q_{hj}^{(2)}\right)
    \prod_{j \notin c} (1-q_{hj}^{(1)})
    +
    \prod_{j \in c} q_{hj}^{(2)}
    \left( \prod_{j \notin c} (1-q_{hj}^{(1)}) - \prod_{j \notin c} (1-q_{hj}^{(2)}) \right),
\end{align*}
and hence the absolute value of $I$ is bounded by 
the sum of the two terms: 
$\left| \prod_{j \in c} q_{hj}^{(1)}-\prod_{j \in c} q_{hj}^{(2)}\right|$
and 
$\left| \prod_{j \notin c} (1-q_{hj}^{(1)}) - \prod_{j \notin c} (1-q_{hj}^{(2)}) \right|$. It is easy to show that the former is bounded by $c_1\sum_{j\in c} | q_{hj}^{(1)}-q_{hj}^{(2)} |$ and the latter is bounded by $c_2\sum_{j\notin c} | q_{hj}^{(1)}-q_{hj}^{(2)} |$ for some $c_1,c_2>0$. So,  $|I| \leq c_3 \sum_{j=1}^J | q_{hj}^{(1)}-q_{hj}^{(2)} |$ for some $c_3>0$.
\end{proof}

\begin{lemma}[Continuity of response probabilities wrt $\theta$]\label{lem:continuity_ccp_wrt_preference}
    Suppose $\mathcal{W}$ is compact. 
    Let $(K,\bm \phi_{1:K})$ and $\bm w\in \mathcal{W}$. Then for each $j \in \mathcal{J}$,
    $\forall \varepsilon>0$ and $\bm \theta^{(1)}=\{ \bm \beta^{(1)},\bm D^{(1)}\}$,
    $\exists \delta>0$ such that for any $\bm \theta^{(2)}=\{\bm \beta^{(2)},\bm D^{(2)}\}$ satisfying $|| \bm \beta^{(1)} - \bm \beta^{(2)} || <\delta$ and \\
    $\sqrt{\text{tr} (\bm D^{(1)^{-1}}\bm D^{(2)}-\bm I)-\log\det (\bm D^{(1)} \bm D^{(2)^{-1}}) } <\delta$, 
    \[
    \left| p(j \vert \bm w; \bm \theta^{(1)}, K, \bm \phi_{1:K} )
    -
    p(j \vert \bm w; \bm \theta^{(2)}, K, \bm \phi_{1:K} )   
    \right|<\varepsilon.
    \]
\end{lemma}

\begin{proof}[Proof of Lemma \ref{lem:continuity_ccp_wrt_preference}]
Recall that for $j\in c$, 
\[
Pr(Y=j\vert \bm \theta, \bm w, c)=\int k_j(\bm w,  \bm \beta, \bm b) \phi(\bm b \vert \bm 0, \bm D) d \bm b,
\]
where we introduced the shorthand notation  for the kernel 
\[
k_j(\bm w,  \bm \beta, \bm b)
=\frac{ e^{\bm x_j'\bm \beta+\bm z_j'\bm b} }{\sum_{\ell \in c}  e^{\bm x_\ell'\bm \beta+\bm z_\ell'\bm b} },
\]
where we suppressed the subscripts with respect to the units $i$ for simplicity of notation. 
We have 
 \begin{align*}
    \left| p(j \vert \bm w; \bm \theta^{(1)}, K, \bm \phi_{1:K} )
    -
    p(j \vert \bm w; \bm \theta^{(2)}, K, \bm \phi_{1:K} )   
    \right|
    &\leq 
    \sum_{c\in \mathcal{C}}
    \pi(c\vert K, \bm \phi_{1:K}) 
    \left| \Pr(j\vert \bm \theta^{(1)},\bm w, c) - \Pr(j\vert \bm \theta^{(2)},\bm w, c)\right|\\
    &=
    \sum_{c: j \in c}
    \pi(c\vert K, \bm \phi_{1:K}) \left| 
    \Pr(j\vert \bm \theta^{(1)},\bm w, c) - \Pr(j\vert \bm \theta^{(2)},\bm w, c)\right|,
 \end{align*} 
 where if there is no $c \in \mathcal{C}$ such that $j \in c$ and $\pi(c\vert K, \bm \phi_{1:K})>0$, the claim is trivially true. Now, 
 \begin{align}
    \left|
    \Pr(j\vert \bm \theta^{(1)},\bm w, c)) - \Pr(j\vert \bm \theta^{(2)},\bm w, c) 
    \right|
     &\leq 
    \left| 
    \Pr(j\vert \{\bm \beta^{(1)}, \bm D^{(1)} \}, \bm w, c) 
    -
    \Pr(j\vert \{\bm \beta^{(2)}, \bm D^{(1)} \}, \bm w, c)    
    \right| \label{bd:lem_continuity_ccp_wrt_prefrence_1}\\
    &+
    \left| 
    \Pr(j\vert \{\bm \beta^{(2)}, \bm D^{(1)} \}, \bm w, c) 
    -
    \Pr(j\vert \{\bm \beta^{(2)},\bm D^{(2)} \}, \bm w, c)    
    \right|.\label{bd:lem_continuity_ccp_wrt_prefrence_2}        
 \end{align}
 
To bound \eqref{bd:lem_continuity_ccp_wrt_prefrence_1}, note that for any $\rho>0$, one can find $M_\rho>0$ such that $\int 1\{||\bm b|| > M_\rho \}\phi(\bm b \vert \bm 0, \bm D^{(1)})d\bm b< \rho$.  The term \eqref{bd:lem_continuity_ccp_wrt_prefrence_1} equals to
\begin{align*}
    &\left| \int \left( 
    k_j(\bm w, \bm \beta^{(1)}, \bm b) 
    -
    k_j(\bm w, \bm \beta^{(2)}, \bm b)    
    \right) 
    \phi(\bm b \vert \bm 0, \bm D^{(1)})d\bm b  \right|  \\
    &\leq 
    \int_{||\bm b|| \leq M_\rho} 
    \left| 
    k_j(\bm w, \bm \beta^{(1)}, \bm b) 
    -
    k_j(\bm w, \bm \beta^{(2)}, \bm b)       
    \right|
    \phi(\bm b \vert \bm 0, \bm D^{(1)})d\bm b    \\
    &+
    \int_{||\bm b|| > M_\rho} 
    \left| 
    k_j(\bm w, \bm \beta^{(1)}, \bm b) 
    -
    k_j(\bm w, \bm \beta^{(2)}, \bm b)       
    \right|
    \phi(\bm b \vert \bm 0, \bm D^{(1)})d\bm b  .  
\end{align*}
Since $k_j(\bm w, \bm \beta, \bm b)$ has  a bounded first derivative with respect to $\bm \beta$ for $||\bm b|| \leq M_\rho$ and under a compact $\mathcal{W}$, there is some $c_1>0$ such that the first term above is bounded by 
$c_1||  \bm \beta^{(1)}- \bm \beta^{(2)} ||$. The second term is bounded by 
$2\int 1\{ ||\bm b|| > M_\rho\}    \phi(\bm b \vert \bm 0, \bm D^{(1)})d\bm b    
< 2\rho$, which can be made smaller than $|| \bm \beta^{(1)} - \bm \beta^{(2)} ||$. 
Hence, \eqref{bd:lem_continuity_ccp_wrt_prefrence_1} is bounded by $c_2||  \bm \beta^{(1)} - \bm \beta^{(2)} ||$
for some constant $c_2>0$.

The term  \eqref{bd:lem_continuity_ccp_wrt_prefrence_2} equals to 
\begin{align*}
    &\left| \int 
    k_j(\bm w, \bm \beta^{(2)}, \bm b)   
    \left(
    \phi(\bm b \vert \bm 0, \bm D^{(1)}) 
    -
    \phi(\bm b \vert \bm 0, \bm D^{(2)})
    \right)d\bm b \right| \\
    &\leq 
    \int 
    \left| 
    \phi(\bm b \vert \bm 0, \bm D^{(1)}) 
    -
    \phi(\bm b \vert \bm 0, \bm D^{(2)})    
    \right|d\bm b \\
    &\leq 
    \sqrt{\text{tr} (\bm D^{(1)^{-1}}\bm D^{(2)}-\bm I)-\log\det (\bm D^{(1)} \bm D^{(2)^{-1}}) } ,
\end{align*}
 where the last inequality is due to a known bound on the total variation distance between normal distributions with a same mean vector but different covariance matrices (\citealp{Devroye2018TVGaussian}).

\end{proof}

\subsection{Intermediate results in the proof of Theorem 2}

The next lemma shows that the response probabilities are continuous for the total variation distance defined as 
\[
d_{TV}(\bm p_{\bm \beta,\bm \pi} , \bm p_{\bm \beta',\bm \pi'} )
=
\int \sum_{\bm y \in \mathcal{J}^T} | p_{\bm \beta,\bm \pi}(\bm y \vert \bm w)g^*(\bm w) -  p_{\bm \beta',\bm \pi'}(\bm y \vert \bm w)g^*(\bm w) | d \bm w.
\]
\begin{lemma}[Continuity of response probabilities]\label{lem:continuity_ccp_total_variation}
    Let $\varepsilon>0$. Then there is $\delta>0$ such that $d((\bm \beta, \bm \pi), (\bm \beta', \bm \pi'))<\delta$ implies that 
    $d_{TV}(\bm p_{\bm \beta,\bm \pi} , \bm p_{\bm \beta',\bm \pi'} )<\varepsilon$.
\end{lemma}

\begin{proof}[Proof of Lemma \ref{lem:continuity_ccp_total_variation}]
\begin{align*}
    &| p_{\bm \beta,\bm \pi}(\bm y \vert \bm w)-  p_{\bm \beta',\bm \pi'}(\bm y \vert \bm w) | 
    \leq 
    | p_{\bm \beta,\bm \pi}(\bm y \vert \bm w)-  p_{\bm \beta',\bm \pi}(\bm y \vert \bm w) | 
+
    | p_{\bm \beta',\bm \pi}(\bm y \vert \bm w)-  p_{\bm \beta',\bm \pi'}(\bm y \vert \bm w) | \\
    &\leq
\sum_c \pi_c 
\left|  
\prod_{t=1}^T \Pr\left( Y_{it} = y_t \vert \bm \beta, \bm w_t, c \bm \right) 
-     
\prod_{t=1}^T \Pr\left( Y_{it} = y_t \vert \bm \beta', \bm w_t, c \bm \right) 
\right|  \\
&+
\sum_c |\pi_c - \pi'_c | 
\prod_{t=1}^T \Pr\left( Y_{it} = y_t \vert \bm \beta', \bm w_t, c \bm \right) 
\\
    &\leq \gamma_1 ||\bm \beta - \bm \beta' ||_2 
    + 
    \gamma_2 ||\bm \pi - \bm \pi' ||_1,
\end{align*}
for some positive constants $\gamma_1$ and $\gamma_2$. 
\end{proof}

\section{Conditional posterior distributions}\label{sec:conditionals}
For the mixture model on the latent consideration sets $\bm C=(\bm{C}_1,\ldots,\bm{C}_n)$, let $S_i \in\{1,2,\ldots\}$ be the latent cluster assignment such that
$C_{ij} \vert S_i=h\sim \text{Bernoulli}(q_{hj})$, independently $j=1,\ldots,J$, for $i=1,\ldots,n$.
We have 
the latent consideration sets $\bm C$, 
the common fixed-effects $\bm \beta$, 
the random effects $\bm b$,
the corresponding covariance matrix $\bm D$,  
the DP parameters $\bm V =(V_1,V_2,\ldots)$ as well as $\bm Q=(\bm q_1,\bm q_2,\ldots)$,  
the DP cluster assignment variables $\bm S =(S_1,\ldots,S_n)$, and 
the DP concentration parameter $\alpha$. 
Then, from the Bayes theorem, we define the posterior density of interest to be
\begin{align}
p\big( \bm C, & \    \bm S, \bm V, \bm Q, \alpha,  \bm \beta, \bm b, \bm D \big\vert \bm y , \bm W \big)
\propto 
p\big(\bm y \big\vert   \bm \beta, \bm b, \bm W, \bm C  \big) 
\cdot 
p( \bm \beta, \bm b, \bm D)
\cdot 
p\big( \bm C,   \bm S, \bm Q,\bm V,  \alpha  \big)  
\nonumber  \\
&=
p\big(\bm y \big\vert  \bm \beta, \bm b, \bm W, \bm C  \big) 
\cdot 
\pi(\bm \beta)p(\bm b \vert \bm D) \pi(\bm D)
\cdot 
p\big( \bm C,   \bm S, \bm Q,\bm V,  \alpha \big),  \label{eq_ap:joint_distribution}
\end{align}
where 
the first term is the likelihood function and
$\pi(\cdot)$ denotes the prior density.  
Only the last term in \eqref{eq_ap:joint_distribution} is associated with the DP model and 
\begin{align}
p\big( \bm C,   \bm S, \bm Q,\bm V,  \alpha \big)
&\propto 
p \big( \bm C \vert \bm Q, \bm S\big)
p\big( \bm Q, \bm V, \bm S, \alpha \big) \nonumber \\
&\propto 
\left[ 
\prod_{i=1}^n p\big( \bm C_i \vert \bm{q}_{S_i}\big) 
p\big(S_i \vert \bm V \big) 
\right]
\cdot 
\left[ 
\prod_{h=1}^\infty p\big( V_h \vert \alpha \big) 
p(\bm q_h \vert \underline{\bm \phi}_q \big) 
\right]
\cdot 
\pi(\alpha) , \label{eq_ap:joint_distribution_mixture}
\end{align}
where 
$p\big( \bm C_i \vert \bm{q}_{S_i}\big)$ is the product of densities for the  independent Bernoulli distributions  $\text{Bernoulli}(q_{S_ij})$ $j=1,\ldots,J$ ,
$p\big(S_i \vert \bm V \big) = \omega_{S_i}$, 
$p\big( V_h \vert \alpha \big)$ is the density of $\text{Beta}(1,\alpha)$,  
$p(\bm q_h \vert \underline{\bm \phi}_q \big) $ is the product of  densities for the independent Beta distributions $\text{Beta}(\underline{a}_{q_j},\underline{b}_{q_j})$ $j=1,\ldots,J$, and 
$\pi(\alpha)$ is the prior density for $\alpha$. 
We apply the slice sampling approach (\citealp{Walker2007}) by augmenting the joint distribution with a sequence of auxiliary random variables $\bm u = (u_1,\ldots,u_n)$ that follow the uniform distribution on $(0,1)$, $u_i\sim \mathcal{U}(0,1)$, $i=1,\ldots,n$:
\begin{equation}
p(\bm  C,\bm  S, \bm Q,\bm  V, \bm u, \alpha) 
\propto 
\left[ 
\prod_{i=1}^n p\big( \bm C_i \vert \bm{q}_{S_i}\big) 
 I(u_i \leq \omega_{S_i} )
\right]
\cdot 
\left[ 
\prod_{h=1}^\infty p\big( V_h \vert \alpha \big) 
p(\bm q_h \vert \underline{\bm \phi}_q \big) 
\right]
\cdot 
\pi(\alpha).
\label{eq_ap:joint_distribution_mixture_withU}
\end{equation}
It is easy to show that we can recover \eqref{eq_ap:joint_distribution_mixture} by integrating out $\bm u$ from \eqref{eq_ap:joint_distribution_mixture_withU}. 
However, by introducing $\bm u$, one only has to choose labels $S_i$ in the finite set $\{h: \omega_h \geq u_i\}$.
See the Supplementary Material for  discussion on hyperparameter selections.

Our MCMC algorithm proceeds by cycling through various conditional
distributions, where these distributions are conditioned on the most
recent values of the remaining unknowns. Specifically, 
given the current draw at the $g$th iteration $\{ u_i^{(g)} \}, \{ V_h^{(g)} \}, 
\{ \bm q_h^{(g)} \}, \{ S_i^{(g)} \},$ $\alpha^{(g)}$, $\big\{ \bm C_i^{(g)} \big\}, 
\bm \delta^{(g)}$, $\bm \beta^{(g)}, \big\{\bm b_i^{(g)} \big\}$, and  $\bm D^{(g)}$, 
the next draw in the sequence is obtained by simulating 
\begin{align*}
\bm \beta^{(g+1)} &\text{ from }
\bm \beta \big\vert 
\big\{y_{it} \big\}, 
\big\{\bm C_i^{(g)} \big\},
\bm \delta^{(g)}, 
\big\{\bm b_i^{(g)} \big\},
\\
\big\{\bm b_i^{(g+1)} \big\} &\text{ from }
\big\{\bm b_i \big\}  \big\vert 
\big\{y_{it} \big\}, 
\big\{\bm C_i^{(g)} \big\},
\bm \delta^{(g)}, 
\bm \beta^{(g+1)},
\bm D^{(g)}, 
\\
\bm D^{(g+1)}&\text{ from } 
\bm D \big\vert \big\{\bm b_i^{(g+1)} \big\} ,
\\
\bm \delta^{(g+1)} &\text{ from }
\bm \delta \big\vert 
\big\{y_{it} \big\}, 
\big\{\bm C_i^{(g)} \big\},
\bm \beta^{(g+1)}, 
\big\{\bm b_i^{(g+1)} \big\},\\
\big\{ \bm C_i^{(g+1)} \big\} &\text{ from } 
\big\{  \bm C_i \big\} \big\vert 
\big\{y_{it} \big\}, 
\bm \delta^{(g+1)}, 
\bm \beta^{(g+1)}, \big\{\bm b_i^{(g+1)} \big\},
\{ \bm q_h^{(g)} \},
\{S_i^{(g)} \},
\\
\{ V_h^{(g+1)} \} &\text{ from } 
\{ V_h  \} \big\vert  \{ S_i^{(g)} \}, \alpha^{(g)}, \\
\{ \bm q_h^{(g+1)} \} &\text{ from } 
\{ \bm q_h  \} \big\vert 
\big\{ \bm C_i^{(g+1)} \big\}, 
\{S_i^{(g)} \},
\\
\{ u_i^{(g+1)} \} &\text{ from } 
\{ u_i \} \big\vert \{ S_i^{(g)} \},\{ V_h^{(g+1)} \}, \\
\{ S_i^{(g+1)} \} &\text{ from } 
\{ S_i \} \big\vert \{ u_i^{(g+1)} \}, \{ \bm q_h^{(g+1)} \}, \{ V_h^{(g+1)} \}, \big\{ \bm C_i^{(g+1)} \big\},  \\
\alpha^{(g+1)} &\text{ from } 
\alpha \big\vert \{ V_h^{(g+1)} \}, \{ S_i^{(g+1)} \}.
\end{align*}
Repeating this procedure $G$ times (beyond a suitable burn-in) produces a sample from the posterior distribution.

The main paper illustrates how the consideration sets are simulated. In this section, we show the conditional posterior distributions of the remaining parameters.
Let $K^*=\min \{h: \sum_{\ell=1}^h \omega_h > 1-u^* \}$, where $u^*=\min (u_1,\ldots,u_n).$
Define $n_h = \sum_{i=1}^n I(S_i=h)$. 
Let  the dot $\bullet$ denote all other parameters and the data. 

\subsection{Simulation of $q_h$}
From \eqref{eq_ap:joint_distribution_mixture}, we have  that
\[
p(\bm q_h \vert\bullet) \propto 
    p(\bm q_h \vert \underline{\bm \phi}_q \big)  \cdot    \prod_{i: S_i=h} \prod_{j=1}^J 
q_{hj}^{C_{ij}} \left( 1-q_{hj}\right)^{1-C_{ij}},
\]
where 
$p(\bm q_h \vert \underline{\bm \phi}_q \big) $ is the product of 
densities for  Beta distributions 
$Beta(\underline{a}_{q_j},\underline{b}_{q_j})$, independently over $j=1,\ldots,J$. 
Then

\[
q_{hj} \vert \bullet \sim \text{Beta}\left( 
\underline{a}_{q_j} + \sum_{i: S_i=h}C_{ij},
\underline{b}_{q_j} + \sum_{i: S_i=h}(1-C_{ij}) \right),
\]
independently over $j=1,\ldots,J$ for $h=1,2,\ldots,K^*$. 
If component $h \leq K^*$ does not contain any observations, then the corresponding 
$\bm q_h$ is drawn from the prior.

\subsection{Simulation of $V_h$}

From \eqref{eq_ap:joint_distribution_mixture}, the conditional distribution of $\bm V$ is  independent and the marginal conditional distributions are
\[
V_h \vert\bullet \sim 
    \text{Beta} \left( 1+n_h, \alpha + \sum_{\ell>h} n_\ell \right),
\]
for $h=1,2,\ldots,K^*$. 
If component $h \leq K^*$ is empty, then the corresponding 
$V_h$ is drawn from the prior. 
\subsection{Simulation of $u_i$}
From \eqref{eq_ap:joint_distribution_mixture_withU}, it is easy to see that 
\[
u_i \vert \bullet \overset{ind}{\sim} \mathcal{U}[0, \omega_{S_i} ] , \quad i=1,\ldots,n.
\]

\subsection{Simulation of $S_i$}
From \eqref{eq_ap:joint_distribution_mixture_withU}, we can see that for $h=1,2,\ldots,K^*$, 
\[
Pr\left( S_i = h \vert \bm C, \bm u, \bm V, \bm Q \right)
=
\frac{
I\left( u_i \leq \omega_h \right) 
\prod_{j=1}^J q_{hj}^{C_{ij}} \left( 1-q_{hj}\right)^{1-C_{ij}}   }{ 
\sum_\ell  
I\left( u_i \leq \omega_\ell \right) 
\prod_{j=1}^J 
q_{\ell j}^{C_{ij}} \left( 1-q_{\ell j}\right)^{1-C_{ij}} 
}.
\]
Note that $Pr(S_i=h \vert \bullet)=0$ for $h>K^*$.
\subsection{Simulation of $\alpha$}

The conditional posterior of $\alpha$ is 
\[
p(\alpha \vert \bullet) \propto p(\bm S \vert \alpha) \pi(\alpha).
\]
Following \cite{EscobarWest1995}, this distribution is sampled by first 
generating $\eta$ conditional on $\alpha$ from the Beta distribution 
\[
\eta \vert \alpha, \bm S \sim \text{Beta}(\alpha+1,n),
\]
and then sampling $\alpha$ conditional on $\eta$ from the Gamma mixture 
\begin{align*}
p(\alpha \vert \eta, \bm S) &= 
\frac{\underline{a}_\alpha+G-1}{\underline{a}_\alpha+G-1+n(\underline{b}_\alpha-\log(\eta))}
\text{Gamma}(\underline{a}_\alpha +G, \underline{b}_\alpha-\log(\eta))\\
&+
\frac{n(\underline{b}_\alpha-\log(\eta))}{\underline{a}_\alpha+G-1+n(\underline{b}_\alpha-\log(\eta))}
\text{Gamma}(\underline{a}_\alpha +G-1, \underline{b}_\alpha-\log(\eta)),
\end{align*}
where $G$ is the total number of existing clusters. 

\subsection{Simulation of $ \beta$}
From Bayes theorem, 
\[
\pi(\bm \beta \vert \bullet ) \propto   \pi(\bm \beta)  \cdot    \prod_{i=1}^n \prod_{t=1}^{T_i} \Pr(Y_{it}=y_{it} \vert \bm \delta, \bm \beta, \bm b_i, \bm w_{it}, \mathcal{C}_i ), 
\] 
where 
$\Pr(Y_{it}=y_{it} \vert \bm \delta, \bm \beta, \bm b_i, \bm w_{it}, 
 \mathcal{C}_i )=\frac{\exp \left(V_{iy_{it}t} \right)}{\sum_{\ell \in \mathcal{C}_i }\exp \left(V_{i\ell t} \right)}$
and  $V_{ijt} =  \delta_j+\bm{x}_{ijt}' \bm \beta +\bm{z}_{ijt}'\bm{b}_i $. 

We use a tailored Metropolis–Hastings (M-H) algorithm to sample $\bm \beta$ \citep{ChibGreenberg1995understandingMH}.
Define the conditional log-likelihood of $\bm \beta$ given 
$\bm \delta$, 
$\{ \bm{b}_i \}$, and
$\{ \bm{C}_i \}$:
$
\log L(\bm \beta \vert \bullet )=\sum_{i=1}^n \sum_{t=1}^{T_i} \log \Pr(Y_{it}=y_{it} \vert \bm \delta, \bm \beta, \bm b_i,\mathcal{C}_i ).
$
At iteration $g$, let $\bm{\beta}^{(g)}$ be the value of $\bm \beta$.
A candidate value is drawn as
\[
\tilde{\bm{\beta}} \sim N_{d_x}\left( \hat{\bm{\beta}}, \hat{\bm{V}}_\beta \right),
\]
where 
\[
\hat{\bm{\beta}}=\arg\max_{\bm \beta} \log L(\bm \beta \vert \bullet ) \pi(\bm \beta), \quad \  
 \hat{\bm{V}}_\beta^{-1} = - \frac{\partial^2}{\partial \bm \beta \partial \bm \beta'}\log L(\bm \beta \vert \bullet ) \pi(\bm \beta) \bigg\vert_{\bm \beta=\hat{\bm{\beta}}}, 
\]
which is accepted with probability 
\[
\min 
\left\{ 
\frac{
\pi(\tilde{\bm \beta} \vert \bullet) \phi(\bm \beta^{(g)} \vert \hat{\bm{\beta}}, \hat{\bm{V}}_\beta )
}{
\pi( \bm \beta^{(g)} \vert \bullet) \phi(\tilde{\bm \beta} \vert \hat{\bm{\beta}}, \hat{\bm{V}}_\beta )
}, 
1  \right\},
\]
where $\phi( \ \ )$ denotes the density of normal distribution. 
The conditional posterior mode $\hat{ \bm \beta}$ is computed using the Newton-Raphson method. 
The likelihood is known to be concave with respect to $ \bm \beta$ under the Gumbel error distribution, so the convergence to $\hat{ \bm \beta}$ is fast and only requires a few iterations in many cases. 
In the empirical application, we multiply the variance of the proposal distribution by $10^{-2}$ in order to achieve desirable acceptance rates.  
\subsection{Simulation of $b_i$}
The full conditional of $ \bm b_i$ (for each $i$) is proportional to 
\[
\pi( \bm b_i \vert \bullet) \propto \phi( \bm b_i\vert  \bm 0, \bm D) \cdot \prod_{t=1}^{T_i} \Pr(Y_{it}=y_{it} \vert \bm \delta, \bm \beta, \bm b_i, \bm w_{it},  \mathcal{C}_i) .
\]
We use a symmetric random-walk M-H to draw from the conditional distribution. 
Define the conditional log-likelihood of  $ \bm b_i$ given 
$\bm \delta$, 
$ \bm \beta $, and
$\{ \bm{C}_i \}$:
$
\log L( \bm b_i \vert \bullet )=\sum_{t=1}^{T_i} \log \Pr(Y_{it}=y_{it} \vert \bm \delta, \bm \beta, \bm  b_i, \bm w_{it}, \mathcal{C}_i).
$
At iteration $g$, let $ \bm b_i^{(g)}$ be the value of $ \bm b_i$.
A candidate value is drawn as
\[
\tilde{ \bm b}_i \sim N_{d_z}\left(  \bm b_i^{(g)},  \bm D^{(g)} \right),
\]
which is accepted with probability 
\[
\min \left\{ \frac{\pi( \tilde{ \bm b}_i \vert \bullet) }{\pi( \bm b_i^{(g)} \vert \bullet) }, 1  \right\}.
\]

The updating step for $b_i$ is independent over $i$, so it can be easily parallelized in a modern computer.

\subsection{Simulation of $D$}
We simulate $ \bm D$ by first simulating $ \bm D^{-1}$ and then taking the inverse of the simulated draw. 
This is because it can be shown that 
\[
 \bm D^{-1} \vert \bullet \sim \text{Wishart} \left( \underline{v}+ n, \left[\underline{ \bm R}^{-1} + \sum_{i=1}^n  \bm b_i  \bm b_i' \right]^{-1} \right).
\]

\subsection{Simulation of $\delta$}\label{sec:sampling_delta}
In princple, we could treat $\bm \delta$ as a part of $\bm \beta$ and sample from the conditional distribution altogether using a tailored M-H algorithm. 
However, the involved optimization step could be slow  when $J$ is large, which is exactly our  focus of the current paper.
Hence, we sample $\bm \delta$ separately from $\bm \beta$. 
Specifically, we use a tailored Metropolis–Hastings (M-H) algorithm to sample $\delta_k$  for $k=1,\ldots,J-1,$ one after another.

From Bayes theorem,
\[
\pi(\delta_k \vert \bm \delta_{\setminus k}, \bm \beta, \bm b, \bm W, \bm C ) \propto   \pi(\delta_k)  \cdot    \prod_{i=1}^n \prod_{t=1}^{T_i} \Pr(Y_{it}=y_{it} \vert \bm \delta, \bm \beta, \bm b_i, \bm w_{it}, \mathcal{C}_i ),
\] 
where $\bm \delta_{\setminus k}$ denotes $\bm \delta$ except for the $k$th element. 
Define the conditional log-likelihood of $\delta_k$ given 
$\bm \delta_{\setminus k}$, 
$\bm \beta$, 
$\{ \bm{b}_i \}$, and
$\{ \bm{C}_i \}$:
$
\log L(\delta_k \vert \bullet )=\sum_{i=1}^n \sum_{t=1}^{T_i} \log \Pr(Y_{it}=y_{it} \vert \bm \delta, \bm \beta, \bm b_i, \bm w_{it}, \mathcal{C}_i ).
$
At iteration $g$, let $\delta_k^{(g)}$ be the value of $\delta_k$.
A candidate value is drawn as
\[
\tilde{\delta}_k \sim N_{1}\left( \hat{\delta}_k, \hat{\sigma}^2_{\delta_k} \right),
\]
where 
\[
\hat{\delta}_k=\arg\max_{\delta_k} \log L(\delta_k \vert \bullet ) \pi(\delta_k), \quad \  
 \hat{\sigma}^{-2}_{\delta_k} = - \frac{\partial^2}{\partial \delta_k^2}\log L(\delta_k \vert \bullet ) \pi(\delta_k) \bigg\vert_{ \delta_k=\hat{\delta}_k}, 
\]
which is accepted with probability 
\[
\min 
\left\{ 
\frac{
\pi(\tilde{\delta}_k \vert \bullet) \phi(\delta_k^{(g)} \vert \hat{\delta}_k, \hat{\sigma}^2_{\delta_k} )
}{
\pi( \delta_k^{(g)} \vert \bullet) \phi(\tilde{\delta}_k \vert \hat{\delta}_k, \hat{\sigma}^2_{\delta_k} )
}, 
1  \right\}.
\]
We randomize the order of updating $\delta_k$, $k=1,\ldots,J-1$.

\section{Prior on the distribution of attention probabilities}\label{sec:prior_on_cs}
\subsection{Remarks on hyperparameters}
In the fitting, we set the parameters of the prior as follows: 
for the product-specific fixed-effects, $\delta_j \sim N(0,2)$ independently for $j=1,\ldots,J$, 
for the common fixed-effect, $\beta_k \sim N(0,3)$ independently for $k=1,\ldots,d_x$, 
for the variance of the random effects, $\bm D^{-1}\sim \text{Wishart}(9,(1/9)\bm I_{d_z})$,
and 
for the DP concentration parameter, 
$\alpha \sim \text{Gamma}(\underline{a}_\alpha, \underline{b}_\alpha)=(1/4,1/4)$ to produce 
$\Pr(H_0: \omega^*>1-\varepsilon) \approx 0.5$.
The prior of the attention probabilities is
$q_{hj}\sim \text{Beta}(\underline{a}_{q_j},\underline{b}_{q_j})$, independently over $j=1,\ldots,J$ for $h=1,\ldots,\infty$.
The choice of hyperparameters, $(\underline{a}_{q_j},\underline{b}_{q_j})$, is important, as it controls the sparsity of the consideration sets. 
We set
$(\underline{a}_{q_j},\underline{b}_{q_j})=(s\cdot r, s \cdot (1-r))$,
where $s>0$ and $r$ is a small prior expectation of $q_{hj}$ (that is, $r<0.5$), for example, 
$r=\frac{r_0}{J}$, where $r_0$ is a positive integer.  
We call this a sparsity-supporting prior because the prior probability is smaller for consideration sets with
larger cardinality.

\subsection{Illustration}
When $J$ is small, we can examine the impact of the hyperparameters on the implied prior probability distribution on consideration
sets by simulating from the prior. First, fix a large positive integer $K$. Second, generate draws from the prior by drawing
\begin{align*}
    \alpha &\sim \text{Gamma}(\underline{a}_\alpha, \underline{b}_\alpha),\\
    V_h \vert \alpha &\overset{ind}{\sim}\text{Beta}(1,\alpha) \text{ for } h=1,\ldots,K,\\
    \omega_h &=V_h \prod_{\ell>h} (1-V_\ell) \text{ for } h=1,\ldots,K,\\
    q_{hj} &\overset{ind}{\sim}\text{Beta}(\underline{a}_{q_j},\underline{b}_{q_j}) \text{ for } j=1,\ldots,J, \quad h=1,\ldots,K.
\end{align*}
Finally, given these draws, calculate the probability of each possible consideration set using the representation in Lemma 1; that is,
\[
\pi_c 
=\sum_{h=1}^K \omega_h 
\left\{ 
\prod_{j \in c} q_{h j} \prod_{ j \notin c} \big(1-q_{h j} \big)
\right\}. 
\]
For example, when $J=4$, 
$\Pr(\mathcal{C}_i = \{ 2, 4\})
=\sum_{h=1}^K \omega_h 
\left\{ 
q_{h2}
q_{h4}
\big(1-q_{h1} \big)
\big(1-q_{h3} \big)
\right\}$.

\begin{figure}[h!] 
\begin{subfigure}[b]{0.45\linewidth}
\includegraphics[width=\linewidth]{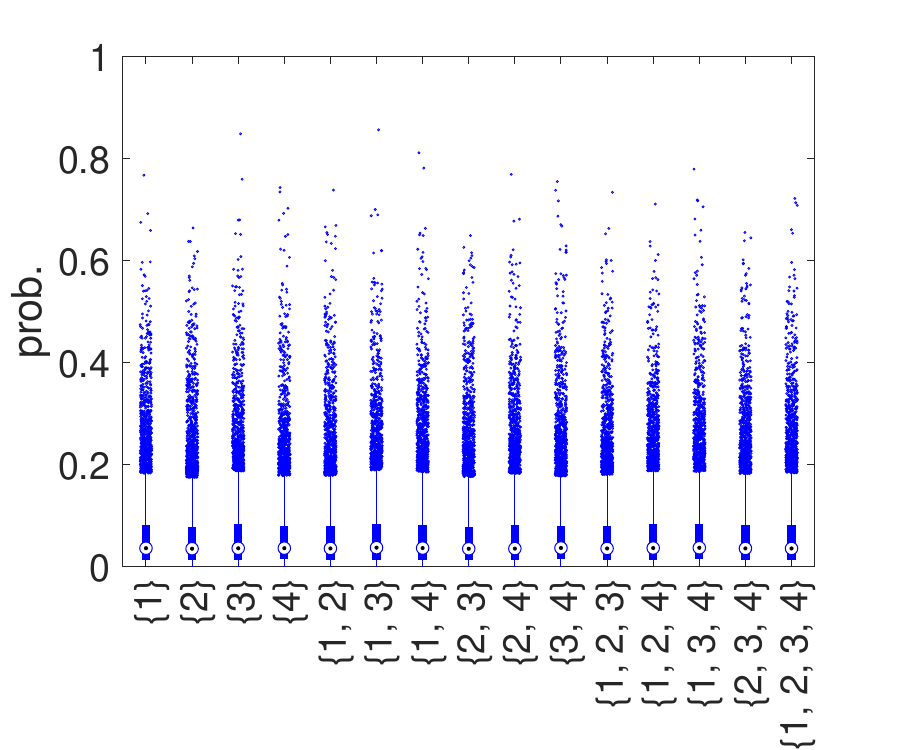} 
\caption{\small{ Uniform prior}}
\end{subfigure} \hfill
\begin{subfigure}[b]{0.45\linewidth}
\includegraphics[width=\linewidth]{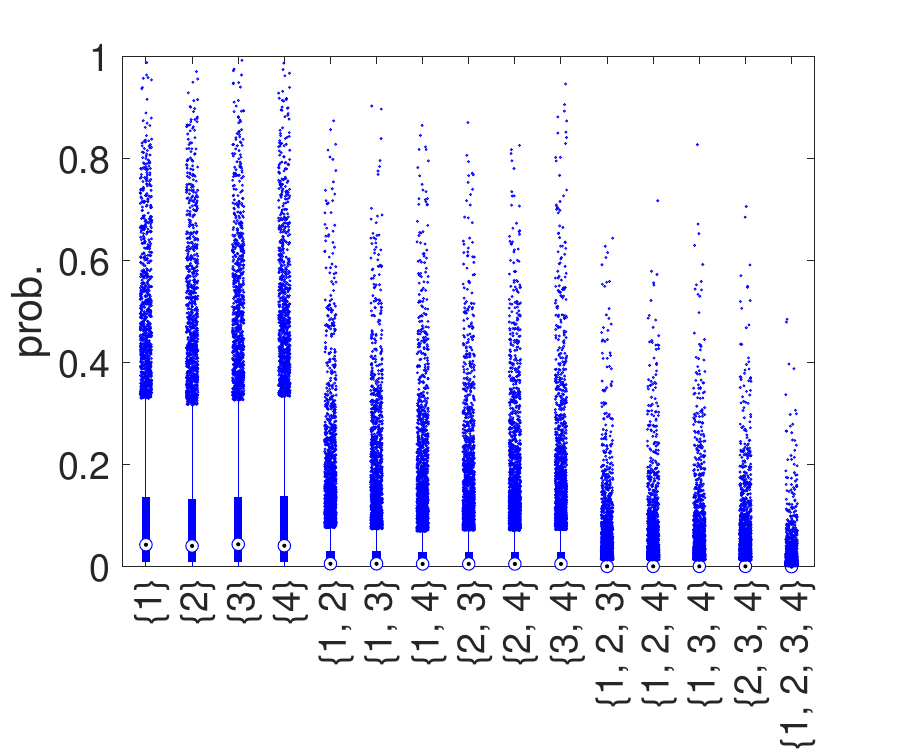} 
\caption{\small{A sparsity-supporting prior    }}
\end{subfigure}
\caption{\small{
Implied prior distribution over consideration sets (box plots) for two different priors on $q_{hj}$. 
Uniform prior (a) with $(\underline{a}_{q_j},\underline{b}_{q_j})=(1,1)$ and sparsity supporting prior (b) with $(\underline{a}_{q_j},\underline{b}_{q_j})=(s r, s  (1-r))$ with $r=\frac{1}{J}$, $s=1$. 
$K=20.$ 
$(\underline{a}_\alpha, \underline{b}_\alpha)=(1/4,1/4)$. 
and 10,000 draws from the prior.}}
\label{fig:simulation_priorCS}
\end{figure}

Panel (a) in Figure \ref{fig:simulation_priorCS} shows the implied prior distribution over the consideration sets under the uniform prior on $q_{hj}$ when $J=4$. 
Under the uniform prior, the prior expectation of $q_{hj}=0.5$, so the prior is the same across all consideration sets and is centered around $0.5^4=0.0625$. 
Panel (b) gives results under our sparsity-supporting prior. In this case, the prior distribution shrinks to $0$ as the cardinality of the consideration set increases. 

The preceding shows that the prior on the attention probabilities $\{ q_{hj} \}$ induces quite different prior distributions on consideration sets. As the number of consideration sets increase exponentially in $J$, it is crucial to apply regularization to the parameter space. 
Our sparsity-supporting prior promotes this regularization. It favors smaller consideration sets, while maintaining positive probabilities on larger sets.

\section{Additional material for the simulation}\label{sec:simulation_additional}
Section \ref{sec:simulation_test_dep_consid} introduces a test for dependence of considerations and illustrates its performance. 
Sections 
\ref{sec:simulation_random_effects}, 
\ref{sec:simulation_autocorrelation}, and 
\ref{sec:simulation_time_varying_cs} show the simulation studies under random effects, auto-correlated covariate, and time-varying consideration sets, respectively.

\subsection{Testing for dependent consideration}\label{sec:simulation_test_dep_consid}

From the MCMC output, it is possible to assess the degree of consideration dependence using the method proposed by \cite{DunsonXing2009}, but now applied to the latent consideration sets. The null hypothesis tests for independent consideration, formulated as $H_0: \omega_1 = 1$.
We utilize the interval null of $H_0:\omega^*>1-\varepsilon$ with $\omega^*=\max \{\omega_k : k=1,\ldots,k^* \}$ and $\varepsilon>0$ is a small value.  
The Bayes factor in favor of the alternative hypothesis, $H_1:\omega^*\leq 1-\varepsilon$, is defined as 
$\frac{\Pr(H_1 \vert \bm D^n) \Pr(H_1)}{\Pr(H_0 \vert \bm D^n) \Pr(H_0)}$, which 
can be estimated using $\hat{\Pr}(H_1 \vert \bm D^n)$, the portion of the posterior sample such that $\omega^*\leq 1-\varepsilon$, and 
$\hat{\Pr}(H_0 \vert \bm D^n)=1-\hat{\Pr}(H_1 \vert \bm D^n)$. 
In the simulations and the application,  $\underline{a}_\alpha=\underline{b}_\alpha=1/4$ is fixed to produce
$\Pr(H_0) \approx 0.5$.
\begin{figure}[ht]
\centering
  \begin{subfigure}[b]{0.45\linewidth}
    \centering
    \includegraphics[width=\linewidth]{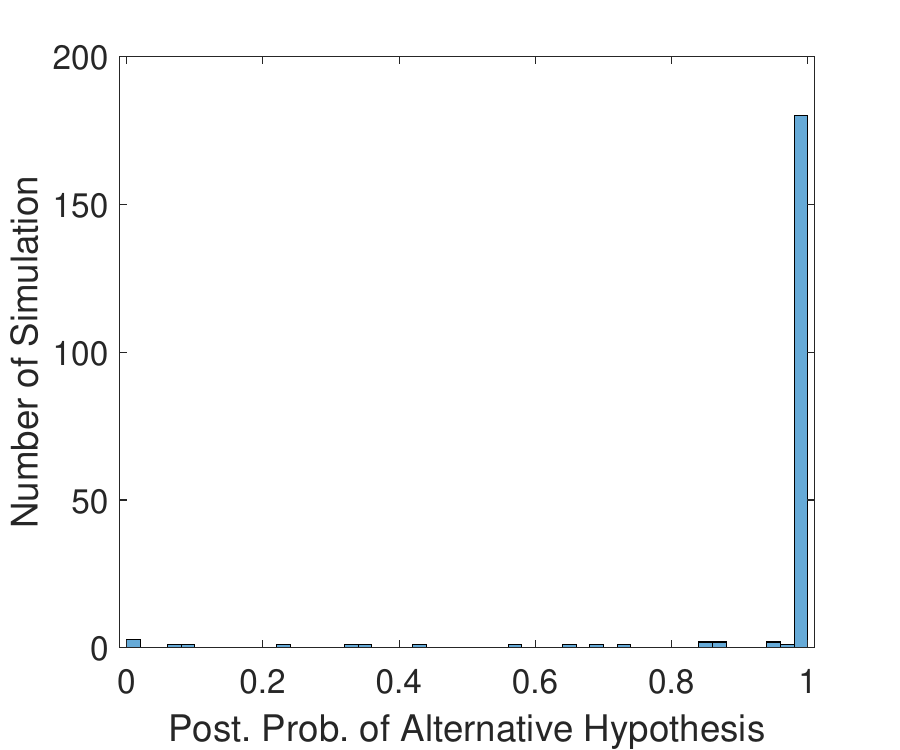} 
\caption{\small{ Case 1 (dependent consid.) }}
  \end{subfigure}
  \begin{subfigure}[b]{0.45\linewidth}
    \centering
    \includegraphics[width=\linewidth]{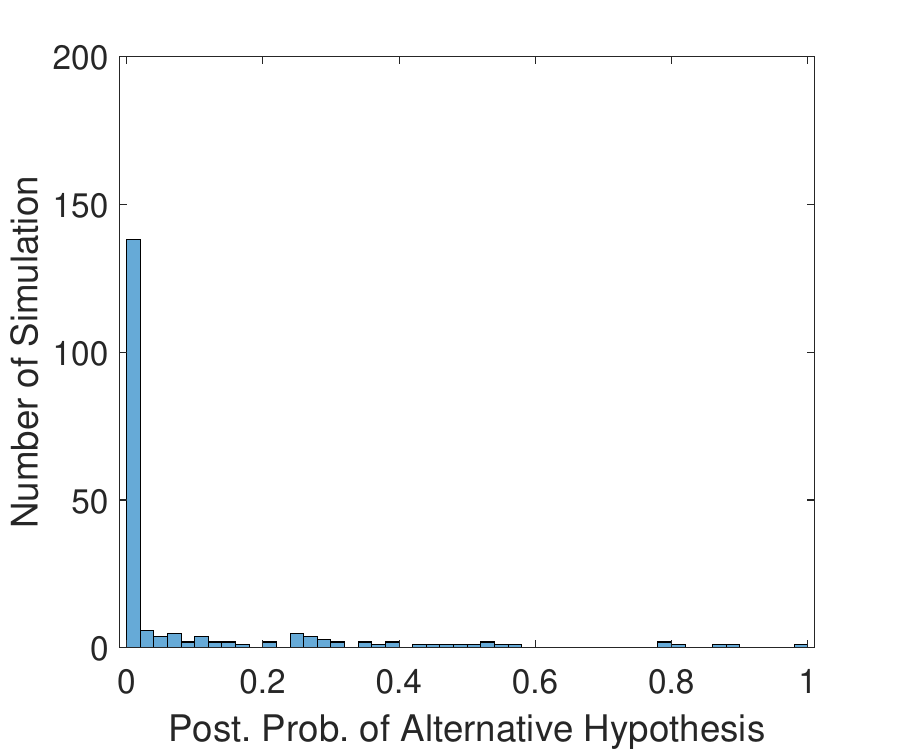} 
\caption{\small{ Case 2 (independent consid.) }}
  \end{subfigure}
\caption{\small{
Histograms of estimated posterior probabilities of $H_1$ in each of the 200 simulations under (a) case 1 (dependent consideration - $H_1$ is true) and (b) case 2 (independent consideration - $H_0$ is true). $\varepsilon=0.1, n=50$. $T=5$.
 }}
\label{fig:simulation_SmallJ_Test}
\end{figure}

We use the current data-generating process as the first case (dependent consideration). In the second case, the consideration is independent. We generate $C_{ij} \overset{iid}{\sim} \text{Bernoulli}(\gamma_j)$, for $j=1,2,3$  
with $(\gamma_1, \gamma_2,\gamma_3)=(0.2,0.15,0.35)$ and fix $C_{i4}=1$, for $i=1,\ldots,n$. 
Figure \ref{fig:simulation_SmallJ_Test} (a) provides a histogram showing the estimated posterior probabilities of $H_1$ under the first case ($H_1$ is true) across the 200 simulated data sets using $\varepsilon=0.1$. The method appropriately assigns a value close to one to $\Pr(H_1 \vert \bm D^n)$ in most cases, with only $9/100$ having an estimated $\Pr(H_1 \vert \bm D^n)<0.5$. 
Figure \ref{fig:simulation_SmallJ_Test} (b) provides the results for case 2. The posterior probability assigned to $H_1$ is close to zero for most simulations.  We find similar results with random effects, as shown below.


\subsection{Simulation results with random effects}\label{sec:simulation_random_effects}
\begin{figure}[ht] 
\centering
  \begin{subfigure}[b]{0.25\linewidth}
    \centering
    \includegraphics[width=\linewidth]{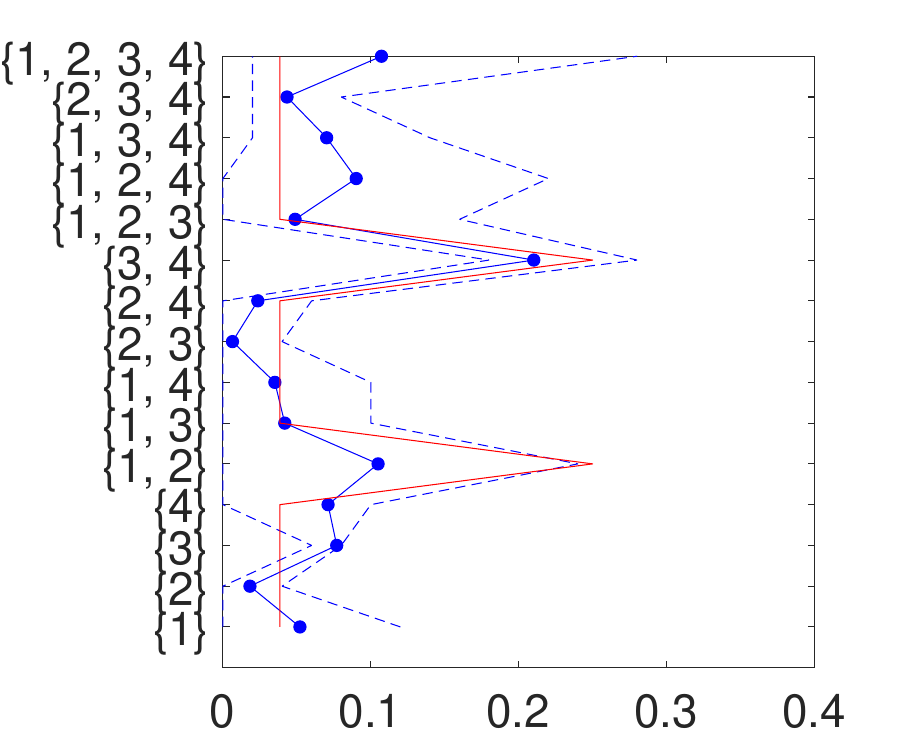} 
\caption{\small{ $K=\infty, n=50$ }}
  \end{subfigure}
  \begin{subfigure}[b]{0.25\linewidth}
    \centering
    \includegraphics[width=\linewidth]{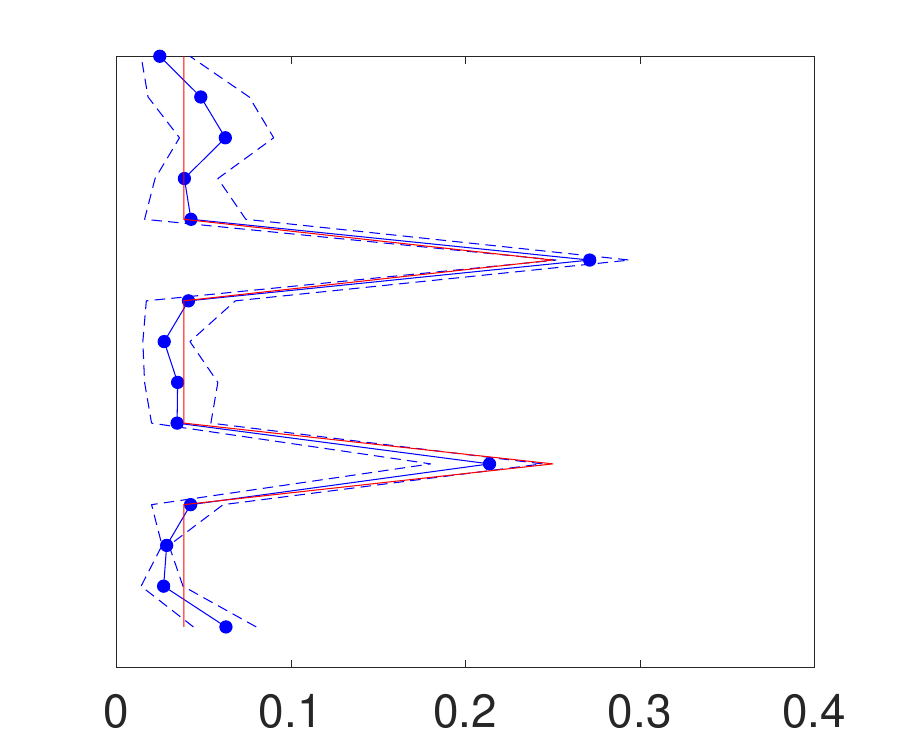} 
\caption{\small{ $K=\infty, n=500$ }}
  \end{subfigure}
  \begin{subfigure}[b]{0.25\linewidth}
    \centering
    \includegraphics[width=\linewidth]{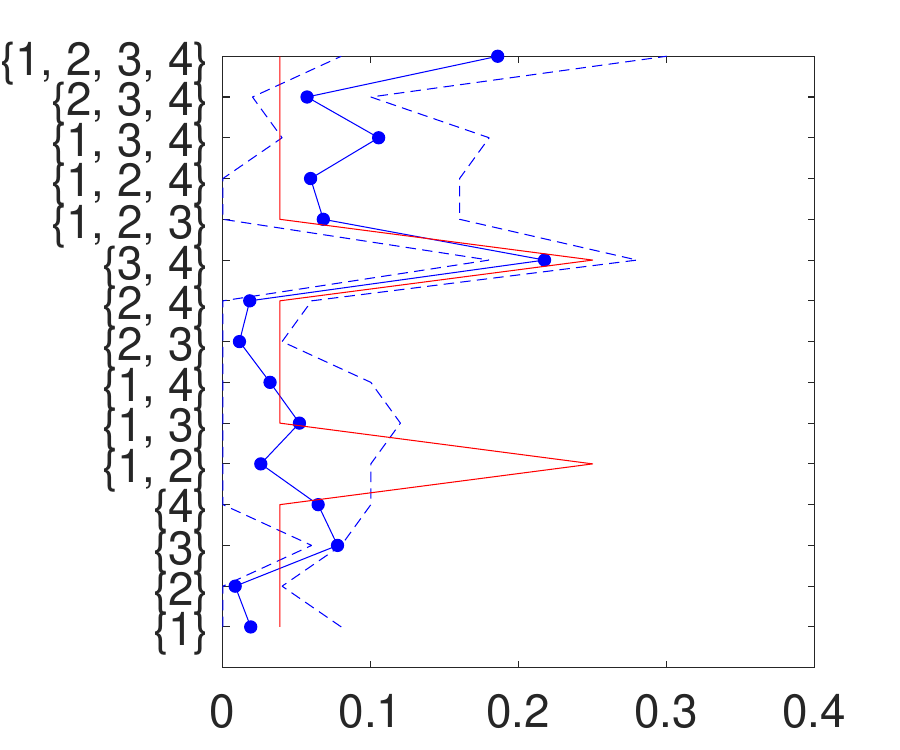} 
\caption{\small{ $K=1, n=50$ }}
  \end{subfigure}
  \begin{subfigure}[b]{0.25\linewidth}
    \centering
    \includegraphics[width=\linewidth]{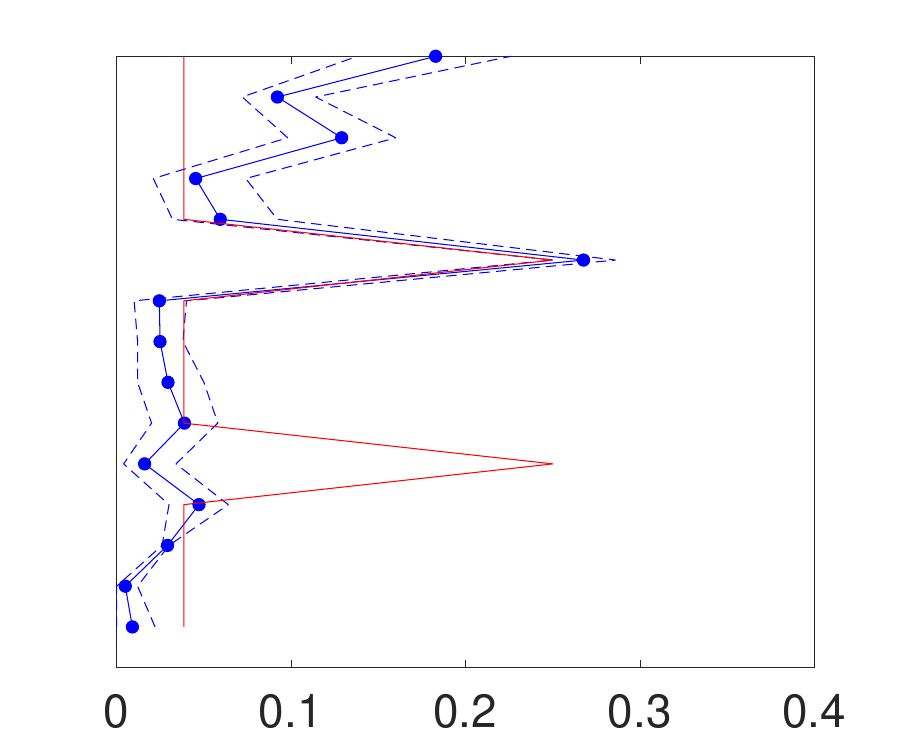} 
\caption{\small{ $K=1, n=500$ }}
  \end{subfigure}
  
\caption{\small{
The true distribution over consideration sets (solid, red), 
posterior mean (solid with dots, blue),  95\% equal-tailed credible interval (dashed, blue). 
Each plot is based on  one realization of simulated data. $J=4$, $T=5$, with random effects.
 }}
\label{fig:simulation_SmallJ_marginalCS_RE}
\end{figure}

We repeat the simulation study now with preference heterogeneity. We generate the consideration sets as before, but now use the random effects logit with the specification
$V_{ijt} =\delta^*_j+ (\beta^*+b_i)  x_{ijt}, $
where $b_i\sim N(0,D^*)$ with $D^*=1$. 
We fit the random effects logit with the proposed flexible approach for the distribution of consideration sets.   
The results are presented in Table \ref{tab:simulation_SmallJ_pref_RE_depCS}. 
We see that as $n$ increases,  RMSEs/L1-errors/SDs  tend to decrease. 
However, this is not the case when the independent consideration is imposed, i.e.\ $K=1$. 
Also, RMSEs/L1-errors are larger in general than for the proposed flexible approach. 
In addition, there are distortions of the coverage of the credible intervals under $K=1$. 
In Figure \ref{fig:simulation_SmallJ_marginalCS_RE}, for $K=\infty$, we see the posterior on $\bm \pi$ approaches to the truth while it does not under $K=1$, due to the mis-specification.

\FloatBarrier
\begin{table}[ht]
\caption{Simulation results  with $J=4$ (random effects)}
\centering
\resizebox{1.1\columnwidth}{!}{%
\begin{threeparttable}
\begin{tabular}{ l  l | l l l    l l l     l l l    l l l    l l l  l l l   l  }

$(K,T)$ & $n$ &   \multicolumn{3}{l}{$\beta$}  &  \multicolumn{3}{l}{$\delta_1$} &  \multicolumn{3}{l}{$\delta_2$} &  \multicolumn{3}{l}{$\delta_3$} &  \multicolumn{3}{l}{$D^{-1/2}$}  &  \multicolumn{3}{l}{$\pi$} & Time \\ 

\hline

 &  &   
\multicolumn{3}{l}{RMSE (MCE) \  \ SD (ESD)  \ \ \ \ \ \ Cov}  &  
\multicolumn{3}{l}{RMSE (MCE) \  \ SD (ESD)  \ \ \ \ \ \ Cov} &  
\multicolumn{3}{l}{RMSE (MCE) \  \ SD (ESD)  \ \ \ \ \ \ Cov} &  
\multicolumn{3}{l}{RMSE (MCE) \  \ SD (ESD)  \ \ \ \ \ \ Cov}  &  
\multicolumn{3}{l}{RMSE (MCE) \  \ SD (ESD)  \ \ \ \ \ \ Cov}  &  
\multicolumn{3}{l}{L1-error (MCE)   SD (ESD)   \ \ \ \ \ \ Cov} & \\

\hline

\multirow{2}{*}{$(\infty, 5)$} 
&$50$ & 0.363 ( 0.011 ) & 0.14 ( 0.195 ) & 0.45 & 0.448 ( 0.025 ) & 0.42 ( 0.445 ) & 0.93 & 0.49 ( 0.028 ) & 0.43 ( 0.462 ) & 0.92 & 0.357 ( 0.018 ) & 0.32 ( 0.289 ) & 0.9 & 0.845 ( 0.006 ) & 0.06 ( 0.133 ) & 0.05 & 0.455 ( 0.009 ) & 0.03 ( 0.028 ) & 0.96 & 2.7 \tabularnewline
&$300$ & 0.124 ( 0.006 ) & 0.1 ( 0.102 ) & 0.85 & 0.221 ( 0.012 ) & 0.19 ( 0.207 ) & 0.93 & 0.212 ( 0.01 ) & 0.19 ( 0.204 ) & 0.95 & 0.151 ( 0.008 ) & 0.16 ( 0.151 ) & 0.97 & 0.215 ( 0.012 ) & 0.12 ( 0.154 ) & 0.72 & 0.268 ( 0.003 ) & 0.02 ( 0.015 ) & 0.93 & 20.97 \tabularnewline

\hline 
\multirow{2}{*}{$(\infty, 15)$} 
&$50$ & 0.194 ( 0.01 ) & 0.16 ( 0.183 ) & 0.87 & 0.205 ( 0.01 ) & 0.21 ( 0.205 ) & 0.95 & 0.188 ( 0.01 ) & 0.21 ( 0.187 ) & 0.97 & 0.172 ( 0.009 ) & 0.18 ( 0.17 ) & 0.97 & 0.315 ( 0.014 ) & 0.15 ( 0.198 ) & 0.66 & 0.356 ( 0.005 ) & 0.03 ( 0.022 ) & 0.97 & 3.69 \tabularnewline
&$300$ & 0.075 ( 0.004 ) & 0.07 ( 0.074 ) & 0.95 & 0.086 ( 0.004 ) & 0.09 ( 0.086 ) & 0.95 & 0.079 ( 0.004 ) & 0.09 ( 0.079 ) & 0.96 & 0.066 ( 0.003 ) & 0.07 ( 0.066 ) & 0.99 & 0.081 ( 0.004 ) & 0.07 ( 0.067 ) & 0.9 & 0.193 ( 0.003 ) & 0.01 ( 0.012 ) & 0.91 & 57.4 \tabularnewline

\hline
\multirow{2}{*}{$(1, 5)$} 
&$50$ & 0.384 ( 0.011 ) & 0.13 ( 0.184 ) & 0.35 & 0.786 ( 0.024 ) & 0.37 ( 0.463 ) & 0.54 & 0.911 ( 0.029 ) & 0.38 ( 0.491 ) & 0.47 & 0.364 ( 0.018 ) & 0.32 ( 0.303 ) & 0.88 & 0.849 ( 0.005 ) & 0.06 ( 0.1 ) & 0.04 & 0.843 ( 0.006 ) & 0.03 ( 0.025 ) & 0.47 & 2.4 \tabularnewline
&$300$ & 0.192 ( 0.005 ) & 0.09 ( 0.085 ) & 0.47 & 1.00 ( 0.015 ) & 0.16 ( 0.214 ) & 0.01 & 1.058 ( 0.016 ) & 0.17 ( 0.223 ) & 0.00 & 0.166 ( 0.008 ) & 0.15 ( 0.162 ) & 0.94 & 0.261 ( 0.009 ) & 0.1 ( 0.134 ) & 0.48 & 0.862 ( 0.005 ) & 0.01 ( 0.013 ) & 0.00 & 19.7 \tabularnewline

\hline
\multirow{2}{*}{$(1, 15)$} 
&$50$ & 0.193 ( 0.01 ) & 0.16 ( 0.181 ) & 0.87 & 0.288 ( 0.016 ) & 0.23 ( 0.246 ) & 0.9 & 0.272 ( 0.018 ) & 0.23 ( 0.23 ) & 0.9 & 0.177 ( 0.01 ) & 0.19 ( 0.177 ) & 0.94 & 0.315 ( 0.014 ) & 0.15 ( 0.2 ) & 0.68 & 0.712 ( 0.002 ) & 0.02 ( 0.018 ) & 0.89 & 3.4 \tabularnewline
&$300$ & 0.076 ( 0.004 ) & 0.07 ( 0.074 ) & 0.96 & 0.159 ( 0.007 ) & 0.09 ( 0.102 ) & 0.72 & 0.15 ( 0.007 ) & 0.09 ( 0.095 ) & 0.77 & 0.085 ( 0.004 ) & 0.08 ( 0.069 ) & 0.92 & 0.078 ( 0.004 ) & 0.07 ( 0.066 ) & 0.93 & 0.704 ( 0.001 ) & 0.01 ( 0.008 ) & 0.63 & 56.98 \tabularnewline

\end{tabular}
\begin{tablenotes}
\small
\item 
For $\beta$, $\delta$, and $D^{1/2}$, for each case, 
we show  the estimated root mean squared error (RMSE), using 
the posterior means as  point estimator. 
In parenthesis, 
the jackknife estimate of Monte Carlo Error (MCE) for the RMSE is presented. 
Next, the average of the posterior standard deviations (SD) is shown with  the empirical standard deviation (ESD) of the posterior mean in the parenthesis.
Third, the empirical coverage (Cov) of 95\% credible interval is given. 
\item For $\pi$, we show the average of $L_1$ norm between the posterior mean and $\bm \pi^*$ (L1-error). 
In the parenthesis, we show its jackknife estimate of MCE. 
The SDs, ESDs, and Covs are averaged over the 15 elements in $\pi$. 

\item Time is the average seconds taken for sampling 1,000 MCMC draws in Matlab on a desktop with a 4.9GHz processor and 64GB RAM. 
The study is based on $R=200$ replications. 
2,000 MCMC draws are obtained for each replication. The average of the inefficiency factors is around 9.5 with standard deviation 1.4.
\end{tablenotes}
\end{threeparttable}
}
\label{tab:simulation_SmallJ_pref_RE_depCS}
\end{table}
\FloatBarrier

\begin{figure}[ht] 
\centering
  \begin{subfigure}[b]{0.45\linewidth}
    \centering
    \includegraphics[width=\linewidth]{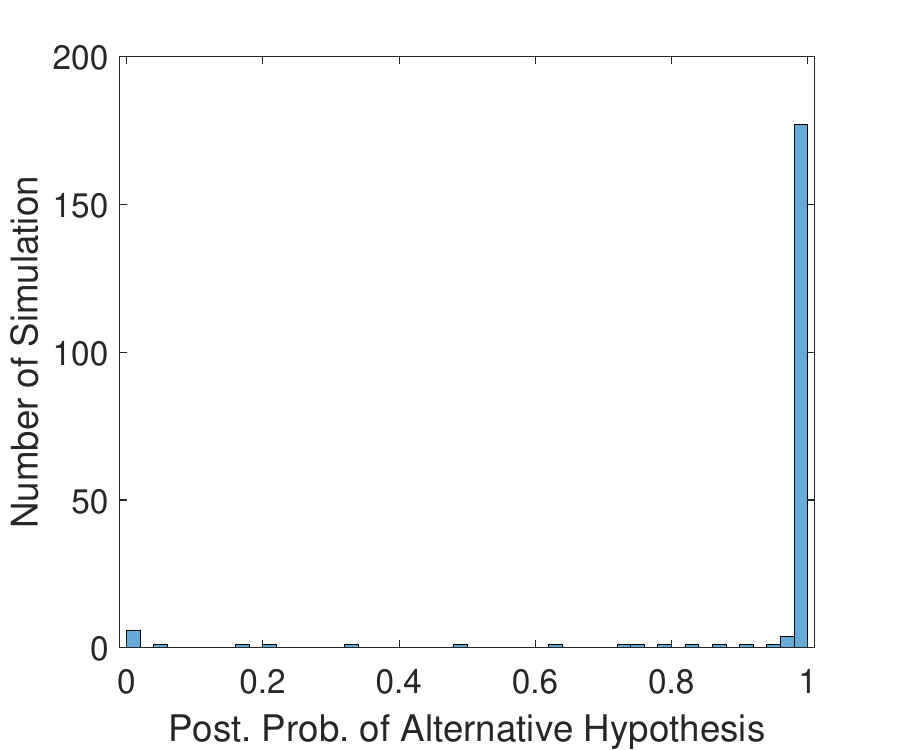} 
\caption{\small{ Case 1 (dependent consideration) }}
  \end{subfigure}
  \begin{subfigure}[b]{0.45\linewidth}
    \centering
    \includegraphics[width=\linewidth]{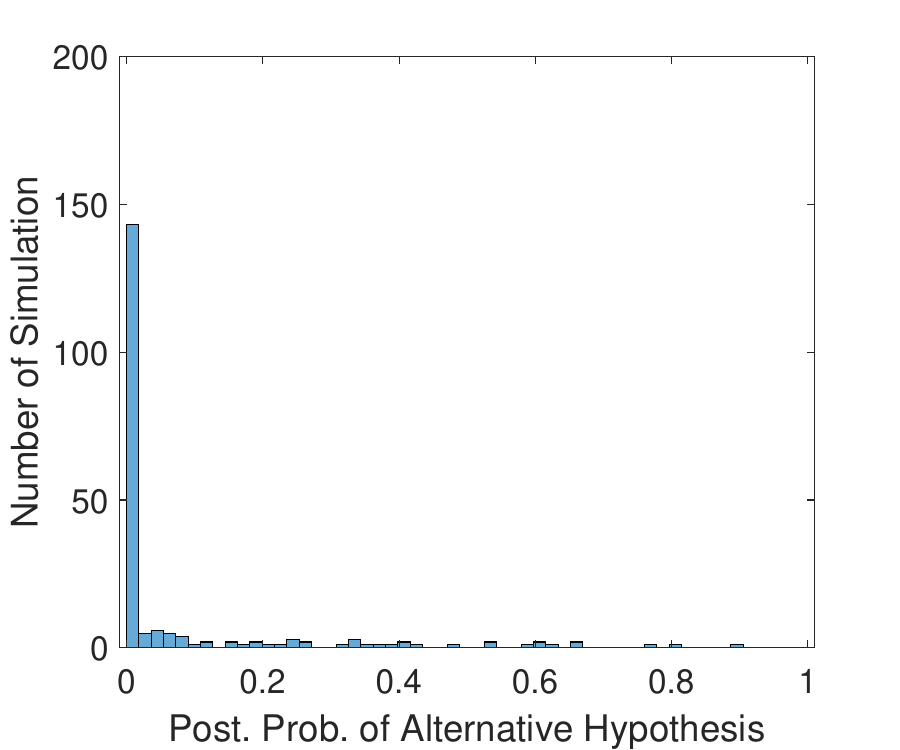} 
\caption{\small{ Case 2 (independent consideration) }}
  \end{subfigure}
\caption{\small{
Histograms of estimated posterior probabilities of $H_1$ in each of the 200 simulations under (a) case 1 (dependent consideration - $H_1$ is true) and (b) case 2 (independent consideration - $H_0$ is true). $\varepsilon=0.1, n=50$. $T=5$. With random effects.
 }}
\label{fig:simulation_SmallJ_Test_RE}
\end{figure}

Figure \ref{fig:simulation_SmallJ_Test_RE} (a) shows a histogram of the estimated posterior probability of $H_1$ (dependent consideration) when $H_1$ is true. The method appropriately assigns values close to one for the majority of the simulations. 
Figure \ref{fig:simulation_SmallJ_Test_RE} (b) shows the result when $H_0$ is true. The posterior probability assigned to $H_1$ is close to zero for the majority of the simulations. 

In summary, even with random effects, our proposed method can deliver consistent estimates of the preference parameters i.e. $\bm \beta$ and $\bm D$ as well as the distribution of consideration sets $\bm \pi$. In addition, our method can be used to test whether latent consideration is independent or dependent. 

\FloatBarrier

\subsection{Simulation results with auto-correlated covariate}\label{sec:simulation_autocorrelation}
We generate a set of auto-correlated covariates as follows: $x_{ijt}=\rho x_{ijt-1}+N(0,1)$ for $t=1,\ldots,T$ with $\rho=0.9$ and  $x_{ij1}\sim N(0,1)$. The rest of the simulation design is the same as in Section 5.1 of the paper. Table \ref{tab:simulation_SmallJ_autocorrelation} shows the results. Overall, the results are similar to the case with no correlation.

\begin{table}[ht]
\caption{Simulation results  with $J=4$ (Auto-correlated covariate)}
\centering
\resizebox{1.1\columnwidth}{!}{%
\begin{threeparttable}
\begin{tabular}{ l  l | l l l    l l l     l l l    l l l    l l l    l  }
$(K,T)$ & $n$ &   \multicolumn{3}{c}{$\beta$}  &  \multicolumn{3}{c}{$\delta_1$} &  \multicolumn{3}{c}{$\delta_2$} &  \multicolumn{3}{c}{$\delta_3$}  &  \multicolumn{3}{c}{$\pi$} & Time \\ 
\hline

 &  &   
 \multicolumn{3}{l}{RMSE (MCE) \  \ SD (ESD)  \ \ \ \ \ \ Cov}  &  
 \multicolumn{3}{l}{RMSE (MCE) \  \ SD (ESD)  \ \ \ \ \ \ Cov} &  
 \multicolumn{3}{l}{RMSE (MCE) \  \ SD (ESD)  \ \ \ \ \ \ Cov} &  
 \multicolumn{3}{l}{RMSE (MCE) \  \ SD (ESD)  \ \ \ \ \ \ Cov}  &  
 \multicolumn{3}{l}{L1-error (MCE)   SD (ESD)   \ \ \ \ \ \ Cov} & \\

\hline
\multirow{2}{*}{$(\infty, 5)$} 
&$50$ & 0.15 ( 0.009 ) & 0.15 ( 0.143 ) & 0.96 & 0.473 ( 0.022 ) & 0.46 ( 0.473 ) & 0.94 & 0.437 ( 0.022 ) & 0.46 ( 0.435 ) & 0.96 & 0.341 ( 0.018 ) & 0.36 ( 0.342 ) & 0.97 & 0.434 ( 0.009 ) & 0.03 ( 0.027 ) & 0.97 & 1.84 \tabularnewline
&$100$ & 0.107 ( 0.006 ) & 0.1 ( 0.103 ) & 0.95 & 0.327 ( 0.017 ) & 0.33 ( 0.315 ) & 0.96 & 0.347 ( 0.02 ) & 0.32 ( 0.334 ) & 0.92 & 0.27 ( 0.013 ) & 0.26 ( 0.269 ) & 0.93 & 0.347 ( 0.005 ) & 0.02 ( 0.021 ) & 0.95 & 3.4 \tabularnewline

\hline 
\multirow{2}{*}{$(\infty, 15)$} 
&$50$ & 0.082 ( 0.004 ) & 0.08 ( 0.081 ) & 0.96 & 0.252 ( 0.013 ) & 0.24 ( 0.25 ) & 0.93 & 0.243 ( 0.013 ) & 0.24 ( 0.242 ) & 0.96 & 0.204 ( 0.012 ) & 0.2 ( 0.204 ) & 0.95 & 0.368 ( 0.005 ) & 0.03 ( 0.024 ) & 0.96 & 2.58 \tabularnewline
&$100$ & 0.062 ( 0.003 ) & 0.06 ( 0.059 ) & 0.94 & 0.177 ( 0.01 ) & 0.17 ( 0.175 ) & 0.96 & 0.166 ( 0.008 ) & 0.17 ( 0.164 ) & 0.96 & 0.144 ( 0.008 ) & 0.14 ( 0.143 ) & 0.96 & 0.301 ( 0.003 ) & 0.02 ( 0.018 ) & 0.93 & 5.09 \tabularnewline

\hline
\multirow{2}{*}{$(1, 5)$} 
&$50$ & 0.155 ( 0.009 ) & 0.15 ( 0.152 ) & 0.95 & 0.673 ( 0.028 ) & 0.48 ( 0.556 ) & 0.82 & 0.675 ( 0.032 ) & 0.48 ( 0.544 ) & 0.84 & 0.364 ( 0.019 ) & 0.37 ( 0.363 ) & 0.97 & 0.77 ( 0.005 ) & 0.03 ( 0.024 ) & 0.78 & 1.56 \tabularnewline
&$100$ & 0.109 ( 0.006 ) & 0.11 ( 0.109 ) & 0.95 & 0.698 ( 0.029 ) & 0.33 ( 0.413 ) & 0.6 & 0.733 ( 0.031 ) & 0.34 ( 0.456 ) & 0.6 & 0.283 ( 0.014 ) & 0.27 ( 0.28 ) & 0.94 & 0.772 ( 0.005 ) & 0.02 ( 0.019 ) & 0.49 & 2.82 \tabularnewline

\hline
\multirow{2}{*}{$(1, 15)$} 
&$50$ & 0.085 ( 0.004 ) & 0.08 ( 0.081 ) & 0.96 & 0.292 ( 0.014 ) & 0.25 ( 0.268 ) & 0.9 & 0.281 ( 0.014 ) & 0.25 ( 0.259 ) & 0.92 & 0.208 ( 0.012 ) & 0.2 ( 0.209 ) & 0.96 & 0.711 ( 0.002 ) & 0.02 ( 0.019 ) & 0.89 & 2.27 \tabularnewline
&$100$ & 0.065 ( 0.004 ) & 0.06 ( 0.06 ) & 0.92 & 0.225 ( 0.012 ) & 0.17 ( 0.192 ) & 0.88 & 0.214 ( 0.01 ) & 0.17 ( 0.182 ) & 0.87 & 0.148 ( 0.008 ) & 0.14 ( 0.145 ) & 0.97 & 0.705 ( 0.001 ) & 0.01 ( 0.014 ) & 0.85 & 4.29 \tabularnewline

\end{tabular}
\begin{tablenotes}
\small
\item 
For $\beta$ and $\delta$, for each case, 
we show  the estimated root mean squared error (RMSE), using 
the posterior means as  point estimator. 
In parenthesis, 
the jackknife estimate of Monte Carlo Error (MCE) for the RMSE is presented. 
Next, the average of the posterior standard deviations (SD) is shown with  the empirical standard deviation (ESD) of the posterior mean in the parenthesis.
Third, the empirical coverage (Cov) of 95\% credible interval is given. 
\item For $\pi$, we show the average of $L_1$ norm between the posterior mean and $\bm \pi^*$ (L1-error). 
In the parenthesis, we show its jackknife estimate of MCE. 
The SDs, ESDs, and Covs are averaged over the 15 elements in $\pi$. 

\item Time is the average seconds taken for sampling 1,000 MCMC draws in Matlab on a desktop with a 4.9GHz processor and 64GB RAM. 
The study is based on $R=200$ replications. 
2,000 MCMC draws are obtained for each replication. The average of the inefficiency factors is around 6.8 with standard deviation 1.2.
\end{tablenotes}
\end{threeparttable}
}
\label{tab:simulation_SmallJ_autocorrelation}
\end{table}

\subsection{Simulation results with time-varying consideration sets}\label{sec:simulation_time_varying_cs}
As described in the paper, the assumption of time invariant consideration sets facilitates theoretical study and computation. However, the actual consideration sets might have some dynamics over time. In this simulation, we study how sensitive our proposed method is with respect to a violation of the time invariance assumption. 

In order to generate time-varying consideration sets, we first draw the consideration sets in the first  period  as before from $\pi^*$. They remain unchanged in period 2. At period $t=3$, the units learn about items outside their consideration sets and 50\% of them add a new item randomly to the sets. The consideration sets are unchanged after this period. 

The rest of the simulation design is the same as in Section 5.1 of the paper. Table \ref{tab:simulation_SmallJ_time_varying_cs} shows the simulation result. 
\cite{CrawfordGriffithIaria2021} show that incorrectly adding items to the consideration sets lead to biased estimates, which is reflected in the increased RMSEs, especially when $n$ is large, compared to Table 2 of the paper where the time invariance assumption holds. In addition, there are distortions of the coverage.

We still see that the RMSEs/L1-errors/SDs clearly decrease in $n$ under the relatively large $T=15$. This is because the sample with $T=15$ has more periods with stable consideration sets than $T=5$. Thus, although the violation of the time-invariant consideration sets leads to issues known in the literature, and we are not an exception, our approach delivers reasonable result when there are enough periods during which the invariant assumption holds.

\begin{table}[ht]
\caption{Simulation results  with $J=4$ (Time-varying consideration sets)}
\centering
\resizebox{1.1\columnwidth}{!}{%
\begin{threeparttable}
\begin{tabular}{ l  l | l l l    l l l     l l l    l l l    l l l    l  }
$(K,T)$ & $n$ &   \multicolumn{3}{c}{$\beta$}  &  \multicolumn{3}{c}{$\delta_1$} &  \multicolumn{3}{c}{$\delta_2$} &  \multicolumn{3}{c}{$\delta_3$}  &  \multicolumn{3}{c}{$\pi$} & Time \\ 
\hline

 &  &   
 \multicolumn{3}{l}{RMSE (MCE) \  \ SD (ESD)  \ \ \ \ \ \ Cov}  &  
 \multicolumn{3}{l}{RMSE (MCE) \  \ SD (ESD)  \ \ \ \ \ \ Cov} &  
 \multicolumn{3}{l}{RMSE (MCE) \  \ SD (ESD)  \ \ \ \ \ \ Cov} &  
 \multicolumn{3}{l}{RMSE (MCE) \  \ SD (ESD)  \ \ \ \ \ \ Cov}  &  
 \multicolumn{3}{l}{L1-error (MCE)   SD (ESD)   \ \ \ \ \ \ Cov} & \\

\hline

\multirow{2}{*}{$(\infty, 5)$} 
&$50$ & 0.184 ( 0.008 ) & 0.13 ( 0.143 ) & 0.81 & 0.483 ( 0.02 ) & 0.33 ( 0.366 ) & 0.78 & 0.387 ( 0.019 ) & 0.34 ( 0.324 ) & 0.91 & 0.407 ( 0.025 ) & 0.34 ( 0.373 ) & 0.92 & 0.724 ( 0.012 ) & 0.03 ( 0.029 ) & 0.9 & 1.78 \tabularnewline
&$100$ & 0.163 ( 0.006 ) & 0.09 ( 0.095 ) & 0.67 & 0.409 ( 0.015 ) & 0.23 ( 0.227 ) & 0.69 & 0.364 ( 0.014 ) & 0.24 ( 0.218 ) & 0.8 & 0.29 ( 0.014 ) & 0.24 ( 0.23 ) & 0.91 & 0.688 ( 0.008 ) & 0.02 ( 0.022 ) & 0.94 & 3.33 \tabularnewline

\hline
\multirow{2}{*}{$(\infty, 15)$} 
&$50$ & 0.091 ( 0.004 ) & 0.07 ( 0.077 ) & 0.91 & 0.178 ( 0.008 ) & 0.15 ( 0.16 ) & 0.89 & 0.167 ( 0.008 ) & 0.16 ( 0.154 ) & 0.94 & 0.173 ( 0.01 ) & 0.18 ( 0.164 ) & 0.93 & 0.707 ( 0.009 ) & 0.02 ( 0.024 ) & 0.87 & 2.51 \tabularnewline
&$100$ & 0.073 ( 0.003 ) & 0.05 ( 0.055 ) & 0.84 & 0.143 ( 0.006 ) & 0.11 ( 0.108 ) & 0.86 & 0.134 ( 0.005 ) & 0.11 ( 0.105 ) & 0.88 & 0.116 ( 0.005 ) & 0.12 ( 0.112 ) & 0.97 & 0.656 ( 0.006 ) & 0.02 ( 0.018 ) & 0.93 & 4.79 \tabularnewline

\end{tabular}
\begin{tablenotes}
\small
\item 
For $\beta$ and $\delta$, for each case, 
we show  the estimated root mean squared error (RMSE), using 
the posterior means as  point estimator. 
In parenthesis, 
the jackknife estimate of Monte Carlo Error (MCE) for the RMSE is presented. 
Next, the average of the posterior standard deviations (SD) is shown with  the empirical standard deviation (ESD) of the posterior mean in the parenthesis.
Third, the empirical coverage (Cov) of 95\% credible interval is given. 
\item For $\pi$, we show the average of $L_1$ norm between the posterior mean and $\bm \pi^*$ (L1-error). 
In the parenthesis, we show its jackknife estimate of MCE. 
The SDs, ESDs, and Covs are averaged over the 15 elements in $\pi$. 

\item Time is the average seconds taken for sampling 1,000 MCMC draws in Matlab on a desktop with a 4.9GHz processor and 64GB RAM. 
The study is based on $R=200$ replications. 
2,000 MCMC draws are obtained for each replication. The average of the inefficiency factors is around 6.4 with standard deviation 1.1.
\end{tablenotes}
\end{threeparttable}
}
\label{tab:simulation_SmallJ_time_varying_cs}
\end{table}

\subsection{Additional simulation results for $J=100$}

\begin{figure}[h!] 
\centering
\begin{subfigure}[b]{0.31\linewidth}
\includegraphics[width=\linewidth]{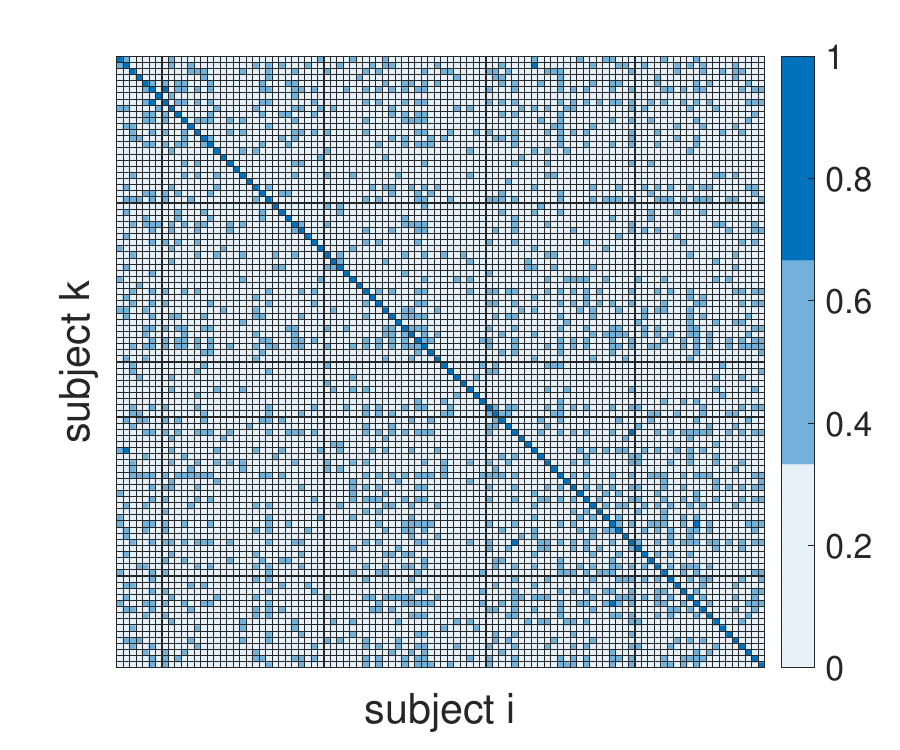}  
\caption{\small{ $T=1$ }}
\end{subfigure}
\begin{subfigure}[b]{0.31\linewidth}
\includegraphics[width=\linewidth]{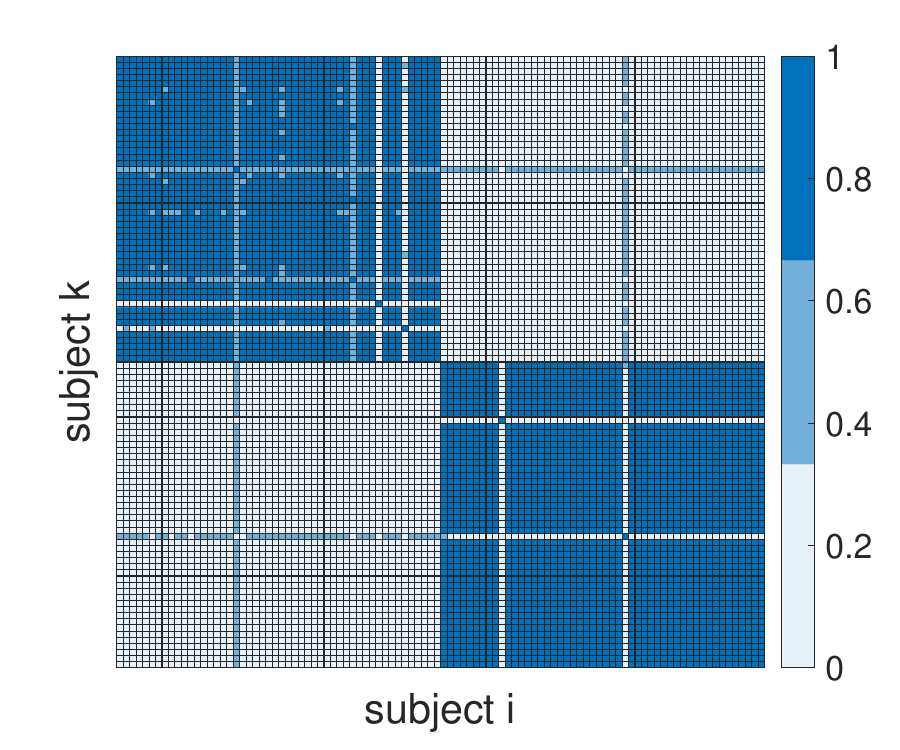}   
\caption{\small{ $T=5$ }}
\end{subfigure}
\begin{subfigure}[b]{0.31\linewidth}
\includegraphics[width=\linewidth]{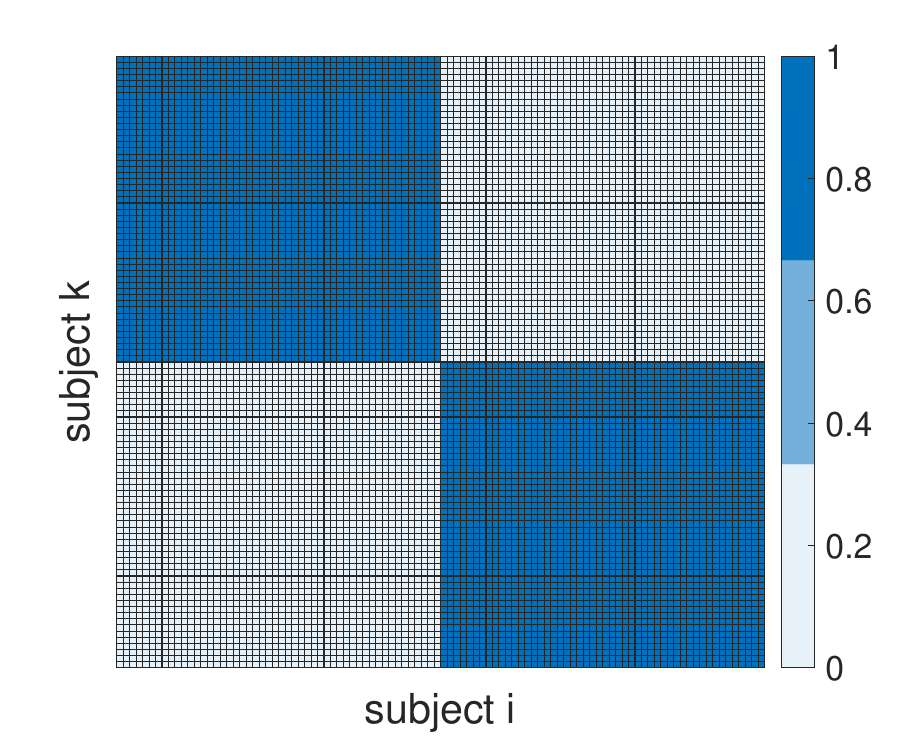}   
\caption{\small{ $T=200$ }}
\end{subfigure} 
\caption{\small{
Similarity matrices. 
The results are based on one replication of simulated data with $J=100$, $T=\{1,5,200\}$, and $n=100$. 
In the true clustering,  the first subpopulation contains the units $1$ to $50$ and the second contains the remaining units. 
}}
\label{fig:simulation_CS_largeJ}
\end{figure}
\FloatBarrier

In Section 5.2, we present the simulation results for the high dimensional case with $J=100$. 
Although it is not possible to show the results for  $\bm \pi$ due to the  $2^{100}-1$ support points, we illustrate its estimation result in Figure \ref{fig:simulation_CS_largeJ} that shows the $n \times n$ ``similarity matrices'' based on one replicate of the simulation data. These give the posterior probability that a given subject in a particular row $k$ is in the same cluster as another subject at a specific column $i$ which is computed as the posterior probability of the event $\{S_k = S_i \}$. This probability ranges from zero (light blue) to one (dark blue). 
By $T = 5$, the  similarity matrix roughly aligns with the true clustering structure, leading to accurate estimates of $\bm \pi$. For $T = 200$,  the clustering structure is recovered with high accuracy.

\section{Additional material for the application}\label{sec:additional_application}

\subsection{Data description}
We combine two sources of the data sets obtained from Nielsen, a store data and a purchase data, in order to prepare a panel data set. 
The preference and consideration patterns might have affected during the pandemic, so we chose the year 2019, which is the earliest year available before the pandemic. 
In the store data, we first choose a retailer, whose identity is not revealed in the Nielsen data, which consistently had over 100 cereal brands available at the majority of its stores. 
There are 239 stores under this retailer, operating mainly in the Midwest of the United States. 
See Figure 3 for the locations of the stores and percentages of the purchases. 
The store data contains product information  at UPC (universal product code) level such as price and size (ounce). 
A `brand' can consist of multiple UPCs. 
Brand-level prices are defined as size-weighted averages of UPC prices. 
We first pick the top 135 cereal brands in terms of the availability at these stores, which are responsible for over 90 percentages of the purchases in the purchase data at these stores in 2019. 
In more than $95\%$ of the store-week combinations, the price information of the 135 brands is available, but if it is missing, 
we impute the value with the average of the prices of the same brand at the other stores in the same week. 
We then defined the top 100 to be the inside options and the rest to be the `other' option. 
In the purchase data, we removed the households who made less than 3 units of cereal. 
When households purchased multiple units of cereal at one shopping trip, we treat them as separate purchases. 
This leaves us a sample with 
$J=101$ brands (see Appendix for a complete list of the brand names). The data contains 
$n=1880$ households, 
25849 purchases at 239 stores of the same retailer throughout 52 weeks.  


\subsection{Hyperparameters}
We set the hyper prior parameters as follows: a sparsity-supporting prior for the attention probabilities 
$q_{hj}\sim \text{Beta}(\underline{a}_{q_j},\underline{b}_{q_j})$, independently over $j=1,\ldots,J$ for $h=1,\ldots,\infty$, with 
$(\underline{a}_{q_j},\underline{b}_{q_j})=(s \cdot r, s \cdot (1-r))$, $r=\frac{r_0}{J}$ with $s=5$ and $r_0=30$, which implies that   the prior mean of $q_{hj}$ is about 0.44.
For the DP concentration parameter, 
$\alpha \sim \text{Gamma}(1/4,2)$.
The priors for $\bm \delta$ and $\beta$ are independent  normal distributions with zero mean and variance $3$. 
The prior for $D$ is an inverse-Wishart distribution with hyper-parameters $(\underline{v}, \underline{ R})=(9,(1/9)I)$.

\subsection{Additional estimation results and discussion}


\subsubsection{Estimated parameters in the response model}
\paragraph{Brand-specific fixed-effects.}
The estimated brand-specific fixed-effects are shown in Table 4 of the paper. 
The number of brand-specific fixed-effects whose 95\% credible intervals do not include 0 is larger for MNL than MNL\_C and for MNL\_R than MNL\_RC. 
This phenomenon was also observed by \cite{ChiangChibNarasimhan1998}. To explain this, we note that under MNL\_C and MNL\_RC, the estimated consideration sets $\{\mathcal{C}_i\}$ are much smaller than the set with all brands. If, for example, there is a brand that is almost never chosen by any household, the estimated $\{\mathcal{C}_i\}$ tends to exclude such a brand. 
The standard logit model does not account for such nonconsideration and instead assumes that every household considers all brands. As a result, the magnitudes (absolute value) of brand-specific fixed effects tend to be overestimated. 
Under the full specification, for 67 out of 100 of them, the corresponding 95\% credible interval does not include 0. Note that we fixed $\delta_J=0$ for identification.

\subsubsection{Estimated parameters in the mixture model}
\begin{figure}[ht] 
\centering
\includegraphics[width=0.5\linewidth]{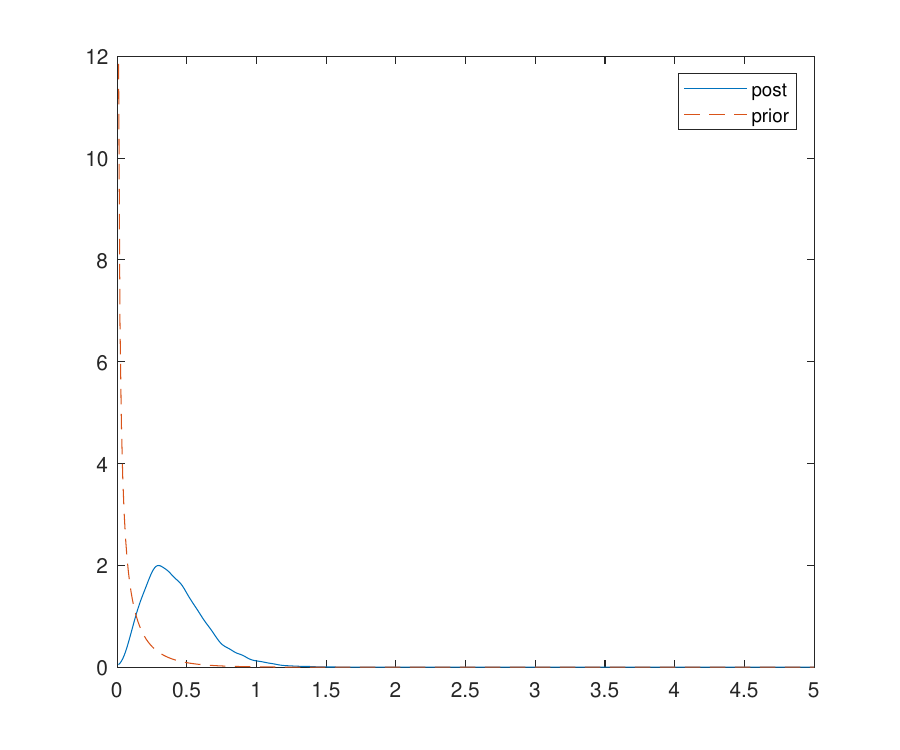} 
\caption{\small{pdf's of $\alpha$: prior (dashed) and posterior (solid) from the empirical application. MNL\_RC.}}
\label{fig:application_DPconcentration}
\end{figure}

\begin{figure}[ht] 
\centering
    \includegraphics[width=0.5\linewidth]{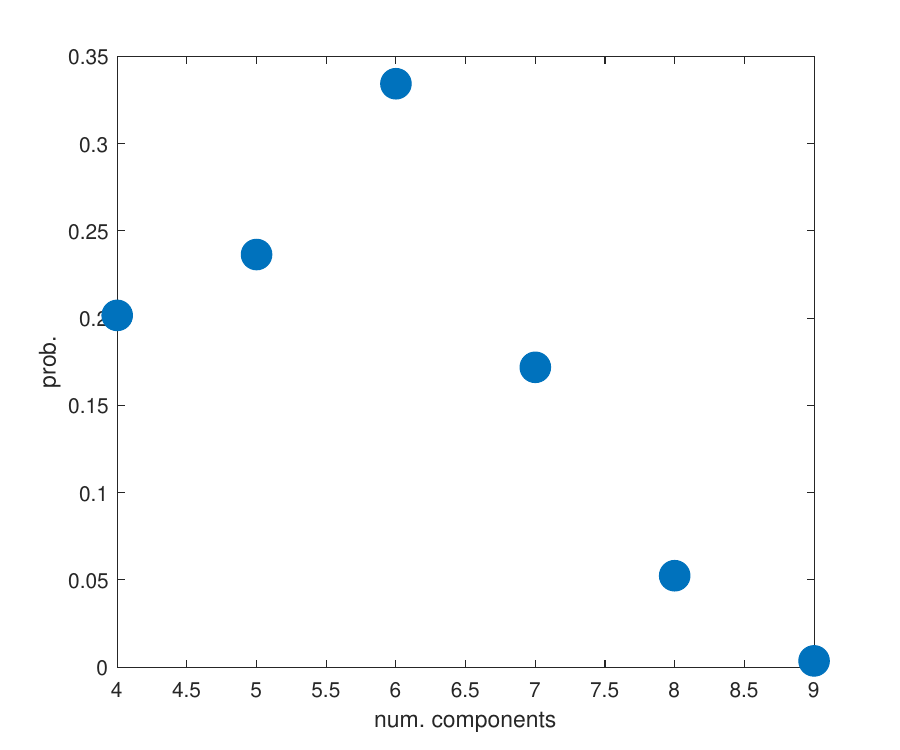} 
\caption{\small{
Posterior probability mass function of the number of nonempty components from the empirical application. MNL\_RC.
}}
\label{fig:application_NumComp_pmf}
\end{figure}

In the following, we present additional estimation results concerning the parameters in the mixture model under the MNL\_RC specification. 
Figure \ref{fig:application_DPconcentration} compares the prior and posterior densities of the DP concentration parameter $\alpha$. 
The vague prior density  $\alpha \sim Gamma (1/4,1/4 )$, suggested by \cite{DunsonXing2009}, 
 is shown as the dashed line and the posterior 
as the solid line. 
\FloatBarrier
\graphicspath{{Figures/Application_Nielsen/}}
Figure \ref{fig:application_NumComp_pmf} shows 
the posterior probability mass function of the non-empty mixture components. 
The posterior mode of the number of non-empty components is six.

\FloatBarrier

\begin{figure}[ht] 
\centering
\includegraphics[width=0.55\linewidth]{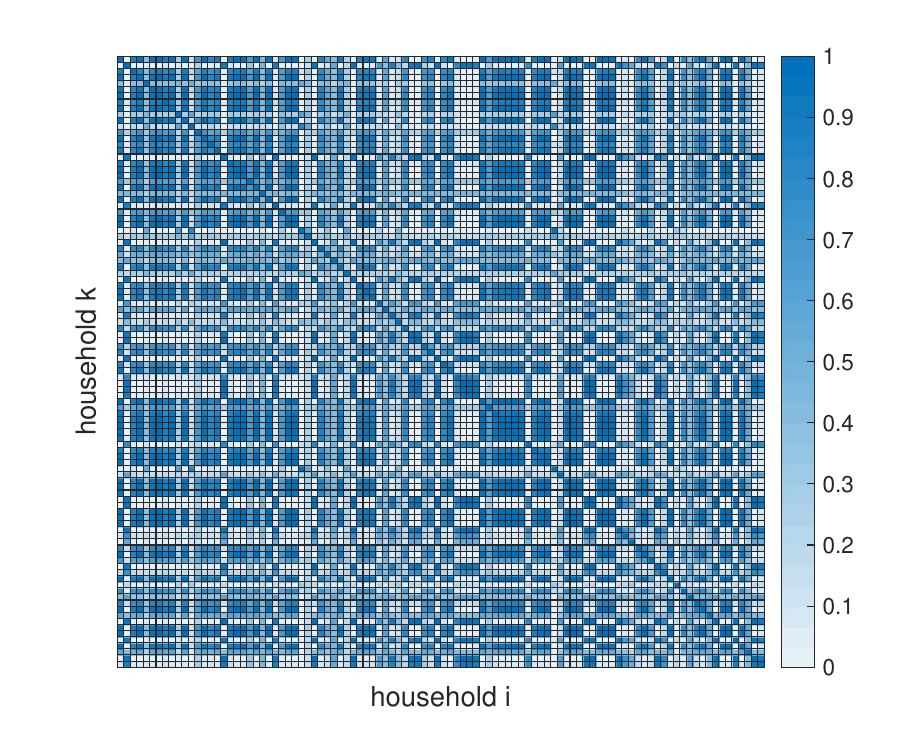} 
\caption{\small{The similarity matrix of a sample of 100 households (out of 1880).}}
\label{fig:application_SimilarityMatrix}
\end{figure}

The similarity matrix is shown in Figure \ref{fig:application_SimilarityMatrix}. 
Each entry of the matrix shows the posterior probability that 
a given pair of households $(k,i)$ are clustered together i.e.\ $S_k = S_i$, 
ranging from zero (light blue) to one (dark blue). 

\subsubsection{Additional material on the prediction}

\begin{figure}[ht] 
\centering
\includegraphics[width=0.85\linewidth]{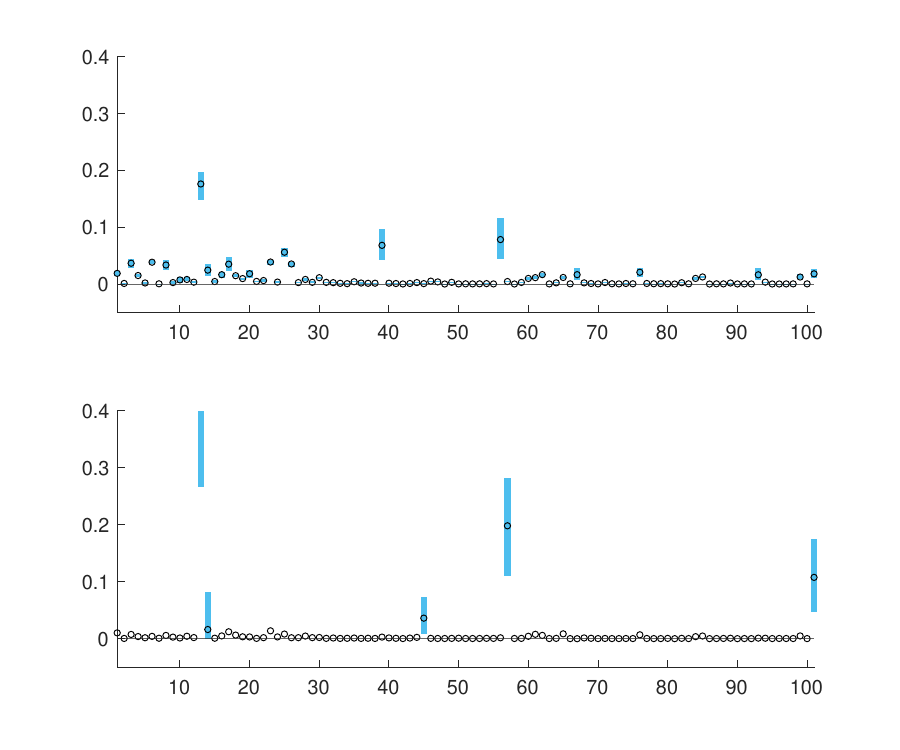} 
\caption{\small{Estimated predictive response probabilities $\Pr(Y_{i,T_{i}+1}=j)$ for $i=3$. 90\% credible intervals (bars) and means (circles). The horizontal axis represents the brands $j\in \{1,\ldots,101 \}$. The actual out-of-sample purchase was brand 13.}}
\label{fig:application_PredCCP}
\end{figure}

To investigate why MNL\_RC outperforms MNL\_R in prediction, we compare the  predictive response probabilities between the two models. 
For each $i \in \mathcal{O}$, we can compute the marginal posterior of $\Pr(Y_{iT_i+s}=j)$, for each alternative $j$ and forecasting horizon $s=1,\ldots,h_i$. 
Figure \ref{fig:application_PredCCP} presents the estimated response probabilities for the household $i=3$ in the first out-of-sample period, $s=1.$
This household repeatedly purchased brands 13, 45, 57, and 101 in the estimation sample: $\{  45,    13,    13,   101,    13,    13 ,   13,    57 ,   57 ,   57,    57\}$, in the order of the purchases. In the first out-of-sample week, the household purchased brand 13.
The figure shows  the 90\% credible intervals (vertical bars) as well as the mean of the estimated response probabilities (circles). 
Clearly, the estimated response probabilities are much sparser for MNL\_RC (lower panel) than MNL\_R (upper). 
The traditional MNL approach necessarily implies a positive probability for every alternative. 
In contrast, the consideration set model allows many alternatives to actually receive zero predictive probabilities. 
Thus, incorporating consideration set heterogeneity can improve predictive performance due to the sparsity in the predictive response probabilities when the time-invariant consideration set assumption is appropriate, which seems to be the case in this data set. 

\FloatBarrier

\subsubsection{Estimated consideration dependence}
\begin{figure*}[ht]
    \centering
    \begin{subfigure}[t]{0.5\textwidth}
        \centering
        \includegraphics[scale=0.5]{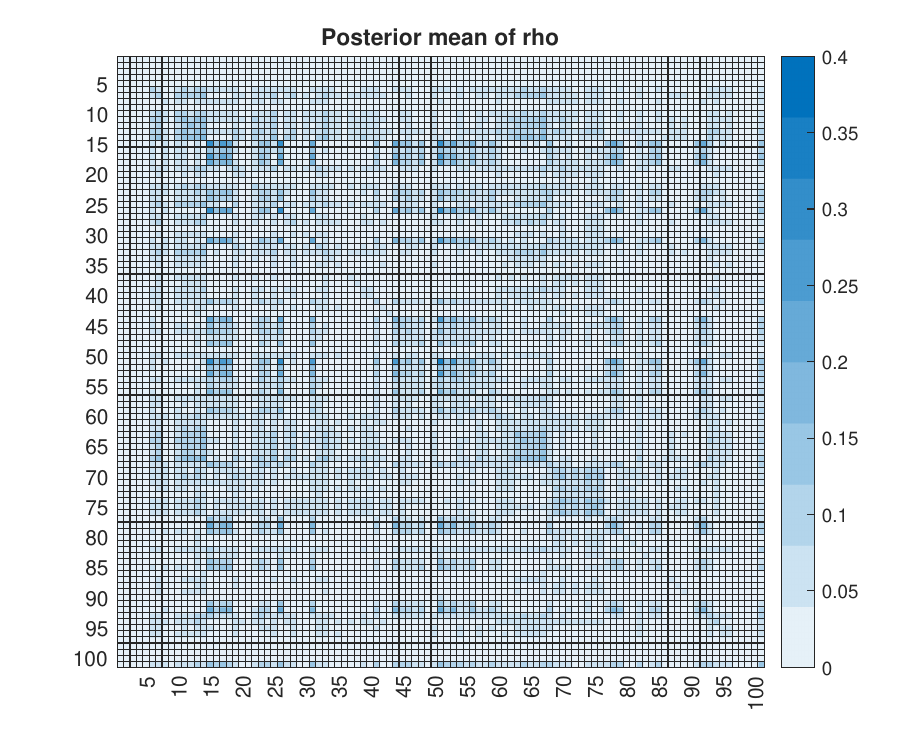}
        \caption{Posterior means of $\rho_{j,\ell}$}
        \label{fig:application_HeatMap_rho_postmean}
    \end{subfigure}%
    ~
    \begin{subfigure}[t]{0.5\textwidth}
        \centering
        \includegraphics[scale=0.5]{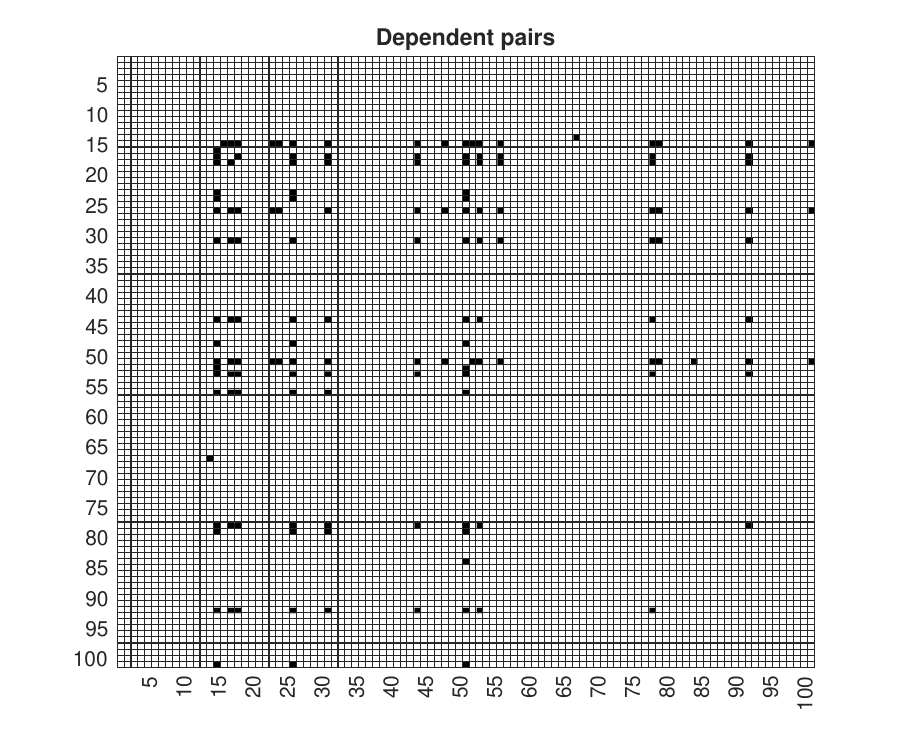}
        \caption{Dependent brands}
        \label{fig:application_HeatMap_deppairs}        
    \end{subfigure}     
    \caption{Consideration dependence in the 2019 Midwest cereal consumption data.}
\end{figure*}

We conduct the test for independent consideration introduced in Section 6. 
The estimated posterior probability of the alternative hypothesis is very close to one i.e. $\Pr(H_1 \vert \bm D^n)\approx 1$, and we conclude that the considerations of cereal products in this particular market are dependent. 
Furthermore, to investigate brand pair-level dependence, consider a hypothetical consumer $i$ whose consideration set is drawn from the true unknown distribution. 
Define the marginal probability that brand $j$ is (and not) considered: $\pi^{(j)}_1=\Pr(C_{ij}=1)$ and $\pi^{(j)}_0=\Pr(C_{ij}=0)$. 
Also define the  probability that a pair of brands $(j,\ell)$ is considered jointly as 
$\pi^{(j,\ell)}_{11}=\Pr(C_{ij}=1 \text{ and } C_{i\ell}=1)$, and similarly define the probabilities for the remaining three cases: 
$\pi^{(j,\ell)}_{01}=\Pr(C_{ij}=0 \text{ and } C_{i\ell}=1)$,
$\pi^{(j,\ell)}_{10}=\Pr(C_{ij}=1 \text{ and } C_{i\ell}=0)$, and $\pi^{(j,\ell)}_{00}=\Pr(C_{ij}=0 \text{ and } C_{i\ell}=0)$. 
We employ the model-based Cramer's V statistics 
as a measure of consideration dependence between brands $j$ and $\ell$ as:
$
\rho^2_{j,\ell}=\sum_{s=0}^1 \sum_{m=0}^1 \left( \pi^{(j,\ell)}_{(s,m)} - \pi^{(j)}_{(s)} \pi^{(\ell)}_{(m)} \right)^2 / \pi^{(j)}_{(s)} \pi^{(\ell)}_{(m)},
$
which ranges from 0 to 1, and $\rho^2_{j,\ell}\approx 0$ indicates that the consideration of the two brands $(j,\ell)$ is nearly independent. 
These probabilities are approximated as functions of the model parameters, for example, 
$\pi^{(j)}_1=\sum_{h=1}^{k^*} \omega_h q_{hj}$, 
$\pi^{(j)}_0=\sum_{h=1}^{k^*} \omega_h (1-q_{hj})$,
and 
$\pi^{(j,\ell)}_{10}=\sum_{h=1}^{k^*} \omega_h q_{hj}(1-q_{h\ell})$, and so on.
Figure \ref{fig:application_HeatMap_rho_postmean} shows the posterior means of $\{\rho_{j,\ell}\}$. 
Figure \ref{fig:application_HeatMap_deppairs} shows the brand pairs $(j,\ell)$ for which the posterior probability that $\rho_{j,\ell}>0.1$ is greater than 0.95.
Based on this criteria, we identified 72 brand pairs (shown in black). 

\end{document}